

\input gr-qc
%
%
%
\newdimen\rotdimen
\def\vspec#1{\special{ps:#1}}
\def\rotstart#1{\vspec{gsave currentpoint currentpoint translate
   #1 neg exch neg exch translate}}
\def\rotfinish{\vspec{currentpoint grestore moveto}}
%
%
\def\rotr#1{\rotdimen=\ht#1\advance\rotdimen by\dp#1%
   \hbox to\rotdimen{\hskip\ht#1\vbox to\wd#1{\rotstart{90 rotate}%
   \box#1\vss}\hss}\rotfinish}
%
%
\def\rotl#1{\rotdimen=\ht#1\advance\rotdimen by\dp#1%
   \hbox to\rotdimen{\vbox to\wd#1{\vskip\wd#1\rotstart{270 rotate}%
   \box#1\vss}\hss}\rotfinish}%
%
%
\def\rotu#1{\rotdimen=\ht#1\advance\rotdimen by\dp#1%
   \hbox to\wd#1{\hskip\wd#1\vbox to\rotdimen{\vskip\rotdimen
   \rotstart{-1 dup scale}\box#1\vss}\hss}\rotfinish}%
%
%
\def\rotf#1{\hbox to\wd#1{\hskip\wd#1\rotstart{-1 1 scale}%
   \box#1\hss}\rotfinish}%
%
%

 \ifx\MYUNDEFINED\BoxedEPSF
   \let\temp\relax
 \else
   \message{}
   \message{ !!! BoxedEPS %
         or BoxedArt macros already defined !!!}
   \let\temp 
 \fi
  \temp
 
 \chardef\CatAt\the\catcode`\@
 \catcode`\@=11
 \chardef\C@tColon\the\catcode`\:
 \chardef\C@tSemicolon\the\catcode`\;
 \chardef\C@tQmark\the\catcode`\?
 \chardef\C@tEmark\the\catcode`\!

 \def\PunctOther@{\catcode`\:=12
   \catcode`\;=12 \catcode`\?=12 \catcode`\!=12}
 \PunctOther@

 \let\wlog@ld\wlog 
 \def\wlog#1{\relax} 

 \newif\ifIN@
 \newdimen\XShift@ \newdimen\YShift@ 
 \newtoks\Realtoks
 
  %
 \newdimen\Wd@ \newdimen\Ht@
 \newdimen\Wd@@ \newdimen\Ht@@
 \newdimen\TT@
 \newdimen\LT@
 \newdimen\BT@
 \newdimen\RT@
 \newdimen\XSlide@ \newdimen\YSlide@ 
 \newdimen\TheScale  
 \newdimen\FigScale  
 \newdimen\ForcedDim@@

 \newtoks\EPSFDirectorytoks@
 \newtoks\EPSFNametoks@
 \newtoks\BdBoxtoks@
 \newtoks\LLXtoks@  
 \newtoks\LLYtoks@

 \newif\ifNotIn@
 \newif\ifForcedDim@
 \newif\ifForceOn@
 \newif\ifForcedHeight@
 \newif\ifPSOrigin

 \newread\EPSFile@ 
 
  \def\ms@g{\immediate\write16}

 \newif\ifIN@\def\IN@{\expandafter\INN@\expandafter}
  \long\def\INN@0#1@#2@{\long\def\NI@##1#1##2##3\ENDNI@
    {\ifx\m@rker##2\IN@false\else\IN@true\fi}%
     \expandafter\NI@#2@@#1\m@rker\ENDNI@}
  \def\m@rker{\m@@rker}

  \newtoks\Initialtoks@  \newtoks\Terminaltoks@
  \def\SPLIT@{\expandafter\SPLITT@\expandafter}
  \def\SPLITT@0#1@#2@{\def\TTILPS@##1#1##2@{%
     \Initialtoks@{##1}\Terminaltoks@{##2}}\expandafter\TTILPS@#2@}


  \newtoks\Trimtoks@

 \def\ForeTrim@{\expandafter\ForeTrim@@\expandafter}
 \def\ForePrim@0 #1@{\Trimtoks@{#1}}
 \def\ForeTrim@@0#1@{\IN@0\m@rker. @\m@rker.#1@%
     \ifIN@\ForePrim@0#1@%
     \else\Trimtoks@\expandafter{#1}\fi}

  \def\Trim@0#1@{%
      \ForeTrim@0#1@%
      \IN@0 @\the\Trimtoks@ @%
        \ifIN@ 
             \SPLIT@0 @\the\Trimtoks@ @\Trimtoks@\Initialtoks@
             \IN@0\the\Terminaltoks@ @ @%
                 \ifIN@
                 \else \Trimtoks@ {FigNameWithSpace}%
                 \fi
        \fi
      }


   \newtoks\pt@ks
   \def \getpt@ks 0.0#1@{\pt@ks{#1}}
   \dimen0=0pt\expandafter\getpt@ks\the\dimen0@

  \newtoks\Realtoks
  \def\Real#1{%
    \dimen2=#1%
      \SPLIT@0\the\pt@ks @\the\dimen2@
       \Realtoks=\Initialtoks@
            }

   \newdimen\Product
   \def\Mult#1#2{%
     \dimen4=#1\relax
     \dimen6=#2%
     \Real{\dimen4}%
     \Product=\the\Realtoks\dimen6%
        }

 \newdimen\Inverse
 \newdimen\hmxdim@ \hmxdim@=8192pt
 \def\Invert#1{%
  \Inverse=\hmxdim@
  \dimen0=#1%
  \divide\Inverse \dimen0%
  \multiply\Inverse 8}

   \def\Rescale#1#2#3{
              \divide #1 by 100\relax
              \dimen2=#3\divide\dimen2 by 100 \Invert{\dimen2}%
              \Mult{#1}{#2}%
              \Mult\Product\Inverse 
              #1=\Product}

  \def\Scale#1{\dimen0=\TheScale %
      \divide #1 by  1280 
      \divide \dimen0 by 5120 %
      \multiply#1 by \dimen0 
      \divide#1 by 10   
     }
 

 \newbox\scrunchbox

 \def\Scrunched#1{{\setbox\scrunchbox\hbox{#1}%
   \wd\scrunchbox=0pt
   \ht\scrunchbox=0pt
   \dp\scrunchbox=0pt
   \box\scrunchbox}}

 \def\Shifted@#1{%
   \vbox {\kern-\YShift@
       \hbox {\kern\XShift@\hbox{#1}\kern-\XShift@}%
           \kern\YShift@}}


 \def\cBoxedEPSF#1{{}\leavevmode 
   \ReadNameAndScale@{#1}%
   \SetEPSFSpec@
   \ReadEPSFile@ \ReadBdB@x  
     \TrimFigDims@ 
     \CalculateFigScale@  
     \ScaleFigDims@
     \SetInkShift@
   \hbox{$\mathsurround=0pt\relax
         \vcenter{\hbox{%
             \FrameSpider{\hskip-.4pt\vrule}%
             \vbox to \Ht@{\offinterlineskip\parindent=\z@%
                \FrameSpider{\vskip-.4pt\hrule}\vfil 
                \hbox to \Wd@{\hfil}%
                \vfil
                \InkShift@{\EPSFSpecial{\EPSFSpec@}{\FigSc@leReal}}%
             \FrameSpider{\hrule\vskip-.4pt}}%
         \FrameSpider{\vrule\hskip-.4pt}}}%
     $}%
    \CleanRegisters@ 
    \ms@g{ *** Box composed for the %
         EPSF file \the\EPSFNametoks@}%
    }
 
 \def\tBoxedEPSF#1{\setbox4\hbox{\cBoxedEPSF{#1}}%
     \setbox4\hbox{\raise -\ht4 \hbox{\box4}}%
     \box4
      }

 \def\bBoxedEPSF#1{\setbox4\hbox{\cBoxedEPSF{#1}}%
     \setbox4\hbox{\raise \dp4 \hbox{\box4}}%
     \box4
      }

  \let\BoxedEPSF\cBoxedEPSF

   %

   %
  \def\gLinefigure[#1scaled#2]_#3{%
        \BoxedEPSF{#3 scaled #2}}
    
   %

  \def\EPSFxsize{\afterassignment\ForceW@\ForcedDim@@}
      \def\ForceW@{\ForcedDim@true\ForcedHeight@false}
  
  \def\EPSFysize{\afterassignment\ForceH@\ForcedDim@@}
      \def\ForceH@{\ForcedDim@true\ForcedHeight@true}

  %
 \def\ReadNameAndScale@#1{\IN@0 scaled@#1@
   \ifIN@\ReadNameAndScale@@0#1@%
   \else \ReadNameAndScale@@0#1 scaled\DefaultMilScale @
   \fi}
  
 \def\ReadNameAndScale@@0#1scaled#2@{
    \let\OldBackslash@\\%
    \def\\{\OtherB@ckslash}%
    \edef\temp@{#1}%
    \Trim@0\temp@ @%
    \EPSFNametoks@\expandafter{\the\Trimtoks@ }%
    \FigScale=#2 pt%
    \let\\\OldBackslash@
    }
 
 \def\SetDefaultEPSFScale#1{%
      \global\def\DefaultMilScale{#1}}

 \SetDefaultEPSFScale{1000}

  %
 \def \SetBogusBbox@{%
     \global\BdBoxtoks@{ BoundingBox:0 0 100 100 }%
     \global\def\BdBoxLine@{ BoundingBox:0 0 100 100 }%
     \ms@g{ !!! Will use placeholder !!!}%
     }

 \def\ReadEPSFile@{
     \openin\EPSFile@\EPSFSpec@
     \relax  
  \ifeof\EPSFile@
     \ms@g{}%
     \ms@g{ !!! EPS FILE \the\EPSFDirectorytoks@
       \the\EPSFNametoks@\ WAS NOT FOUND !!!}
     \SetBogusBbox@
  \else
   \begingroup
   \catcode`\%=12\catcode`\:=12\catcode`\!=12
   \catcode`\G=14\catcode`\\=14\relax
   \global\read\EPSFile@ to \BdBoxLine@
   \IN@0!PS@\BdBoxLine@ @%
   \ifIN@
     \NotIn@true 
     \loop   
       \ifeof\EPSFile@\NotIn@false 
         \ms@g{}%
         \ms@g{ !!! BoundingBox NOT FOUND IN %
            \the\EPSFDirectorytoks@\the\EPSFNametoks@\ !!! }%
         \SetBogusBbox@
       \else\global\read\EPSFile@ to \BdBoxLine@
       \fi
       \global\BdBoxtoks@\expandafter{\BdBoxLine@}%
       \IN@0BoundingBox:@\the\BdBoxtoks@ @%
       \ifIN@\NotIn@false\fi%
     \ifNotIn@\repeat
   \else
         \ms@g{}%
         \ms@g{ !!! \the\EPSFNametoks@\ not PS!\  !!!}%
         \SetBogusBbox@
   \fi
  \endgroup\relax
  \fi
  \closein\EPSFile@ 
   }

  \def\ReadBdB@x{
   \expandafter\ReadBdB@x@\the\BdBoxtoks@ @}
  
  \def\ReadBdB@x@#1BoundingBox:#2@{
    \ForeTrim@0#2@%
    \IN@0atend@\the\Trimtoks@ @
       \ifIN@\Trimtoks@={0 0 100 100 }
         \ms@g{}%
         \ms@g{ !!! BoundingBox not found in %
         \the\EPSFDirectorytoks@\the\EPSFNametoks@\space !!!}%
         \ms@g{ !!! It must not be at end of EPSF !!!}%
         \ms@g{ !!! Will use placeholder !!!}%
       \fi
    \expandafter\ReadBdB@x@@\the\Trimtoks@ @%
   }
    
  \def\ReadBdB@x@@#1 #2 #3 #4@{
      \Wd@=#3bp\advance\Wd@ by -#1bp%
      \Ht@=#4bp\advance\Ht@ by-#2bp%
       \Wd@@=\Wd@ \Ht@@=\Ht@ 
       \LLXtoks@={#1}\LLYtoks@={#2}
      \ifPSOrigin\XShift@=-#1bp\YShift@=-#2bp\fi 
     }

   %
   \def\G@bbl@#1{}
   \bgroup
     \global\edef\OtherB@ckslash{\expandafter\G@bbl@\string\\}
   \egroup

  \def\SetEPSFDirectory{
           \bgroup\PunctOther@\relax
           \let\\\OtherB@ckslash
           \SetEPSFDirectory@}

 \def\SetEPSFDirectory@#1{
    \edef\temp@{#1}%
    \Trim@0\temp@ @
    \global\toks1\expandafter{\the\Trimtoks@ }\relax
    \egroup
    \EPSFDirectorytoks@=\toks1
    }

 \def\SetEPSFSpec@{%
     \bgroup
     \let\\=\OtherB@ckslash
     \global\edef\EPSFSpec@{%
        \the\EPSFDirectorytoks@\the\EPSFNametoks@}%
     \global\edef\EPSFSpec@{\EPSFSpec@}%
     \egroup}

  %
 \def\TrimTop#1{\advance\TT@ by #1}
 \def\TrimLeft#1{\advance\LT@ by #1}
 \def\TrimBottom#1{\advance\BT@ by #1}
 \def\TrimRight#1{\advance\RT@ by #1}

 \def\TrimFigDims@{%
    \advance\Wd@ by -\LT@ 
    \advance\Wd@ by -\RT@ \RT@=\z@
    \advance\Ht@ by -\TT@ \TT@=\z@
    \advance\Ht@ by -\BT@ 
    }

  %
  \def\ForceWidth#1{\ForcedDim@true
       \ForcedDim@@#1\ForcedHeight@false}
  
  \def\ForceHeight#1{\ForcedDim@true
       \ForcedDim@@=#1\ForcedHeight@true}

  \def\ForceOn{\ForceOn@true}
  \def\ForceOff{\ForceOn@false\ForcedDim@false}
  
  \def\epsfxsize{\afterassignment\ForceW@\ForcedDim@@}
      \def\ForceW@{\ForcedDim@true\ForcedHeight@false}
  
  \def\epsfysize{\afterassignment\ForceH@\ForcedDim@@}
      \def\ForceH@{\ForcedDim@true\ForcedHeight@true}
  
  \def\CalculateFigScale@{%
     \ifForcedDim@\FigScale=1000pt
           \ifForcedHeight@
                \Rescale\FigScale\ForcedDim@@\Ht@
           \else
                \Rescale\FigScale\ForcedDim@@\Wd@
           \fi
     \fi
     \Real{\FigScale}%
     \edef\FigSc@leReal{\the\Realtoks}%
     }
   
  \def\ScaleFigDims@{\TheScale=\FigScale
      \ifForcedDim@
           \ifForcedHeight@ \Ht@=\ForcedDim@@  \Scale\Wd@
           \else \Wd@=\ForcedDim@@ \Scale\Ht@
           \fi
      \else \Scale\Wd@\Scale\Ht@        
      \fi
      \ifForceOn@\relax\else\global\ForcedDim@false\fi
      \Scale\LT@\Scale\BT@  
      \Scale\XShift@\Scale\YShift@
      }
      

 \let\HideDisplacementBoxes\HideReservedBoxes  
 \let\ShowDisplacementBoxes\ShowReservedBoxes

  \ShowDisplacementBoxes
 
 \def\hSlide#1{\advance\XSlide@ by #1}
 \def\vSlide#1{\advance\YSlide@ by #1}
 
  \def\SetInkShift@{%
            \advance\XShift@ by -\LT@
            \advance\XShift@ by \XSlide@
            \advance\YShift@ by -\BT@
            \advance\YShift@ by -\YSlide@
             }
  \def\InkShift@#1{\Shifted@{\Scrunched{#1}}}
 
   %
  \def\CleanRegisters@{%
      \globaldefs=1\relax
        \XShift@=\z@\YShift@=\z@\XSlide@=\z@\YSlide@=\z@
        \TT@=\z@\LT@=\z@\BT@=\z@\RT@=\z@
      \globaldefs=0\relax}

 
 \def\SetTexturesEPSFSpecial{\PSOriginfalse
  \gdef\EPSFSpecial##1##2{\relax
    \edef\specialthis{##2}%
    \SPLIT@0.@\specialthis.@\relax
    \special{illustration ##1 scaled
                        \the\Initialtoks@}}}
 
  \def\SetUnixCoopEPSFSpecial{\PSOrigintrue 
   \gdef\EPSFSpecial##1##2{%
      \dimen4=##2pt
      \divide\dimen4 by 1000\relax
      \Real{\dimen4}
      \edef\Aux@{\the\Realtoks}%
      \includegraphics{##1\space}}}

  \def\SetBechtolsheimRokickiEPSFSpecial{\PSOrigintrue 
   \gdef\EPSFSpecial##1##2{%
      \dimen4=##2pt
      \divide\dimen4 by 1000\relax
      \Real{\dimen4}
      \edef\Aux@{\the\Realtoks}%
      \special{ps: psfiginit}%
      \special{ps: literal 1 1 0 0 1 1 startTexFig
           \the\mag\space 1000 div \Aux@\space mul 
           \the\mag\space 1000 div \Aux@\space mul scale}%
      \special{ps: include  ##1}%
      \special{ps: literal endTexFig}%
        }}

  \def\SetLisEPSFSpecial{\PSOrigintrue 
   \gdef\EPSFSpecial##1##2{%
      \dimen4=##2pt
      \divide\dimen4 by 1000\relax
      \Real{\dimen4}
      \edef\Aux@{\the\Realtoks}%
      \special{pstext="1 1 0 0 1 1 startTexFig\space
           \the\mag\space 1000 div \Aux@\space mul 
           \the\mag\space 1000 div \Aux@\space mul scale}%
      \includegraphics{##1}%
      \special{pstext=endTexFig}%
        }}

  \def\SetRokickiEPSFSpecial{\PSOrigintrue 
   \gdef\EPSFSpecial##1##2{%
      \dimen4=##2pt
      \divide\dimen4 by 10\relax
      \Real{\dimen4}
      \edef\Aux@{\the\Realtoks}%
      \includegraphics{##1}}}

  \def\SetInlineRokickiEPSFSpecial{\PSOrigintrue 
   \gdef\EPSFSpecial##1##2{%
      \dimen4=##2pt
      \divide\dimen4 by 1000\relax
      \Real{\dimen4}
      \edef\Aux@{\the\Realtoks}%
      \special{ps::[begin] 1 1 0 0 1 1 startTexFig\space
           \the\mag\space 1000 div \Aux@\space mul 
           \the\mag\space 1000 div \Aux@\space mul scale}%
      \special{ps: plotfile ##1}%
      \special{ps::[end] endTexFig}%
        }}

  \def\SetOzTeXEPSFSpecial{\PSOriginfalse 
  \gdef\EPSFSpecial##1##2{
     \special{##1\space 
       ##2 1000 div \the\mag\space 1000 div mul
       ##2 1000 div \the\mag\space 1000 div mul scale
       \the\LLXtoks@\space neg \the\LLYtoks@\space neg translate
             }}} 
 

 \def\SetArborEPSFSpecial{\PSOriginfalse 
   \gdef\EPSFSpecial##1##2{%
     \edef\specialthis{##2}%
     \SPLIT@0.@\specialthis.@\relax 
     \special{ps: epsfile ##1\space \the\Initialtoks@}}}

 \def\SetClarkEPSFSpecial{\PSOriginfalse 
   \gdef\EPSFSpecial##1##2{%
     \Rescale {\Wd@@}{##2pt}{1000pt}%
     \Rescale {\Ht@@}{##2pt}{1000pt}%
     \special{dvitops: import 
           ##1\space\the\Wd@@\space\the\Ht@@}}}



 \def\SetStandardEPSFSpecial{%
   \gdef\EPSFSpecial##1##2{%
     \ms@g{}
     \ms@g{%
       !!! Sorry! There is still no standard for \string%
       \special\ EPSF integration !!!}%
     \ms@g{%
      --- So you will have to identify your driver using a command}%
     \ms@g{%
      --- of the form \string\Set...EPSFSpecial, in order to get}%
     \ms@g{%
      --- your graphics to print.  See BoxedEPS.doc.}%
     \ms@g{}
     \KillEPSFSpecial
     }}

  \def\KillEPSFSpecial{\gdef\EPSFSpecial##1##2{}}

  \SetStandardEPSFSpecial 
 
 \let\wlog\wlog@ld 

 \catcode`\:=\C@tColon
 \catcode`\;=\C@tSemicolon
 \catcode`\?=\C@tQmark
 \catcode`\!=\C@tEmark

 \catcode`\@=\CatAt

 %
 %
 %
 %
 %
\HideDisplacementBoxes
\SetRokickiEPSFSpecial
\def\stomp#1{\setbox0=\hbox{#1}\dp0=0pt\ht0=0pt\box0}
\newdimen\stepx
\newdimen\stepy
\newdimen\basex
\newdimen\basey
\def\SetBase(#1,#2){\basex=#1pt\basey=#2pt\relax}
%
%
\def\at(#1,#2)#3{\vbox to 0pt{\kern#2cm%
                 \hbox to 0pt{\kern#1cm\stomp{#3}\hss}\vss}\nointerlineskip}
\def\atpt(#1,#2)#3{\vbox to 0pt{\kern#2pt%
                   \hbox to 0pt{\kern#1pt\stomp{#3}\hss}\vss}\nointerlineskip}
\def\atptalt(#1,#2)#3{\stepx=#1pt\advance\stepx-\basex\relax
                      \stepy=\basey\advance\stepy-#2pt\relax
                      \vbox to 0pt{\kern\stepy%
                      \hbox to 0pt{\kern\stepx\stomp{#3}\hss}\vss}\nointerlineskip}

\def\SetNoBorder{\def\Border{}\HideDisplacementBoxes}
\SetNoBorder

\overfullrule=0pt

\ShortCitations

\openin\tmpfile=\jobname.itm
\ifeof\tmpfile\def\readtmp{\message{Can't find \jobname.itm}}\else
              \def\readtmp{\input \jobname.itm\relax}\fi
\closein\tmpfile
\readtmp\def\readtmp{\relax}

\def\Order(#1){{\cal O}(#1)}

\def\l{{\hbox{\tenpoint\tt l}}}
\def\o{{\hbox{\tenpoint\tt o}}}
\def\r{{\hbox{\tenpoint\tt r}}}
\def\p{{\hbox{\tenpoint\tt +}}}
\def\m{{\hbox{\tenpoint\tt -}}}

%
%
\def\Displaylines#1{\halign{\hbox to\displaywidth{%
$\hfil\displaystyle##\hfil $}&\llap{##}\crcr #1\crcr}}

%
%
\def \figRNCThroat {1}
\def \figRNCLadder {2}
\def \figRNCCell {3}

\def \figLxxRunCD {5}

\def \figHamCnstrntRunCD {11}
\def \figMomCnstrntRunCD {12}
\def \figLapseRunCD {13}
\def \figResolution {14}

\def \figHorizon {16}

\title{%
{Long term stable integration of a maximally sliced}\cr
{Schwarzschild black hole using a smooth lattice method}\cr}

\address{%
\cr
{\rm Leo Brewin}\cr
\cr
{\sl Department of Mathematics \& Statistics}\cr
{\sl Monash University}\cr
{\sl Clayton, Vic. 3800}\cr
{\sl Australia}\cr
{\rm 22-Jun-2001}\cr}

\beginabstract

We will present results of a numerical integration of a maximally sliced
Schwarzschild black hole using a smooth lattice method. The results show no
signs of any instability forming during the evolutions to $t=1000m$. The
principle features of our method are i) the use of a lattice to record the
geometry, ii) the use of local Riemann normal coordinates to apply the 1+1 ADM
equations to the lattice and iii) the use of the Bianchi identities to assist
in the computation of the curvatures. No other special techniques are used.
The evolution is unconstrained and the ADM equations are used in their
standard form.

PACS numbers: 04.25.Dm, 04.60.Nc, 02.70.-c

\endabstract

\beginsection{Introduction}

Recent studies
\setbox0=\hbox{%
\cite{AlcubierreEtalB}
\cite{AlcubierreEtalC}
\cite{Baumgarte}%
\cite{KellyEtal}%
\cite{ScheelB}%
\cite{Shibata}%
\cite{Yoneda}}
[\citeAlcubierreEtalB--\citeYoneda]
have shown that the stability of numerical integrations of
the Einstein field equations can depend on the formulation of the evolution
equations. Subtle changes in the structure of the evolution equations have been
shown to have a dramatic effect on the long term stability of the integrations.
These are relatively new investigations and thus at present there is no precise
mathematical explanation as to what is the root cause of the instabilities or
how best they can be avoided or minimised. What we have at present is a growing
set of examples which suggests that the standard ADM evolution equations may
not be the most suitable equations for numerical relativity.
Consequently many people are looking at alternative formulations such as the
hyperbolic formulations of Einstein's equations
\setbox0=\hbox{%
\cite{Reula}%
\cite{AlcubierreEtalD}%
\cite{ScheelA}}
[\citeReula--\citeScheelA]
and the conformal ADM equations of Shibita
and Nakamura \cite{Shibata} and Baumgarte and Shapiro \cite{Baumgarte}.

One alternative is the smooth lattice approach which we presented in two
earlier papers [\citeLeoSLGR,\citeLeoADMSLGR]. This is a method which uses a
lattice similar to that used in the Regge calculus but differing significantly
in the way the field equations are imposed on the lattice. In the smooth
lattice method we employ a series of local Riemann normal coordinates in which
the connection vanishes at  the origin of each such frame. Collectively these
frames enable us to obtain point estimates of the curvatures in terms of the
lattice data (in particular the leg lengths). The upshot is that the 3+1 ADM
equations can be applied directly to the lattice. This is clearly a radically
different approach to that normally used in numerical relativity. It is thus
interesting to explore its stability properties against those for traditional
techniques.

In our first paper \cite{LeoSLGR} we showed how the smooth lattice method could be used
to obtain the initial data for a Schwarzschild spacetime. In the second paper
\cite{LeoADMSLGR} we showed how the 3+1 ADM equations could be applied to a lattice using
the Kasner spacetime as a test case. In both papers the results were very
encouraging. In this paper we return to the Schwarzschild spacetime, this time
to study the stability of its evolution in a maximal guage.

Studies of a maximally sliced Schwarzschild spacetimes in spherical symmetry
were quite popular some years ago (see for example
[\citeBernsteinEtalB,\citeBernsteinEtalC]). These studies showed that the
evolutions were invariably unstable. The source of the instability was
attributed to the stretching of the grid as grid points nearer the black hole
were drawn into the black hole quicker than those further out. With the
consequent loss of resolution the estimates for the derivatives were seriously
in error and the non-linear feedback in the equations quickly drove the
solution into exponential overflow.

We will repeat these calculations using our smooth lattice method so that we
can address the simple questions : When will the loss of resolution become
apparent, what impact will it have on the subsequent evolution and will it
trigger an unstable evolution?

We should point out that this paper is not an attempt to revive the use
of maximal slicing as a preferred slicing condition. We are, instead, using
it solely as a test of the smooth lattice method.

We will try to stay as close as possible to the earlier work of Bernstein,
Hobill and Smarr \cite{BernsteinEtalC}. Thus we shall not be using any of the
modern techniques, such as apparent horizon boundary conditions
\cite{AnninosEtalD}, conformal differentiation \cite{AnninosEtalB} and using
the Hamiltonian constraint to stabilise the maximal slicing equation
\cite{BernsteinEtalB}. We will, however, employ some techniques of our own, in
particular we will use a lattice to record the metric, we will cover the
lattice with a series of local frames and we will use the Bianchi identities in
computing the curvatures.

As this method may be unfamiliar to many readers we have included more details
of the derivations than might normally be included. However, to spare the reader
we have relegated the bulk of the derivations to a (large) appendix. The
sections that precede the appendix contain all of the important results with
few derivations. We begin with a brief review of the smooth lattice method,
followed by a description of the particular lattice used for the Schwarzschild
spacetime. We then present the details of the 1+1 ADM equations, the results
of the integrations and finally we review what we have found.

\beginsection{Riemann Normal Coordinates and Smooth Lattices}\secdef{RNCIntro}

The concept of a smooth lattice was first introduced by Brewin
[\citeLeoSLGR,\citeLeoADMSLGR] as a method by which a discrete lattice could be
coupled to a family of Riemann normal frames in such a way as to provide
smooth estimates for the metric and curvature on the lattice. 

The principle features of the smooth lattice method are
\begnarrow
\item{\MyMark} The lattice is a finite collection of vertices and legs
               connected in any suitable fashion (e.g. a simplicial lattice).
\item{\MyMark} The data recorded on the lattice is purely geometrical, such
               as the leg lengths and angles between pairs of legs.
\item{\MyMark} To each vertex their is assigned a small neighbourhood
               in which local Riemann normal coordinates are employed.
               Each such neighbourhood is called a computational cell. For
               a simplicial lattice the computational cell can be chosen as
               the set of simplicies attached to the central vertex.
\item{\MyMark} The legs of a computational cell are taken as geodesic
               segments of a locally smooth metric.
\item{\MyMark} The leg lengths are small compared to the curvature length
               scales, that is, $RL^2<\!<1$ where $R$ and $L$ typical values for
               the curvature and leg lengths respectively.
\endnarrow

There are a number of equivalent definitions of Riemann normal coordinates
[\citeLeoRNC,\citePetrovRNC,\citeEisenhartRNC,\citeMTWRNC]. One definition has them to be
the coordinates which, for a given point $P$, the geodesics through $P$ are all
of the form $x^\mu(s) = sa^\mu$ where $s$
is an affine parameter and the $a^\mu$ are constants (i.e. they are ``straight'' lines). The coordinates of a
point $Q$ near $P$ are then taken as  $x^\mu(s)$ for the geodesic that joins
$P$ to $Q$. Clearly this restricts the region in which these coordinates can be
used to that for which there is a unique geodesic which joins $P$ to $Q$. In
larger regions it is possible that pairs of geodesics may cross and thus the
coordinates at the intersection would not be unique. This problem is avoided
in a smooth lattice by requiring the leg lengths to be small when compared
to the curvature length scales, i.e. $RL^2<\!<1$.

An equivalent definition of Riemann normal coordinates is that the connection
and its symmetric first derivative vanish at $P$,
$$
\eqalignno{%
0&=\Gamma^\mu{}_{\alpha\beta}
&\eqndef{RNCdefineA}\cr
0&=\Gamma^\mu{}_{\alpha\beta,\lambda}
  +\Gamma^\mu{}_{\beta\lambda,\alpha}
  +\Gamma^\mu{}_{\lambda\alpha,\beta}
&\eqndef{RNCdefineB}\cr
}
$$
In a Riemann normal frame it is relatively easy to show that the metric, when
expanded about the point $P$, takes the form
$$
g_{\mu\nu}(x) = g_{\mu\nu} 
               - {1\over3}R_{\mu\alpha\nu\beta}x^\alpha x^\beta
               + \Order(x^3)
\eqno\eqndef{RNCMetric}
$$
where $g_{\mu\nu}$ are constants that can always be chosen to be ${\rm
diag}(1,1,1)$.  This prescription does not uniquely determine the coordinates
as rotations around $P$ are still allowed. This guage freedom can be used to
orient the axes to preferred edges of the lattice.

From this form of the metric it is relatively easy to compute various
geometrical quantities such as the geodesic length $L_{ij}$ between two
vertices $i$ and $j$
$$
L^2_{ij} = g_{\mu\nu}\Delta x^\mu_{ij}\Delta x^\nu_{ij}
         -{1\over3}R_{\mu\alpha\nu\beta}x^\mu_i x^\nu_i x^\alpha_j x^\beta_j
         + \Order(\epsilon^5)
\eqno\eqndef{RNCLsq}
$$
and the angle $\theta_i$ subtended at the vertex $i$ in the geodesic triangle
$(ijk)$
$$
2L_{ij}L_{ik}\cos\theta_i = L^2_{ij}+L^2_{ik}-L^2_{jk}
                          - {1\over3}R_{\mu\alpha\nu\beta}
                            \Delta x^\mu_{ij} \Delta x^\nu_{ij}
                            \Delta x^\alpha_{ik} \Delta x^\beta_{ik}
                          + \Order(\epsilon^5)
\eqno\eqndef{RNCCos}
$$
In these equations $x^\mu_i$ are the coordinates of vertex $i$, 
$\Delta x^\mu_{ij} = x^\mu_i - x^\mu_j$ and $\epsilon$ is a typical (small)
length scale. See \cite{LeoSLGR} for a derivation of these equations.

The principle advantages of using a smooth lattice method over other lattice
methods are
\begnarrow
\item{\MyMark} The metric is smooth and differentiable.
\item{\MyMark} Point estimates of the curvatures are easy to compute.
\item{\MyMark} The equivalence principle is explicitly used in each
               computational cell.
\item{\MyMark} The vanishing of the connection at a vertex greatly simplifies
               many equations (i.e. covariant differentiation reduces
               to partial differentiation).
\endnarrow

\beginsection{The lattice}\secdef{RNCLattice}

The smooth lattice method could equally be applied to the 4-dimensional
spacetime or, in a 3+1 ADM context, to each spacelike Cauchy surface. We will
adopt the second approach simply because it reduces the complexity of the
bookkeeping (a 3-d computational cell has far less complexity than its 4-d
counterpart). The 3+1 ADM equations will be used to evolve the 3-d smooth
lattice.

For a Schwarzschild spacetime it is reasonable to choose the Cauchy surface to
be a lattice built upon concentric 2-spheres. Each two sphere could be
subdivided into a set of cells (e.g. triangles) using the same pattern on every
2-sphere. The successive 2-spheres can be joined by short radial legs connecting
pairs of similar vertices. This construction is not unique as it does allow
for a creeping rotation to occur between successive 2-spheres. This can be
eliminated by demanding that the radial legs be perpendicular to the 2-spheres. 

As part of the smooth lattice approach we require each leg in the lattice to be
a geodesic segment of the 3-metric. Thus, as the radial legs are required to be
normal to each 2-sphere, we see that each sequence of connected radial legs
also forms a global geodesic of the 3-metric.

What flexibility do we have in choosing the leg lengths? Since the space must
by spherically symmetric we can only adjust the overall scale of each two sphere
and the distance between successive 2-spheres. These two pieces of information
can be recorded by specifying the leg lengths between a pair of 2-spheres, $L_{zz}$,
and the typical length within a 2-sphere, $L_{xx}$. By this means we can reduce
the complexity of a full 3-dimensional lattice to a simple 2-dimensional
ladder as indicated in figure \figrfr{RNCLadder}.

\beginsubsection{The computational cell}

Each Riemann normal frame was chosen to cover the region between three
consecutive 2-spheres. Thus each Riemann normal frame will include the three
rungs $L^\p_{xx}$, $L^\o_{xx}$, $L^\m_{xx}$ and the two struts $L^\p_{zz}$ and $L^\m_{zz}$.
We also chose to align each Riemann normal frame so that i) the ladder
was confined to the $xz$-plane, ii) the $z$-axis coincided with the familiar
radial axis and iii) the $z$-axis threaded the mid-points of each rung
(see figures (\figRNCThroat,\figRNCLadder,\figRNCCell)). It follows that the
coordinates of the vertices must be of the following form
\bgroup

\def\A#1{\hbox to 8mm{$\displaystyle\hfil#1$}}
\def\B#1{\hbox to 5mm{$\displaystyle\hfil#1$}}
\def\C#1{\hbox to 6mm{$\displaystyle\hfil#1$}}
\def\D#1{\hbox to 6mm{$\displaystyle\hfil#1$}}
\def\E#1{\hbox to 4mm{$\displaystyle\hfil#1$}}
\def\F#1{\hbox to 5mm{$\displaystyle\hfil#1$}}
%
%
\def\maindivider{\leaders\hrule height 1.5pt depth 0pt\hfil}
\def\medmdivider{\leaders\hrule height 1.0pt depth 0pt\hfil}

\def\Stroke{\vrule width 1.0pt}
\def\MyStroke{\omit\Stroke}
%
%

%
%
%
%
%
{\offinterlineskip
\vskip 18pt plus 5pt minus 5pt
\centerline{\hfil\vrule width 0pt%
\vtop{\tabskip 0.0cm\halign{%
\vrule height 16pt depth 7pt width 0pt#\tabskip=0.0cm&%
\hskip0.5cm\hfil{#}\hfil&#\vrule width 0pt&%
\hfil{#}\hfil\hskip0.5cm&#\vrule width 0pt&%
\hskip0.5cm\hfil{#}\hfil&#\vrule width 0pt&%
\hfil{#}\hfil\hskip0.5cm&#\vrule width 0pt&%
\hskip0.5cm\hfil{#}\hfil&#\vrule width 0pt&%
\hfil{#}\hfil\hskip0.5cm&%
\vrule height 16pt depth 7pt width 0pt#\tabskip=0.0cm\cr
\multispan{13}\maindivider\cr
&\multispan{11}
\hfil \bf Table \tbldef{VertexCoords}. Coordinates for the vertices in
Figure \figrfr{RNCCell}\hfil&\cr
\multispan{13}\maindivider\cr
&\bf Vertex\ &&$(x,y,z)$&\MyStroke%
&\bf Vertex\ &&$(x,y,z)$&\MyStroke%
&\bf Vertex\ &&$(x,y,z)$&\cr
\multispan{13}\medmdivider\cr
&$1^\m$&&$(\A{u^\m},\E{0},\D{v^\m})$&\MyStroke%
&$1$&&$(\A{u^\o},\E{0},\E{0})$&\MyStroke%
&$1^\p$&&$(\A{u^\p},\E{0},\D{v^\p})$&\cr
&$2^\m$&&$(\A{-u^\m},\E{0},\D{v^\m})$&\MyStroke%
&$2$&&$(\A{-u^\o},\E{0},\E{0})$&\MyStroke%
&$2^\p$&&$(\A{-u^\p},\E{0},\D{v^\p})$&\cr
\multispan{13}\maindivider\cr}}\vrule width 0pt\hfil}}

\egroup

\vskip \parskip

with $v^\m<0$ and $v^\p>0$. Note that the origin of the Riemann normal frame
has been located (by a translation in the radial direction) so that the
$z$ coordinate of vertices 1 and 2 are zero.

We also require that successive pairs of struts form a global radial geodesic,
that is, there can be no kink at the vertex where the radial struts meet.
Thus we demand that
$$
\pi = \theta^\p+\theta^\m
\eqno\eqndef{GeodesicConstrnt}
$$
which we call the geodesic constraint. From these five leg lengths and one
geodesic constraint we need to compute the curvatures and all of the
coordinates (see section \secrfr{AppRNC} for details).

\beginsubsection{The Riemann curvatures}

For a spherically symmetric space there are just two algebraically independent
curvature terms, $R_{xyxy}$ and $R_{xzxz}=R_{yzyz}$ (the equality arises from
the rotational symmetry around the $z$-axis). In Appendix \secrfr{AppRNC} we
show that the system of equations for the leg lengths and the geodesic
constraint can be reduced to a single equation for $R_{xzxz}$
$$
0={2\over L^\p_{zz}+L^\m_{zz}}
  \left({L^\p_{xx}-L^\o_{xx}\over L^\p_{zz}}
       +{L^\m_{xx}-L^\o_{xx}\over L^\m_{zz}}\right)
 +R_{xzxz}L^\o_{xx}
\eqno\eqndef{RNCRxzxz}
$$
which in turn is seen to be a finite difference approximation (on a non-uniform
lattice) to the geodesic deviation equation for nearby radial geodesics,
namely
$$
0 = {d^2 L_{xx}\over dz^2} + R_{xzxz}L_{xx}
\eqno\eqndef{ExactRxzxz}
$$
The remaining smooth lattice equations serve only to determine the
coordinates $u^\o$, $u^\p$, etc. Clearly we need one more equation in order to
compute the second curvature $R_{xyxy}$. This could be obtained by introducing
extra structure into the lattice, such as the diagonal braces joining pairs of
vertices on a 2-sphere. In fact this structure already exists -- when we first
spoke of the lattice we imagined each 2-sphere to be fully triangulated. Only
later latter did we choose one of the legs on which to build our ladder. We
could simply choose the collection of triangles attached to a particular vertex
as a base on which to build a more sophisticated lattice with one ladder built
over each leg of each triangle. This lattice would contain two classes of
ladders -- those sharing the common radial geodesic (that generated by the
central vertex) and a chain of ladders forming a cylinder around the common
radial geodesic. This lattice would therefore contain two classes of rungs --
one for each class of ladder. In principle this lattice should allow us to
compute both curvatures, $R_{xzxz}$ and $R_{xyxy}$. However, there is a further
complexity in that we do not know, a priori, the relationship between the leg
lengths of the two classes of rungs. One might be tempted to set the lengths of
each class of rung in the asymptotically flat region (i.e. far from the throat)
and to then impose the same ratios (between the two classes) for all of the
rungs back down to the throat. This would be correct if the rungs of the
ladders where geodesic segments of each 2-sphere. However, the rungs are
geodesic segments of the full 3-dimensional metric and thus their ratios will
change with distance from the black hole's throat. Rather than pursue a
solution to this problem we chose instead to retain our simple (one ladder)
lattice and to employ the Bianchi identities to compute the second curvature
$R_{xyxy}$. For our spherically symmetric space we can show (see Appendix
\secrfr{AppBianchi}) that there is only one non-trivial Bianchi identity
$$
0 = {\left(L^2_{xx} R_{xyxy}\right)^\p-\left(L^2_{xx} R_{xyxy}\right)^\o
     \over L^\p_{zz}}
  -{1\over2}\left(R^\p_{xzxz}+R^\o_{xzxz}\right)
    {\left(L^2_{xx}\right)^\p-\left(L^2_{xx}\right)^\o
     \over  L^\p_{zz}}
\eqno\eqndef{RNCRxyxy}
$$
which is is a simple forward finite difference approximation to the continuum
equation
$$
0 = {d\left(L^2_{xx} R_{xyxy}\right)\over dz} - R_{xzxz} {dL^2_{xx}\over dz}
\eqno\eqndef{ExactRxyxy}
$$
in which $z$ is the proper distance measured along the radial axis from the
throat.

We solve the coupled equations (\eqnRNCRxzxz,\eqnRNCRxyxy) for the curvatures
given all the leg lengths and a suitable initial value for $R_{xyxy}$ on the
inner boundary of the lattice (see section \secrfr{NumMethod} for details).

\beginsection{ The 1+1 ADM equations}\secdef{ADMEqtns}

\beginsubsection{The evolution equations}

In an earlier paper \cite{LeoADMSLGR} we showed how the 3+1 ADM evolution
equations may be applied to any lattice. For the present problem with zero
shift and drift the evolution equations may be written
(see Appendix \secrfr{AppADMEqtns}) as
$$
\openup10pt
\eqalignno{%
{dL^2_{ij}\over dt} &{}= - 2N K_{\mu\nu}
                              \Delta x^\mu_{ij}\Delta x^\nu_{ij}
&\eqndef{ADMEvolLsq}\cr
{d\over dt}\left(K_{\mu\nu}\Delta x^\mu_{ij}\Delta x^\nu_{ij}\right) &{}=
 \left( - N_{\vert\mu\nu} 
        + N \left( R_{\mu\nu} 
                  + K K_{\mu\nu}
                  -2K_{\mu\alpha}K^\alpha_\nu \right) \right)
 \Delta x^\mu_{ij}\Delta x^\nu_{ij}
&\eqndef{ADMEvolKij}\cr}
$$
which when applied to our lattice (see Appendix \secrfr{AppADMEqtns}) leads to
$$
\eqalignno{%
{dL_{xx}\over dt} & = -NK_{xx}L_{xx}
&\eqndef{RNCEvolLx}\cr
{dL_{zz}\over dt} & = -NK_{zz}L_{zz}
&\eqndef{RNCEvolLz}\cr
{dK_{xx}\over dt} &= -N_{,xx} + N\left(R_{xx} + K K_{xx}\right)
&\eqndef{RNCEvolKxx}\cr
{dK_{zz}\over dt} &= -N_{,zz} + N\left(R_{zz} + K K_{zz}\right)
&\eqndef{RNCEvolKzz}\cr}
$$
where $K=2K_{xx}+K_{zz}$, $R_{xx}:=R^\mu{}_{x\mu x}=R_{xyxy}+R_{xzxz}$ and
$R_{zz}:=R^\mu{}_{z\mu z}=2R_{xzxz}$. The partial derivatives of the lapse
function $N_{,xx}$ and $N_{,zz}$ can be evaluated using the techniques in
Appendix \secrfr{AppLaplacian}, leading to
$$
\eqalignno{%
N_{,xx} &= {1\over L_{xx}}{dL_{xx}\over dz}{dN\over dz}
&\eqndef{RNCNxx}\cr
N_{,zz} &= {d^2 N\over dz^2}
&\eqndef{RNCNzz}\cr}
$$

\beginsubsection{The constraint equations}

For the Schwarzschild spacetime there are only two non-trivial constraint equations,
the Hamiltonian constraint
$$
0 = R + K^2 - K^{\mu\nu} K_{\mu\nu}
\eqno\eqndef{ADMHam}
$$
and the momentum constraint
$$
0 = K{}_{\vert\mu} - K_{\mu}{}^{\nu}{}_{\vert\nu}
\eqno\eqndef{ADMMom}
$$
where $K= K^\mu{}_{\mu}$. These equations are readily adapted to the smooth
lattice (see Appendix \secrfr{AppConstraints}) leading to
$$
\openup+10pt
\eqalignno{%
0 &{}=R_{xyxy} + 2R_{xzxz} + K^2_{xx} + 2K_{xx}K_{zz}
&\eqndef{RNCHam}\cr
0 &{}= {d\left(L_{xx} K_{xx}\right)\over dz} - K_{zz}{dL_{xx}\over dz}
&\eqndef{RNCMom}\cr}
$$

\beginsection{Numerical methods}\secdef{NumMethod}

\beginsubsection{The initial data}

The initial data for the lattice consists of the leg lengths $L_{xx}$,  $L_{zz}$ and
the extrinsic curvatures $K_{xx}$, $K_{zz}$. These can be freely chosen subject
to the two constraints \eqnrfr{RNCHam} and \eqnrfr{RNCMom}. For a time
symmetric slice we must have $0=K_{xx}=K_{zz}$ and consequently the momentum
constraint \eqnrfr{RNCMom} is identically satisfied. This leaves the Hamiltonian
constraint, which now takes the simple form,
$$
0=R_{xyxy}+2R_{xzxz}
\eqno\eqndef{RNCHamTS}
$$
as the one equation to constrain $L_{xx}$ and $L_{zz}$. Following Bernstein,
Hobill and Smarr \cite{BernsteinEtalC} we chose to set $L_{zz}$ while computing
the $L_{xx}$ as a solution of the Hamiltonian constraint.

The radial leg lengths $L_{zz}$ where set as follows. A stretched grid of
isotropic radial coordinates were defined by
$$
r_j = {m\over2}e^{(j\Delta)}
\eqno\eqnnxt
$$
with $j=0$ on the inner boundary (the throat) and $j=N$ on the outer boundary.
The parameters $\Delta$ and $N$ were chosen by Bernstein, Hobill and Smarr so
that the outer boundary was at $r\approx 200m$. They  chose $\Delta=6/N$ and
$N=200$ while for our production runs we chose $N=800$. The $L_{zz}$ were then
chosen as
$$
L_{zz} = \int_{r_j}^{r_{j+1}}\>\left(1+{m\over2r}\right)^2\>dr
\eqno\eqnnxt
$$
for $j=0$ to $j=N-1$.

The $L_{xx}$ were set by re-arranging the coupled system (\eqnRNCRxzxz,\eqnRNCRxyxy)
and (\eqnRNCHamTS) in the form of a radial integration. Starting from the
throat and working outwards,
$$
\eqalignno{%
L^\p_{xx} &{}= L^\o_{xx}
   +{L^\p_{zz}\over L^\m_{zz}}
    \left(L^\o_{xx}-L^\m_{xx}\right)
   +{1\over4}L^\p_{zz}
    \left(L^\p_{zz}+L^\m_{zz}\right)
         \left(L_{xx} R_{xyxy}\right)^\o
&\eqndef{RNCInitLx}\cr
R^\p_{xyxy} &{}= R^\o_{xyxy}
   \left( { 5\left(L^2_{xx}\right)^\o - \left(L^2_{xx}\right)^\p\over
            5\left(L^2_{xx}\right)^\p - \left(L^2_{xx}\right)^\o } \right)
&\eqndef{RNCInitRxyxy}\cr
R^\p_{xzxz} &{}= -{1\over2} R^\p_{xyxy}
&\eqndef{RNCInitRxzxz}\cr}
$$
At the throat we imposed reflection symmetry by setting $L^\p_{xx}=L^\m_{xx}$,
$L^\p_{zz}=L^\m_{zz}$. We also chose $L^\o_{xx}=m/10$ (the equations are linear in
$L^\o_{xx}$ and thus this choice is not crucial) and
$R^\o_{xyxy}=1/(4m^2)=-2R^\o_{xzxz}$ (which we obtained from the analytic
solution). Finally, we set $m=1$. With this information we applied the above
equations to generate all of the lattice data from the throat out to the
outer boundary.

Is it reasonable to be using the analytic solution to assist us in setting the
initial data? The only information borrowed from the analytic solution are the
$L_{zz}$ in each cell and $R_{xyxy}$ at the throat. Where we chose to locate the
successive Riemann cells is up to us, that is, we are free to choose the $L_{zz}$
as we see fit. This is identical to the freedom to choose the lapse function
when evolving the initial data. Thus this use of the analytic solution is not
crucial and is made only to allow us to make direct comparisons with Bernstein,
Hobill and Smarr. But what of the choice of curvature $R_{xyxy}$?  Notice that
the analytic solution depends on just one parameter, the mass $m$. Our lattice
initial data also depends on just one parameter, the value of $R_{xyxy}$ at the
throat. Thus whatever choice we make for $R_{xyxy}$ we are in effect choosing
the ADM mass of our numerical spacetime. We could make some other choice for
$R_{xyxy}$ and then later determine the ADM mass for our numerical spacetime.
To take this approach would be tedious and thus we chose to take the easier
option where we set the ADM mass at the outset. Note that we make no further
use of the analytic solution throughout the subsequent evolution. 

\beginsubsection{The boundary conditions}\secdef{BndryConds}

The standard boundary conditions for Schwarzschild initial data are that the
inner boundary is reflection symmetric and that in the distant regions the data
is asymptotically flat. The reflection symmetry can be imposed by extending the
lattice so as to have a computational cell that straddles the throat and then
to demand that the two halves of this cell be mirror copies of each other. With
this extra computational cell we can apply any of the lattice equations at the
throat.

For a reflection symmetric throat we demand $L^\p_{xx}=L^\m_{xx}$ and $L^\p_{zz}=L^\m_{zz}$
at the throat. This condition was imposed throughout the evolution by setting
$L^\m_{xx}$ and $L^\m_{zz}$ equal to their updated counterparts $L^\p_{xx}$ and
$L^\p_{zz}$ at each stage of the integration (i.e. within each of the four steps
of  the 4th order Runge-Kutta).

At the outer boundary we extended the lattice by half a cell so that we could
apply the evolution equations to the data associated with that cell. The data
for the extra half cell were obtained by cubic extrapolation from the interior.
The only data that needed to be extrapolated was $K_{xx}$, $N$, $N_{zz}$ and
$R_{xzxz}$ for the grid centred scheme (see section (\secrfr{TimeStepping}))
and $R_{xzxz}$ for the standard scheme. In both schemes we set
$0=dL^\p_{xx}/dt$.

\beginsubsection{The lapse function}

A maximally sliced spacetime is defined to be a spacetime for which $K=0$
everywhere. This is normally imposed by setting $K=0$ on the initial Cauchy
surface and then setting $dK/dt=0$ throughout the evolution.  This condition
leads, through the standard ADM evolution equations, to the following elliptic
equation for the lapse function
$$
0 = \nabla^2 N - R N
\eqno\eqndef{ADMMaximal}
$$
Using the results of section \secrfr{AppLaplacian} we can write this as
$$
0 = {d^2N\over dz^2} + {2\over L_{xx}}{dL_{xx}\over dz}{dN\over dz} - R N
\eqno\eqndef{ODEMaximal}
$$
which in turn can be applied to the lattice by replacing each of the
derivatives by finite difference approximations on a non-uniform lattice
(see Appendix \secrfr{AppFiniteDiffs})

The common wisdom is that to obtain a stable evolution, the Hamiltonian
constraint should be used to eliminate the curvatures terms in the above
equation. However, we found (see section \secrfr{Results}) that the evolution
was stable without this modification.

The boundary conditions of reflection symmetry at the throat, $dN/dz=0$, and
asymptotic flatness $\lim_{r\rightarrow\infty}N=1$ were prescribed on our
(finite) lattice as
$$
\openup+5pt
\eqalignno{%
\llap{\hbox{inner boundary :}\hskip 3cm}
0&=N^\p-N^\m
&\eqndef{RNCLapseInnerBC}\cr
\llap{\hbox{outer boundary :}\hskip 3cm}
1&=N^\o
&\eqndef{RNCLapseOuterBC}\cr}
$$
The coupled system (\eqnODEMaximal--\eqnRNCLapseOuterBC) were solved by three
iterations of a shooting method. In each shot we made a guess ${\Tilde N}^\m$
for $N^\m$ on the inner boundary and then used a Thomas algorithm to solve a
modified system of equations consisting of ${\Tilde N}^\m=N^\m$ at the inner
boundary, the main equation \eqnrfr{ODEMaximal} everywhere on the lattice
(except the outer boundary) and the outer boundary equation $1=N^\o$. The two
initial guesses for $N^\m$ were taken as ${\Tilde N}^\m = 0$ and ${\Tilde
N}^\m=1$. After each shot, an estimate for $dN/dz$ at the inner boundary was
formed,
$$
{dN\over dz}={N^\p-N^\m\over 2L_{zz}}
\eqno\eqnnxt
$$
Since the equation \eqnrfr{ODEMaximal} is linear in $N$ it is possible to form a linear
combination of the two solutions (one from each shot) so as to satisfy both the
inner and outer boundary conditions. This leads to the third and final guess
$$
{\Tilde N}^\m_3 = \left({dN\over dz}\right)_1
    \left(\left({dN\over dz}\right)_1-\left({dN\over dz}\right)_2\right)^{-1}
\eqno\eqnnxt
$$
with the subscripts denoting the shot number. This guess was used for the final
shot for the lapse on the lattice.

\beginsubsection{Time stepping}\secdef{TimeStepping}

In each Riemann normal frame we can use the evolution equations to compute the
time derivatives of the lattice data. However some of the lattice data are
shared between neighbouring frames, $L_{zz}$ for example, and thus we will obtain
multiple estimates for their time derivatives. How then should we compute a
single estimate for the time derivatives of the lattice data? Though there are
many schemes which could be imposed, we chose two related schemes. 

In the first scheme, which we will refer to as the standard scheme, we formed
simple averages over all of the time derivatives for each component of the
lattice data. For example, as each leg joins two vertices there will be two
estimates available for its time derivative, one from each vertex. For some
legs, such as $L_{xx}$, the two vertices are in equivalent frames (i.e. the two
frames are at the same distance from the throat) and thus yield identical
values for the time derivatives. Thus, for these legs it was sufficient to
compute just the one time derivative. However, for legs such as $L_{zz}$, the
two Riemann frames are distinct and both time derivatives must be computed. In
the standard scheme no averaging was performed for the time derivatives of the
$K_{\mu\nu}$.

In the second scheme we treat $L_{zz}$ and $K_{zz}$ as if they are defined
at the centre of their associated legs. All other data,
$L_{xx},K_{xx},R_{xyxy},R_{xzxz},N,N_{xx},N_{zz}$ are taken to be defined on
the vertices. When the time derivatives are evaluated, interpolations of the
data are required to assemble all terms at the appropriate point. We take the
average of the vertex data when interpolating from a pair of vertices to
the centre of the leg. However, the interpolation of the data from the centres
of the legs to the vertices is slightly more involved. For a function $f$
defined at the centre of the radial legs the interpolation to the common vertex
of two successive radial legs is given by
$$
{\Tilde f} = {1\over L^\p_{zz}+L^\m_{zz}}
             \left(L^\p_{zz}f^\m+L^\m_{zz}f^\p\right)
\eqno\eqnnxt
$$
This formula was only applied to the $K_{zz}$ terms when computing the time
derivatives for $L_{xx}$ and $K_{xx}$. The final step was to take averages as
per the standard scheme. This method will be referred to as the grid centred
scheme.

One could ask why the $L_{xx}$ and $K_{xx}$ where not given a similar centred
treatment. The answer is that, because of the rotational symmetry of the
lattice, the interpolation would produce values identical to that obtained by
assuming the $L_{xx}$ were based on the vertices. So for simplicity we choose
to take $L_{xx}$ and $K_{xx}$ as based on the vertices in the grid centred
scheme.

For all of our production runs we chose a 4-th order Runge-Kutta scheme with a
fixed time step of $\delta t=0.01$. The evolution equations
(\eqnRNCEvolLx--\eqnRNCEvolKzz) were treated as a fully coupled system of
equations. 

In each of the four steps of a single Runge-Kutta cycle we would first compute
all of the $R_{xzxz}$ using \eqnrfr{RNCRxzxz}. From \eqnrfr{RNCRxyxy} we would
then compute the $R_{xyxy}$, from the throat to the outer boundary, using the
Hamiltonian constraint \eqnrfr{RNCHam} as a boundary condition for $R_{xyxy}$
at the throat. Then followed the computation of the lapse function, the time
derivatives and their averages before the partial updates were made. This would
be repeated for the remaining steps in the Runge-Kutta cycle.

\beginsection{Results}\secdef{Results}

The results of our integrations, using the grid centred scheme on a grid with
800 radial legs, are presented in figures (\figLxxRunCD--\figLapseRunCD). 
Notice that all of the curves are smooth and show no signs of any instabilities
to $t=1000m$. We have also marked the location of the apparent horizon on each
curve with a diamond (the procedure for locating the apparent horizon is
described later in section \secrfr{ApparentH}). This clearly shows that the
horizon propagates smoothly and with almost constant speed across the grid. It
is also clear that majority of the dynamics occurs within a very narrow region
straddling the horizon.

The standard scheme produced curves that were qualitatively similar to those
from the grid centred scheme. In particular they showed no signs of any
numerical instabilities out to $t=1000m$. They did, however, show non-trivial
quantitative differences in the later stages of the evolution -- at $t=1000m$
the standard scheme's grid had stretched out to $739m$ as opposed to $826m$
for the grid centred scheme. The grid centred scheme also showed better
constancy in the area of the apparent horizon (see section
(\secrfr{ApparentH})), with only a 4\% change in area from $t=0$ to $t=100m$ as
opposed to a 13\% change for the standard scheme. In the later stages of the
evolution we can expect that the accuracy of the solution will be degraded due
to the loss of resolution near the apparent horizon. This problem will occur
with all numerical methods that do not provide for special treatment over the
highly dynamical regions (e.g. an adaptive grid refinement scheme such as that
due to Berger and Oliger \cite{BergerOliger}).

There are a number of simple checks that have been used by many others
[\citeBernsteinEtalB,\citeBernsteinEtalC,\citeAnninosEtalB] to
check the quality of the numerical solution. These include monitoring the
constraints, the growth of the apparent horizon, the convergence of the maximal
slices to $r=3m/2$ and the so called {\it collapse of the lapse}. We will
discuss each of these checks in turn.

\beginsubsection{Constraints}

Ideally the constraints should remain bounded throughout the evolution. In
practice though the constraints do drift away from their initial values. The
plots in figures (\figHamCnstrntRunCD,\figMomCnstrntRunCD) do show an initial
growth in the constraints but after $t\approx100m$ the constraints remain
bounded.

\beginsubsection{The $r=3m/2$ limit}

It has been shown by Estabrook etal \cite{Estabrook} that the maximal slices of
the Schwarzschild spacetime converge to the limit surface $r=3m/2$ where in
this case $r$ is the standard Schwarzschild radial coordinate. We can use this
as a check on our solution. Those parts of the grid for which the lapse has
collapsed will be frozen on $r=3m/2$. Thus the lattice data should be constant
across those sections. This can be seen in the early stages of the evolution
where inside the apparent horizon all of the data are (approximately) constant.
For the curvature terms we can estimate what these constants should be. Notice
that the geometry inside the apparent horizon is that of a cylinder, i.e.
$S^2\times R$. The scalar 3-curvature will be just that of the 2-curvature of
the 2-dimensional cross sections (i.e of a 2-sphere of radius $r=3m/2$). Thus
we must have $R=2/r^2$. We also know that $R=2(R_{xyxy}+2R_{xzxz})$ and as
$L^2_{xx}$ is constant along the cylinder we find from \eqnrfr{ExactRxzxz} that
$R_{xzxz}=0$. Thus we deduce that $R_{xyxy}=4/(9m^2)=0.\dot4/m^2$ which agrees
well with our numerical value of $0.44445/m^2$ during the early evolution, from
$t=0$ to $t=100m$. However, we loose agreement in the long term evolution,
$t=100m$ to $t=1000m$. This is caused by a loss of resolution near the apparent
horizon due to grid stretching. To show this, we plotted $R_{xyxy}$ at $t=100m$
for four different grid resolutions, with 100, 200, 400 and 800 grid points,
see figure (\figResolution). This clearly shows that the grid resolution has a
significant impact on the accuracy of the solution.

\beginsubsection{Geodesic slicing}

It is well known \cite{MTW} that for a geodesicly sliced Schwarzschild spacetime
(i.e. setting the lapse equal to 1) the throat will collide with the $r=0$
singularity at coordinate time $\pi m$. This provides a simple test -- run
the code and see when it crashes (its also a curious way to compute $\pi$!).
We ran the code for 100, 200, 400 and 800 grid points and found that the
code crashed within two time steps of $t=3.14$ (with a time step of $0.01$).

\beginsubsection{Collapse of the lapse}

Beig \cite{Beig} has shown that for a maximally sliced Schwarzschild
spacetime, the lapse at the throat will die exponentially with time,
$$
\eqalignno{%
N &\sim \beta e^{-\alpha t}+{\cal O}\left(e^{-2\alpha t}\right)
{\quad\rm as\ }t\rightarrow\infty&\eqnnxt\cr}
$$
where
$$
\eqalignno{%
\alpha&={4\over3\sqrt{6}}\approx 0.54433&\eqnnxt\cr
\beta &={4\over3\sqrt{2}}\exp\left({4\gamma\over3\sqrt{6}}\right)\approx 0.83725&\eqnnxt\cr
\gamma&={3\sqrt{6}\over4}\ln(54\sqrt{2}-72)
       -2\ln\left({3\sqrt{3}-5\over 9\sqrt{6}-22}\right)\approx -0.21815&\eqnnxt\cr}
$$
See also the earlier works by Estabrook and others
[\citeEstabrook,
\citePetrich,
\citeReinhart,
\citeSmarr].
This gives us another test where a plot of
$\ln N$ versus $t$ should be a straight line. This we have done in figures
(\figrfr{Collapse}) and for $t=10m$ to $t=100m$ we get a very straight line indeed.
Using a standard least squares method we found $\alpha=0.5474$ and $\beta=1.011$.
Though our estimate of $\alpha$ agrees well with Beig's result, our estimate
for $\beta$ is not so good. This is probably due to a number of factors such
as the use of of an inexact outer boundary condition, the problems of grid
stretching and numerical error. The numerical error in fitting a straight line
could be significant. Notice that for $0<t<100m$  the vertical intercept in
figure (\figrfr{Collapse}) is very small (approximately $0.01$) relative to
the vertical range (approximately
$60$). Thus any small errors, either in the data or in fitting the line may
produce large {\it relative} errors in the vertical intercept, namely
$\ln\beta$. To demonstrate this point we chose to re-compute $\alpha$ and
$\beta$ by constraining the curve to pass through the data point at $t=10m$.
Thus we applied the least squares method to
$\ln N-\ln N(10)=\alpha(t-10)$ and found $\alpha=0.5429$ and
$\beta=0.8112$. This is an improvement over our earlier estimates.

At later times, $t=100m$ to $t=1000m$ the line does bend slightly. This is not
surprising since we know that for these times we have lost accuracy due to
grid stretching. Note that at $t=1000m$ we have $N\sim 10^{-145}$ at the
throat -- the evolution really has been halted.

\beginsubsection{Apparent horizon}\secdef{ApparentH}

An apparent horizon is defined as a closed 2-surface with zero divergence of
its outward pointing null vectors \cite{HawkingEllis}. For the Schwarzschild
spacetime the apparent horizon must be a 2-sphere whose area must be constant
when Lie dragged in the direction of the outward pointing null vectors. Thus
$$
0 = {\cal L}_u A = {dA\over du}
\eqno\eqnnxt
$$
where $A$ is the area of the 2-sphere and $\partial/\partial u$ is the outward
pointing null vector. For our lattice we may put
$$
\openup+5pt
\eqalignno{%
A&=kL^2_{xx}&\eqnnxt\cr
{\partial\over\partial u} &= {1\over N}{\partial\over\partial t}
                           + {\partial\over\partial z}&\eqnnxt\cr}
$$
where $k$ is a pure constant, $z$ is the proper distance measured along the
radial axis and $N$ is the lapse function. Thus from $0=dA/du$ we obtain
the apparent horizon equation
$$
0 = {dL_{xx}\over dz} - L_{xx} K_{xx}
\eqno\eqndef{ApparentH}
$$
Our aim is to solve this equation for two things i) the location of the
apparent horizon and ii) the value of $L^2_{xx}$ at that location. We do so as
follows. First we convert the derivative in $L_{xx}$ into a finite difference
approximation using the methods of section \secrfr{AppFiniteDiffs}. Then we scan the grid while
monitoring the value of the right hand side of \eqnrfr{ApparentH} which we
denote by the function $Q(z)$. When we find two successive grid points for
which $Q$ changes sign, we stop the scan and then use linear interpolation to
predict the true location of the apparent horizon (i.e. $z$ such that $0=Q$)
and any data at that point (e.g. $L^2_{xx}$ and $N$).

Since there is no gravitational radiation in the Schwarzschild spacetime we
know that the area of the apparent horizon must remain constant throughout the
evolution. We have plotted $L^2_{xx}=A/k$ in figure (\figHorizon). For the case
of 800 grid points we see that the area varies by less than 4\% for the first
$100m$ of the evolution. However at later times, around $t=1000m$, the error
has blown out by 700\%. This is very large indeed. We do not believe that this
is an error in our code but rather a very severe consequence of the loss of
resolution due to grid stretching.

Anninos etal. \cite{AnninosEtalB} found that for their 1-dimensional code, with
a constant $\Delta r=0.1m$ and an outer boundary at $r=130m$, that at $t=25m$
their apparent horizon mass (defined by $M_{ah} =(A/16\pi)^{1/2}$) had grown by
about 4\% (see their figures 15 and 16). When we repeated their calculations,
on their grid but using our equations, we found an error of 3\% for the
standard scheme and 1\% for the grid centred scheme. In a related work, Anninos
etal. \cite{AnninosEtalD} found that for a grid with 400 points the error in
the apparent horizon mass was about 25\% at $t=100m$. For our code, with 400
grid points, we found errors of 12\% for the standard scheme and 6\% for the
grid centred scheme.

We should mention here that setting $A=kL^2_{xx}$ is not strictly correct
because the rungs of the ladder are not geodesic segments of the 2-sphere but
rather of the surrounding 3-dimensional space. However if the $L^2_{xx}$ are
small relative to the area of the apparent horizon then there will be little
error in using these legs as approximate geodesics on the apparent horizon.

\beginsection{Discussion}\secdef{Discussion}

The results we have presented are very encouraging. Our most important
observation is that the evolution is very stable out to $t=1000m$. In contrast,
the best results using traditional methods (\citeAnninosEtalD,\citeAnninosEtalB)
developed fatal instabilities by about $t=100$. Thus we are forced to ask the
obvious question : What is it about our method that, in this instance, gave us
a stable evolution? The features that distinguish our method from traditional
methods is that we use a lattice to record the metric, we cover the lattice
with a series of Riemann normal frames and we use the Bianchi identities to
assist in the computation of the curvatures. It is premature at this stage to
identify which, if any, of these ingredients is crucial to the stability, but
we can review each of their roles in the method.

{\bf Riemann normal coordinates.\ }These were chosen not simply because they
allowed us to extract curvatures from the lattice but rather for their
relationship with the Einstein equivalence principle. In its full 4-dimensional
setting the equivalence principle is equivalent to having the 4-connection
vanish in the freely falling frame. Using such frames as a computational tool
must surely bring some advantage to the computations. However, we are actually
using the Riemann normal coordinates only for the 3-geometry and thus this
argument is not so strong. On the other hand, a zero connection does greatly
simplify many of the computations, for example, covariant differentiation
reduces to simple partial differentiation. Furthermore, as the source terms in
the ADM equations have a different structure to that found in Bernstein,
Hobill and Smarr (as a typical example), we can expect that the stability
properties may differ from those for traditional methods.

{\bf The lattice.\ }This is essential for it provides the structure by which
the local Riemann normal frames can be connected together to form a global
coordinate atlas for the spacetime. Each pair of adjacent local Riemann normal
frames has a non trivial overlap and through the scalar data they share on the
lattice the transformation form one frame to another is well defined.

{\bf The Bianchi identities.\ }We do not consider the Bianchi identities as
central to our method for they were introduced only to overcome a limitation in
our highly simplified lattice. We could test their role by redoing our
calculations on a more sophisticated lattice. But we should point out that
recent work by Christodoulou and Klainerman \cite{ChristoKlainA} place great
importance on the Bianchi identities in their proof of long term stability for
weak initial data.

Another option that may explain the stability is that the method may lack
sufficient accuracy, due to dispersion or truncation errors, that the sharp
peaks needed to trigger an unstable mode are never resolved. However, our results
were always at least as good as those obtained by others (\citeAnninosEtalD,
\citeAnninosEtalB) which tends to discount this option.

Clearly more work is required by applying this method to other more
challenging spacetimes. We shall report on these calculations soon.

\beginsection{Appendix}\secdef{Appendix}

\beginsubsection{The Riemann normal frame}\secdef{AppRNC}

It is a simple matter to substitute the coordinates listed in table
\tblrfr{VertexCoords} into the smooth lattice equations \eqnrfr{RNCLsq} and
the geodesic constraint \eqnrfr{GeodesicConstrnt}. This leads directly to the
following equations
$$
\openup+5pt
\Displaylines{%
\left(L^\o_{xx}\right)^2 = \left(2u^\o\right)^2&\eqnnxt\cr
\left(L^\m_{xx}\right)^2 = \left(2u^\m\right)^2 
                      - {1\over3}R_{xzxz}\left(2u^\m v^\m\right)^2&\eqnnxt\cr
\left(L^\p_{xx}\right)^2 = \left(2u^\p\right)^2 
                      - {1\over3}R_{xzxz}\left(2u^\p v^\p\right)^2&\eqnnxt\cr
\left(L^\m_{zz}\right)^2 = \left(v^\m\right)^2 + \left(u^\m-u^\o\right)^2
                      - {1\over3}R_{xzxz}\left(u^\o v^\m\right)^2&\eqnnxt\cr
\left(L^\p_{zz}\right)^2 = \left(v^\p\right)^2 + \left(u^\p-u^\o\right)^2
                      - {1\over3}R_{xzxz}\left(u^\o v^\p\right)^2&\eqnnxt\cr
0 = {1\over L^\p_{zz}}\left(u^\o-u^\p
                       -{1\over3}R_{xzxz}\left(u^\o v^\p\right)^2\right)
  + {1\over L^\m_{zz}}\left(u^\o-u^\m
                       -{1\over3}R_{xzxz}\left(u^\o v^\m\right)^2\right)&\eqnnxt\cr
}
$$
For a given choice of $R_{xzxz}$ the first five equations can be solved for the
coordinates $u^\o$, $u^\p$ etc. The last equation is the geodesic constraint in
the form $0=L^\o_{xx}(\cos\theta^\p+\cos\theta^\m)$. This equation will constrain
the choice of the curvature $R_{xzxz}$.

Though these equations could be solved using a Newton-Raphson method it would
be better if we could find an explicit solution. The simple trick to achieving
this hinges on the fact that these equations serve only as an approximation
to the true continuum metric and curvatures and are valid only when the domain
of the Riemann normal frame is small compared with the curvature lengths
scales. Thus it is sufficient to solve the leg length equations by a
perturbation expansion around flat space.

Consider the first five equations and let $R_{xzxz}=\Order(\epsilon)$ with
$\epsilon$ taken as our expansion parameter. Then the leading order solution
is simply
$$
\eqalignno{%
\left(2u^\o\right)_0 &= L^\o_{xx}&\eqnnxt\cr
\left(2u^\m\right)_0 &= L^\m_{xx}&\eqnnxt\cr
\left(2u^\p\right)_0 &= L^\p_{xx}&\eqnnxt\cr
\left(v^\m\right)^2_0 &= \left(L^\m_{zz}\right)^2 - \left(u^\m-u^\o\right)^2_0&\eqnnxt\cr
\left(v^\p\right)^2_0 &= \left(L^\p_{zz}\right)^2 - \left(u^\p-u^\o\right)^2_0&\eqnnxt\cr}
$$
The next level of approximation is obtained by substituting these back into the
above equations and solving once again for the five coordinates. The result is
$$
\eqalignno{%
\left(2u^\o\right)_1 &= L^\o_{xx}&\eqnnxt\cr
\left(2u^\m\right)_1 &= L^\m_{xx} + {1\over6}R_{xzxz}L^\m_{xx} \left(v^\m\right)^2_0&\eqnnxt\cr
\left(2u^\p\right)_1 &= L^\p_{xx} + {1\over6}R_{xzxz}L^\p_{xx} \left(v^\p\right)^2_0&\eqnnxt\cr
\left(v^\m\right)^2_1 &= \left(L^\m_{zz}\right)^2 - \left(u^\m-u^\o\right)^2_1
                       - {1\over3}R_{xzxz}\left(u^\o v^\m\right)^2_0&\eqnnxt\cr
\left(v^\p\right)^2_1 &= \left(L^\p_{zz}\right)^2 - \left(u^\p-u^\o\right)^2_1
                       - {1\over3}R_{xzxz}\left(u^\o v^\p\right)^2_0&\eqnnxt\cr}
$$
This process could be continued to obtain higher order approximations but
that would be a waste of effort as the smooth lattice equations are valid
only to linear terms in the curvatures, that is to linear terms in $\epsilon$.

Substituting these approximations into the geodesic constraint and retaining
just the terms linear in $\epsilon$ leads to
$$
\openup+5pt
\eqalign{%
0&={2\over L^\p_{zz}+L^\m_{zz}}
  \left({L^\p_{xx}-L^\o_{xx}\over L^\p_{zz}}
       +{L^\m_{xx}-L^\o_{xx}\over L^\m_{zz}}\right)
  +R_{xzxz}L^\o_{xx}\cr
 &\hskip2cm-{L^\o_{xx}\over L^\p_{zz}+L^\m_{zz}} R_{xzxz}
  \left( {\left(L^\p_{xx}-L^\o_{xx}\right)^2\over L^\p_{zz}}
 +       {\left(L^\m_{xx}-L^\o_{xx}\right)^2\over L^\m_{zz}} \right)\cr
}
\eqno\eqndef{RNCRxzxzFirst}
$$
This is easily seen to be the finite difference approximation to the
differential equation
$$
0 = {d^2 L_{xx}\over dz^2} + R_{xzxz}L_{xx}
  - L_{xx} R_{xzxz} \left( {dL_{xx}\over dz} \right)^2
\eqno\eqnnxt
$$
which we recognise, apart from the last term, to be the standard geodesic
deviation equation applied to the two radial geodesics. By making the simple
change of scale $L_{xx}\rightarrow\lambda L_{xx}$ we can see that for $\lambda<\!<1$
(i.e. for very short leg lengths) the last term is insignificant compared to
the remaining terms (at a fixed position on the radial axis). This is to be
expected since for the two radial geodesics to be nearly parallel everywhere
we must have $dL_{xx}/dz <\!<1$. 

Upon deleting this term from the discrete equations we obtain
$$
0={2\over L^\p_{zz}+L^\m_{zz}}
  \left({L^\p_{xx}-L^\o_{xx}\over L^\p_{zz}}
       +{L^\m_{xx}-L^\o_{xx}\over L^\m_{zz}}\right)
 +R_{xzxz}L^\o_{xx}
\eqno\eqnrfr{RNCRxzxz}
$$
as our basic equation from which we can compute the curvature $R_{xzxz}$.

Incidentally this requirement that the two radial geodesics be nearly parallel
everywhere also shows that
$\left(L^\m_{zz}\right)^2 >> \left(u^\m-u^\o\right)^2_0$ and thus to leading order
$\left(v^\m\right)_0 = -\left(L^\m_{zz}\right)$. Proceeding in this fashion we
find the following estimates for the coordinates.
$$
\eqalignno{%
\left(2u^\o\right)_1 &= L^\o_{xx}
&\eqndef{RNCCoordA}\cr
\left(2u^\m\right)_1 &= L^\m_{xx} + {1\over6}R_{xzxz}L^\m_{xx} \left(L^\m_{zz}\right)^2
&\eqndef{RNCCoordB}\cr
\left(2u^\p\right)_1 &= L^\p_{xx} + {1\over6}R_{xzxz}L^\p_{xx} \left(L^\p_{zz}\right)^2
&\eqndef{RNCCoordC}\cr
\left(v^\m\right)_1 &= L^\m_{zz}
                       - {1\over24}R_{xzxz}L^\m_{zz}\left(L^\o_{xx}\right)^2
&\eqndef{RNCCoordD}\cr
\left(v^\p\right)_1 &= L^\p_{zz} 
                       - {1\over24}R_{xzxz}L^\p_{zz}\left(L^\o_{xx}\right)^2
&\eqndef{RNCCoordE}\cr}
$$
These estimates will be of use later in Appendix \secrfr{AppADMEqtns}.

\beginsubsection{The 1+1 ADM evolution equations}\secdef{AppADMEqtns}

In the paper by Brewin \cite{LeoADMSLGR} the 3+1 ADM evolution equations for
a lattice where given as
$$
\openup+10pt
\eqalignno{%
{d^2L^2_{ij}\over dt^2} ={}& -2\left({d\over dt}(N K_{\mu\nu})\right)
                             \Delta x^\mu_{ij}\Delta x^\nu_{ij}
                             + Q_{ij}
&\eqnnxt\cr}
$$
where $Q_{ij}$ represented the terms involving the shift and drift vectors. We
shall immediately set $Q_{ij}=0$ since this corresponds to our lattice which
has zero shift and drift.

For many reasons (in particular, for ease of numerical integration) it is 
customary to express the evolution equations as a system of first order
equations. This will be the main focus of this section.

In \cite{LeoADMSLGR} two important coordinate frames were used. The first was
the Riemann normal frame for a given computational cell in one Cauchy surface
$\Sigma_0$. These coordinates were denoted by $x^\mu$. Some of these
coordinates could be freely chosen (by aligning the coordinate axes) on some of
the vertices of the cell (in particular, at the origin of the cell). For the
remaining vertices the coordinates must be computed during the solution of the
lattice equations
$$
L^2_{ij} = g_{\mu\nu}\Delta x^\mu_{ij}\Delta x^\nu_{ij}
         -{1\over3}R_{\mu\alpha\nu\beta}x^\mu_i x^\nu_i x^\alpha_j x^\beta_j
\eqno\eqnnxt
$$
Since the leg lengths are expected to be functions of time we must also expect
that in the Riemann normal frame the vertex coordinates $x^\mu_i$ will also be
functions of time. Thus in this frame we may choose the shift and drift to be
zero at the origin of the cell but must accept non-zero values elsewhere.

The second frame used in \cite{LeoADMSLGR} was not a Riemann normal frame but
one in which the shift and drift vectors where set to zero everywhere. These
coordinates were denoted by $x''^\mu$. On the initial Cauchy surface the two
coordinate frames are identical, $x^\mu=x''^\mu$ on $\Sigma_0$, but for future
times the coordinates will differ. This establishes a time dependent
transformation between the two coordinate frames.

Consider for the moment the shadow frame with coordinates $x''^\mu$. For the
case of zero shift and drift the standard definition of the extrinsic curvature
is
$$
{dg''_{\mu\nu}\over dt} = -2NK''_{\mu\nu}
\eqno\eqnnxt
$$
The leg lengths can be estimated from
$$
L^2_{ij} = g''_{\mu\nu} \Delta x''^\mu_{ij} \Delta x''^\nu_{ij}
\eqno\eqnnxt
$$
But in this frame $0=dx''^\mu_i/dt$ and thus we have
$$
{d L^2_{ij}\over dt} = -2NK''_{\mu\nu}\Delta x''^\mu_{ij} \Delta x''^\nu_{ij}
\eqno\eqnnxt
$$
which when evaluated on $\Sigma_0$, where the two frames coincide, we have
$$
{d L^2_{ij}\over dt} = -2NK_{\mu\nu}\Delta x^\mu_{ij} \Delta x^\nu_{ij}
\eqno\eqnrfr{ADMEvolLsq}
$$
Now consider the time derivative of this last pair of equations. These may be
written as
$$
\eqalignno{%
{d^2 L^2_{ij}\over dt^2} &= -2N{dK''_{\mu\nu}\over dt}\Delta x''^\mu_{ij} \Delta x''^\nu_{ij}
                            -2{dN\over dt}K''_{\mu\nu}\Delta x''^\mu_{ij} \Delta x''^\nu_{ij}
&\eqnnxt\cr
\noalign{and}                          
{d^2 L^2_{ij}\over dt^2} &= -2N{d\over dt}\left(K_{\mu\nu}\Delta x^\mu_{ij} \Delta x^\nu_{ij}\right)
                            -2{dN\over dt}K_{\mu\nu}\Delta x^\mu_{ij} \Delta x^\nu_{ij}
&\eqnnxt\cr}
$$
which if we now combine with the standard ADM equation
$$
{d K''_{\mu\nu}\over dt} = - N_{\vert\mu\nu} 
                           + N \left( R_{\mu\nu} 
                                     + K K_{\mu\nu}
                                     -2K_{\mu\alpha}K^\alpha_\nu \right)
\eqno\eqnnxt
$$
leads directly to
$$
{d\over dt}\left(K_{\mu\nu}\Delta x^\mu_{ij}\Delta x^\nu_{ij}\right) =
 \left( - N_{,\mu\nu} 
        + N \left( R_{\mu\nu} 
                  + K K_{\mu\nu}
                  -2K_{\mu\alpha}K^\alpha_\nu \right) \right)
 \Delta x^\mu_{ij}\Delta x^\nu_{ij}
\eqno\eqnrfr{ADMEvolKij}
$$
on $\Sigma_0$. We have written $N_{,\mu\nu}$ as opposed to $N_{\vert\mu\nu}$
since in the Riemann normal frame the connection vanishes at the origin and
thus $N_{,\mu\nu}=N_{\vert\mu\nu}$ on the central vertex of the computational
cell.

This completes our stated aim -- to derive a pair of first order evolution
equations -- (\eqnADMEvolLsq) and (\eqnADMEvolKij). These will now be applied
to the lattice. To do so first requires equations (\eqnRNCLsq,\eqnRNCCos) to be
solved for the coordinates $x^\mu$ for each vertex in the computational cell.
This was done in section \secrfr{AppRNC} and lead to equations
(\eqnRNCCoordA--\eqnRNCCoordE). Substituting these in
(\eqnADMEvolLsq,\eqnADMEvolKij) and retaining only the leading order terms we
find
$$
\eqalignno{%
{dL_{xx}\over dt} & = -NK_{xx}L_{xx}
&\eqnrfr{RNCEvolLx}\cr
{dL_{zz}\over dt} & = -NK_{zz}L_{zz}
&\eqnrfr{RNCEvolLz}\cr
{dK_{xx}\over dt} &= -N_{,xx} + N\left(R_{xx} + K K_{xx}\right)
&\eqnrfr{RNCEvolKxx}\cr
{dK_{zz}\over dt} &= -N_{,zz} + N\left(R_{zz} + K K_{zz}\right)
&\eqnrfr{RNCEvolKzz}\cr}
$$

\beginsubsection{The ADM evolution equations : An alternative derivation}

There is another way in which all of the above equations can be easily
developed. The idea is to begin with the usual ansatz for a spherically
symmetric space such as
$$
ds^2 =- N^2(r,t)dt^2 
      + A^2(r,t)dr^2+B(r,t)^2(d\theta^2+\sin^2\theta d\phi^2)
\eqno\eqnnxt
$$
and to build a lattice in the $\theta=\pi/2$ plane by assigning leg
lengths according to
$$
\eqalignno{%
L_{xx} &= B\Delta\phi&\eqnnxt\cr
L_{zz} &= A\Delta r&\eqnnxt\cr}
$$
The functions $A$ and $B$ would be evaluated at the centre of each leg while
$\Delta r$ and $\Delta\phi$ would be chosen as some suitably small numbers.

The standard ADM equations, for a metric with zero shift, are
$$
\eqalignno{%
{dg_{\mu\nu}\over dt} &= -2 N K_{\mu\nu}&\eqnnxt\cr
{dK_{\mu\nu}\over dt} &= -N_{\vert\mu\nu} 
                       + N\left( R_{\mu\nu}
                       +KK_{\mu\nu} 
                       -2K_{\mu\alpha}K^{\alpha}{}_{\nu}
                       \right)&\eqnnxt\cr}
$$
We now ask the simple question: What form do these equations take when applied
to a lattice? The answer is also simple. Since there is no shift vector, the
coordinates of each vertex remain constant throughout the evolution, thus
$0=d\Delta x^\mu_{ij}/dt$ and we therefore have
$$
\eqalignno{%
{d\over dt}\left(g_{\mu\nu}\Delta x^\mu_{ij}\Delta x^\nu_{ij}\right)
     &= -2 N K_{\mu\nu}\Delta x^\mu_{ij}\Delta x^\nu_{ij}&\eqnnxt\cr
{d\over dt}\left(K_{\mu\nu}\Delta x^\mu_{ij}\Delta x^\nu_{ij}\right)
     &= \left( -N_{\vert\mu\nu} 
              + N\left( R_{\mu\nu}
                       +KK_{\mu\nu} 
                       -2K_{\mu\alpha}K^{\alpha}{}_{\nu}
                 \right) \right)\Delta x^\mu_{ij} \Delta x^\nu_{ij}&\eqnnxt\cr}
$$
These differ from the smooth lattice equations (\eqnRNCEvolLx--\eqnRNCEvolKzz)
only in how the terms
$N_{\vert\mu\nu}$ and $R_{\mu\alpha\nu\beta}$ are computed. It is easy to show
that for the above metric
$$
\openup+5pt
\Displaylines{%
N_{\vert rr} = N_{,rr} - {1\over A}A_{,r} N_{,r}&\eqnnxt\cr
N_{\vert\phi\phi} = N_{,\phi\phi} + {B\over A^2} B_{,r}N_{,r}&\eqnnxt\cr
R_{r\phi r\phi} ={B\over A}\left(A_{,r}B_{,r}-AB_{,rr}\right)&\eqnnxt\cr
R_{\theta\phi\theta\phi} =\left(B\over A\right)^2
                           \left(A^2-(B_{,r})^2\right)&\eqnnxt\cr
0 = 2B_{,r}R_{r\phi r\phi} - {BA^2}
   \left({R_{\theta\phi\theta\phi}\over B^2}\right)_{,r}&\eqnnxt\cr}
$$
These equations can be adapted to the lattice by making a coordinate
transformation into a local Riemann normal frame
$$
\openup+5pt
\Displaylines{%
dz = A dr&\eqnnxt\cr
dx = B d\phi&\eqnnxt\cr
N_{,zz} = {1\over A^2}N_{\vert rr}&\eqnnxt\cr
N_{,xx} = {1\over B^2} N_{\vert\phi\phi}&\eqnnxt\cr
R_{xzxz} = {1\over A^2 B^2}R_{r\phi r\phi}&\eqnnxt\cr
R_{xyxy} = {1\over B^4} R_{\theta\phi\theta\phi}&\eqnnxt\cr}
$$
and then eliminating, where possible, $A$ and $B$ in favour of $L_{xx}$ and $L_{zz}$.
This leads directly to the smooth lattice equations (\eqnRNCRxzxz,
\penalty-250\relax\eqnRNCRxyxy, \penalty-250\relax\eqnRNCNxx,
\penalty-250\relax\eqnRNCNzz). The exception is the equation for
$R_{\theta\phi\theta\phi}$ for which there is no direct counterpart in the
smooth lattice equations. For this quantity we find
$$
R_{xyxy} = {1\over L^2_{xx}}\left(\Delta\phi^2 - \left({dL_{xx}\over dz}\right)^2\right)
\eqno\eqnnxt
$$
This equation can not be used on this lattice for two reasons. First, there is
no clear method for determining the parameter $\Delta\phi$ from the lattice
data $L_{xx}$ and $L_{zz}$. Second, this equation simply does not arise as a
consequence of the basic smooth lattice equations (\eqnRNCLsq, \eqnRNCCos).
This problem could be overcome by employing a different lattice. For example,
if we chose a lattice in which each 2-sphere was fully triangulated then we
could reasonably expect that both curvatures could be computed from the smooth
lattice equations without reference to the Bianchi identities. On such a lattice
we should also be able to compute $\Delta\phi$.

This alternative derivation is useful not only in giving us confidence that we
have the correct lattice equations but it also gives us a technique for quickly
adapting a continuum equation directly to the lattice. Indeed we could well
have chosen this as the primary method by which to develop our equations.

\beginsubsection{Covariant differentiation}\secdef{AppCovDiff}

Since the connection vanishes in Riemann Normal Coordinates we have, for any
vector field, that
$$
v^\mu{}_{;\nu} = v^\mu{}_{,\nu}
\eqno\eqnnxt
$$
at the origin of the RNC cell.

There are probably many ways in which the partial derivatives could be
evaluated, however, in this section we shall focus on a method based on
coordinate transformations and finite differences.

The basis of our approach is to import values of the vector field from
neighbouring RNC cells, by simple coordinate transformations such as  rotations
and translations, and to then use this data in a finite difference
approximation. This is exactly the same as parallel transport since, once
again, the connection vanishes, and thus to leading order we can construct the
transformations as if we were in flat space.

We will demonstrate this approach for a spherically symmetric 3-geometry.
Consider a typical RNC cell with its axes oriented as per figure
\figrfr{RNCCell}. This cell will be denoted by $\sigma^\o$ while four of its
six immediate neighbours will be denoted by $\sigma^\l$, $\sigma^\r$, being the
left and right neighbours of $\sigma^o$, and $\sigma^\p$, $\sigma^\m$ being the
cells above and below $\sigma^\o$ along the radial axis. The remaining to cells
lie along the $y$-axis relative to $\sigma^\o$ and we will have no need of
these cells in the following calculations. We will use these superscripts to
denote quantities which are defined relative to the cell with the same
superscript.

Suppose we have a spherically symmetric vector field $v^\mu$. Then each cell on
a 2-sphere will share exactly the same values for $v^\mu$ in their respective
local RNC frames, that is
$$
v^{\mu\l} = v^{\mu\r} = v^{\mu\o}
\eqno\eqnnxt
$$
for each $i,j$. We will now import these values into the central cell
$\sigma^o$. Consider first the cells $\sigma^\o$ and $\sigma^\r$. The RNC
coordinates of these frames are related by a rotation (to align the directions
of their axes) and a translation (to align their origins) which we will write
as
$$
x^{\mu\r} = A^{\mu}{}_{\nu} x^{\mu\o} + B^\mu
\eqno\eqnnxt
$$
The situation is depicted in figure \figrfr{RNCBianchi}.

The components $A^\mu{}_\nu$ can be assembled into a rotation matrix
$$
\eqalign{%
\left[ A^\mu{}_\nu \right] &=
\left[
\matrix{\cos\Delta\theta&0&\sin\Delta\theta\cr
        0&1&0\cr
        -\sin\Delta\theta&0&\cos\Delta\theta\cr}
\right]
\approx
{\rm I} + \Delta\theta
\left[
\matrix{0&0&1\cr
        0&0&0\cr
        -1&0&0\cr}
\right]\cr}
\eqno\eqnnxt
$$
where $\Delta\theta$ is a (small) angle of rotation and where $\mu$ is taken as
the row index. The approximation that $\Delta\theta$ is small is imposed so
that the we have an accurate approximation to the continuum geometry. By
inspection of figure \figrfr{RNCCell} we see that $\Delta\theta = \Delta
L_{xx}/\Delta z$.

Using the standard tensor transformation laws we find that the values of
$v^{\mu\r}$ in the RNC frame of $\sigma^\o$ to be
$$
v'^{\mu\r} = {\partial x^{\mu\r}\over \partial x^{\nu\o}} v^{\nu\r}
           = A^\mu{}_\nu v^{\nu\o}
\eqno\eqnnxt
$$
Thus for cells $\sigma^\l$ and $\sigma^\r$ we find
$$
\eqalignno{%
v'^{x\r} &= v^{x\o} + \Delta\theta v^{z\o}&\eqnnxt\cr
v'^{y\r} &= v^{y\o}&\eqnnxt\cr
v'^{z\r} &= v^{z\o} - \Delta\theta v^{x\o}&\eqnnxt\cr
v'^{x\l} &= v^{x\o} - \Delta\theta v^{z\o}&\eqnnxt\cr
v'^{y\l} &= v^{y\o}&\eqnnxt\cr
v'^{z\l} &= v^{z\o} + \Delta\theta v^{x\o}&\eqnnxt\cr}
$$
Since these two cells reside on opposite sides of the $yz-$plane we can use the
above to form a centred finite difference approximation for $v^\mu{}_{,x}$ at
the origin of $\sigma^\o$, namely
$$
v^\mu{}_{,x} = {v^{\mu\r} - v^{\mu\l}\over \Delta x}
\eqno\eqnnxt
$$
where $\Delta x$ is, to leading order, the distance between the origins of
cells $\sigma^\r$ and $\sigma^\l$. This is easily seen to be $2L_{xx}$, 
leading to
\bgroup
$$
\openup5pt
\eqalignno{%
v^x{}_{;x} &= v^x{}_{,x} = \phantom{-}
                           {1\over L_{xx}} {\Delta L_{xx}\over \Delta z} v^{z\o}&\eqnnxt\cr
v^y{}_{;x} &= v^y{}_{,x} = 0&\eqnnxt\cr
v^z{}_{;x} &= v^z{}_{,x} = -{1\over L_{xx}} {\Delta L_{xx}\over \Delta z} v^{x\o}&\eqnnxt\cr}
$$
\egroup
The same idea can be applied to derivatives in the $y$ direction, leading to
\bgroup
$$
\openup5pt
\eqalignno{%
v^x{}_{;y} &= v^x{}_{,y} = 0&\eqnnxt\cr
v^y{}_{;y} &= v^y{}_{,y} = \phantom{-}
                           {1\over L_{xx}} {\Delta L_{xx}\over \Delta z} v^{z\o}&\eqnnxt\cr
v^z{}_{;y} &= v^z{}_{,y} = -{1\over L_{xx}} {\Delta L_{xx}\over \Delta z} v^{y\o}&\eqnnxt\cr}
$$
\egroup
which clearly could also have been derived by simple symmetry arguments. The
$z-$derivatives are very easy to calculate. Since the RNC frames of the three
cells $\sigma^\o,\sigma^\m$ and $\sigma^\p$ are related by simple
translations along the $z$-axis, the coordinate transformations are trivial
and lead directly to
\bgroup
$$
\openup5pt
\eqalignno{%
v^x{}_{;z} &= v^x{}_{,z} = {\Delta v^x\over\Delta z}&\eqnnxt\cr
v^y{}_{;z} &= v^y{}_{,z} = {\Delta v^y\over\Delta z}&\eqnnxt\cr
v^z{}_{;z} &= v^z{}_{,z} = {\Delta v^z\over\Delta z}&\eqnnxt\cr}
$$
\egroup
It is useful to make one small change to the above. We will replace the finite
difference approximations by their continuum limits (i.e. $\Delta\mapsto d$)
to simplify the presentations in the following sections. 

\beginsubsection{The Laplacian}\secdef{AppLaplacian}

Let $\phi$ be a function which is constant on each 2-sphere of a spherically
symmetric space. If we put $v^\mu = \phi{}^{;\mu}$ then at the origin of each
RNC cell we have $v^\mu = \phi_{,\mu}$ since $g_{\mu\nu} = {\rm diag}(1,1,1)$
and $0=\Gamma^\mu{}_{\alpha\beta}$ at the origin of each RNC frame. Since
$\phi$ is constant on each 2-sphere, we must have, 
$$
\phi_{,x} = 0,\hskip 1cm
\phi_{,y} = 0,\hskip 1cm
\phi_{,z} = {d\phi\over dz}
\eqno\eqnnxt
$$
on the $z-$axis. Using this and the results of the previous section we find
that, on the $z-$axis
\bgroup
$$
\openup5pt
\eqalignno{%
\phi_{,xy} &= v^x{}_{,y} = 0&\eqnnxt\cr
\phi_{,xz} &= v^x{}_{,z} = {dv^x\over dz} = 0&\eqnnxt\cr
\phi_{,yz} &= v^y{}_{,z} = {dv^y\over dz} = 0&\eqnnxt\cr
\phi_{,xx} &= v^x{}_{,x} = {1\over L_{xx}}{dL_{xx}\over dz}{d\phi\over dz}&\eqnnxt\cr
\phi_{,yy} &= v^y{}_{,y} = {1\over L_{xx}}{dL_{xx}\over dz}{d\phi\over dz}&\eqnnxt\cr
\phi_{,zz} &= v^z{}_{,z} = {d^2\phi\over dz^2}&\eqnnxt\cr}
$$
\egroup
and thus
$$
\nabla^2 \phi = {d^2\phi\over dz^2} + {2\over L_{xx}}{dL_{xx}\over dz}{d\phi\over dz}
\eqno\eqnnxt
$$
It is easy to see that this leads to the correct Laplacian for flat space
(i.e. put $L_{xx} = r\Delta\theta$ and $r=z$ where $r$ is the usual radial
coordinate and $z$ is the proper distance measured along the radial axis).

\beginsubsection{The ADM Constraints}\secdef{AppConstraints}

The Hamiltonian constraint
$$
0 = R + K^2 - K^{\mu\nu} K_{\mu\nu}
\eqno\eqnrfr{ADMHam}
$$
can be evaluated directly on the lattice as each term on the right hand side
is known at each vertex of the lattice. Furthermore, since the lattice is
spherically symmetric the only terms which survive are those that contain
$R_{xyxy}, R_{xzxz}, K_{xx}$ and $K_{zz}$. This leads to
$$
0 = R_{xyxy} + 2R_{xzxz} + K^2_{xx} + 2K_{xx}K_{zz}
\eqno\eqnrfr{RNCHam}
$$
The standard form of the ADM momentum constraints are
$$
0 = K{}_{\vert\mu} - K_{\mu}{}^{\nu}{}_{\vert\nu}
\eqno\eqnnxt
$$
where $K= K^\mu{}_{\mu}$. These equations require more care as they contain
covariant derivatives. At the origin of a RNC frame we have $g_{\mu\nu} = {\rm
diag}(1,1,1)$ and $0=\Gamma^\mu_{\alpha\beta}$, and thus the constraints may be
reduced to
$$
0 = K{}_{,\mu} - K_{\mu}{}^{\nu}{}_{,\nu}
\eqno\eqnnxt
$$
Each of the partial derivatives can be evaluated using the methods of the
previous section (though modified for use on a two index tensor). The results
are
$$
\openup5pt
\Displaylines{%
K_{,x} = K_{,y} = 0&\eqnnxt\cr
K_{,z} = {dK_{xx}\over dz} + {dK_{yy}\over dz} + {dK_{zz}\over dz}&\eqnnxt\cr
K_{x}{}^{\mu}{}_{,\mu} = K_{y}{}^{\mu}{}_{,\mu} = 0&\eqnnxt\cr
K_{z}{}^{\mu}{}_{,\mu} = {1\over L_{xx}}{dL_{xx}\over dz}
   \left(2K_{zz}-K_{xx}-K_{yy}\right)
   +{dK_{zz}\over dz}&\eqnnxt\cr}
$$
One also finds that terms such as $K_{xz,x}\not=0$ even though $K_{xz}=0$
at the origin of each RNC cell. This fact has been used in the above results.

The only non-trivial momentum equation is that for
$\mu=z$ and this leads to
$$
0 = {d\left(L_{xx} K_{xx}\right)\over dz} - K_{zz}{dL_{xx}\over dz}
\eqno\eqnrfr{RNCMom}
$$
where we have used the fact that $K_{xx}=K_{yy}$ in our RNC frame.

\beginsubsection{Bianchi Identities}\secdef{AppBianchi}

The Bianchi identities in a RNC frame are just
$$
0 =  R_{\mu\nu\alpha\beta,\rho} 
   + R_{\mu\nu\beta\rho,\alpha}
   + R_{\mu\nu\rho\alpha,\beta}
\eqno\eqnnxt
$$
In a spherically symmetric space there is only one non-trivial Bianchi identity,
namely,
$$
0 = R_{xyxy,z} + R_{xyyz,x} + R_{xyzx,y}
\eqno\eqnnxt
$$
The only non-zero components of the Riemann tensor at the origin of a RNC cell
are $R_{xzxz}=R_{yzyz}$ and $R_{xyxy}$ (and others obtained by standard
symmetries in the indicies). However, like the calculations above for the
momentum constraints, we find that many terms including $R_{xyyz,x}$ and
$R_{xyzx,y}$ are not zero. This fact is a simple consequence of the mixing that
occurs amongst the non-zero $R_{\mu\nu\alpha\beta}$ brought about by the
rotation matrices. Rather than list all of the non-zero derivatives we
shall list only those that we need for the above Bianchi identity. They are
$$
\openup5pt
\eqalignno{%
R_{xyxy,z} &= {dR_{xyxy}\over dz}&\eqnnxt\cr
R_{xyyz,x} &= {1\over L_{xx}}{dL_{xx}\over dz}
              \left( R_{xzxz}-R_{xyxy}\right)&\eqnnxt\cr
R_{xyxz,y} &={1\over L_{xx}}{dL_{xx}\over dz}
              \left( R_{xzxz}-R_{xyxy}\right)&\eqnnxt\cr}
$$
which when substituted into the above equation leads to
$$
0 = {d\left(L^2_{xx} R_{xyxy}\right)\over dz} - R_{xzxz} {dL^2_{xx}\over dz}
\eqno\eqnrfr{ExactRxyxy}
$$

\beginsubsection{Non-uniform finite differences}\secdef{AppFiniteDiffs}

By applying standard Taylor series expansions to a smooth function $f(z)$ it
is easy to derive the following second order accurate finite difference
approximations
$$
\openup+5pt
\eqalignno{%
{df\over dz} &= {1\over L^\p_{zz}+L^\m_{zz}}
  \left(L^\m_{zz}\left({f^\p-f^\o\over L^\p_{zz}}\right)
       +L^\p_{zz}\left({f^\o-f^\m\over L^\m_{zz}}\right)\right)
&\eqndef{FirstDeriv}\cr
{d^2f\over dz^2} &= {2\over L^\p_{zz}+L^\m_{zz}}
              \left({f^\p-f^\o\over L^\p_{zz}}
                   +{f^\m-f^\o\over L^\m_{zz}}\right)
&\eqndef{SecondDeriv}\cr}
$$
for a non-uniform lattice (where $L^\p_{zz} \not= L^\m_{zz}$ are the lattice
spacings). Centred finite differences are not appropriate for two simple
reasons. First, we chose our initial data $L_{zz}$ to be non-uniform. Second, even
if we did choose an initially uniform lattice, the subsequent dynamics
($dL_{zz}/dt\not=0$) would immediately produce a non-uniform lattice.

These approximations are used at various places in the text (e.g. for $dL_{xx}/dz$
and $d^2N/dz^2$).

The only exception to the above was in the discretisation of the Bianchi
identities. This equation was approximated at the centre of the radial struts by
a forward finite difference operator ${df/dz} = {(f^\p-f^\o)/ L^\p_{zz}}$ and
by setting $R_{xzxz} = (R^\p_{xzxz}+R^\o_{xzxz})/2$.

\beginsection{References}

\LongCitations

\bgroup

\def\REF #1!#2!#3!#4!{\vskip 0pt plus .1\vsize\penalty -250%
                      \vskip 0pt plus-.1\vsize%
                      \rm\item{\bf [#1]}
                      \rm\ignorespaces #2\ \penalty+250
                      \sl\ignorespaces #3\ \penalty+250
                      \rm\ignorespaces #4\ \penalty+250
                      \vskip 2pt}


\overfullrule=0pt

\cite {AlcubierreEtalB}
\cite {AlcubierreEtalC}
\cite {Baumgarte}
\cite {KellyEtal}
\cite {ScheelB}
\cite {Shibata}
\cite {Yoneda}
\cite {Reula}
\cite {AlcubierreEtalD}
\cite {ScheelA}
\cite {LeoSLGR}
\cite {LeoADMSLGR}
\cite {BernsteinEtalB}
\cite {BernsteinEtalC}
\cite {AnninosEtalD}
\cite {AnninosEtalB}
\cite {LeoRNC}
\cite {PetrovRNC}
\cite {EisenhartRNC}
\cite {MTWRNC}
\cite {BergerOliger}
\cite {Estabrook}
\cite {MTW}
\cite {Beig}
\cite {Petrich}
\cite {Reinhart}
\cite {Smarr}
\cite {HawkingEllis}
\cite {ChristoKlainA}

\egroup

\ShortCitations

\vfill\eject

\twelvepointsgl
\nopagenumbers
\overfullrule=0pt
\parskip=10pt plus 4pt minus 2pt
%

\ShortCitations

\def\MyBig{\seventeenpointsgl}

\def\eps{\epsilon}
\def\Oeps(#1){{\cal O}(\eps^{#1})}

\def\l{{\hbox{\twelvepoint\tt l}}}
\def\o{{\hbox{\twelvepoint\tt o}}}
\def\r{{\hbox{\twelvepoint\tt r}}}
\def\p{{\hbox{\twelvepoint\tt +}}}
\def\m{{\hbox{\twelvepoint\tt -}}}
%
%
\parindent=0pt
%
{\seventeenpoint
\leftline{\bf Figure captions}}
\vskip 10pt plus 5pt

{\bf Figure \figrfr{RNCThroat}.}\ An embedding diagram for a 2-dimensional
slice of the spherically symmetric 3-geometry of a Schwarzschild black hole.
The lattice is constructed as a ladder, with one end on the throat and the
other in the weak field region of the black hole. The two radial edges of the
ladder are radial geodesics, while each rung is a geodesic segment of the full
3-metric. Thus each rung is not confined to the 2-spheres (except at the
throat).

{\bf Figure \figrfr{RNCLadder}.}\ The ladder on which the numerical solution
was built. Successive computational cells overlap by sharing two successive
rungs of the ladder. In this diagram there are just three computational cells.
The production runs employed 800 rungs and had an outer boundary set at
approximately $r=200M$.

{\bf Figure \figrfr{RNCCell}.}\ A typical computational cell. The coordinate
frame has been oriented so that the $z$-axis points in the usual radial
direction while the origin has been located so that the $z$ coordinate of
vertices $(1)$ and $(2)$ equals zero. The $y-$axis points directly into the
page and thus it has been suppressed.

{\bf Figure \figrfr{RNCBianchi}.}\ The three neighbouring computational cells
used in applying the Binachi identities. For simplicity we have not drawn the
middle rungs in each cell and we have drawn each geodesic segment as a straight
line. Once again the $y$-axis has been suppressed.

{\bf Figure \figrfr{LxxRunCD}--\figrfr{LapseRunCD}.}\ These figures show the
evolution of the basic lattice data for $t=10m$ to $t=100m$ in steps of $10m$
and also from $t=100m$ to $t=1000m$ in steps of $100m$. All of the figures
display a smooth evolution with no signs of any instabilities. In each of these
figures we can clearly see the stretching that occurs in the grid.  On each
curve we have used a diamond to mark the location of the apparent horizon. In
many of the figures for $0<t<100m$ we can clearly see the loss of resolution
brought on by the grid stretching (e.g. the fall off in the plateau of $K_{xx}$
and in the decay in the sharp peak of $K_{zz}$). These effects are much more
pronounced in the long term evolution, $100m<t<1000m$.  Note that at these late
times the radial legs near the apparent horizon have been stretched by almost a
factor of 100 while the rungs have been shrunk be a factor of approximately
100. This is a severe change in shape of the grid and so its not surprising
that the accuracy has been lost. Each of the plots in figures
\figrfr{LxxRunCD}--\figrfr{LapseRunCD} were produced using the grid centred
scheme on a Bernstein, Hobill and Smarr grid with 800 grid points. The plots
for $0<t<100m$ were restricted to a proper distance of $100m$ simply to better
display the changes in the grid. At $t=100m$ the grid extends out to a proper
distance of over $265m$.

{\bf Figure \figrfr{Resolution}.}\ This displays the curvature term $R_{xyxy}$
at $t=100m$ for four different resolutions of 100, 200, 400 and 800 grid points.
This clearly shows that the ability to maintain a flat plateau behind the
horizon is compromised when there is a loss of resolution near the apparent
horizon.

{\bf Figure \figrfr{Collapse}.}\ This shows the exponential collapse of the
lapse at the throat. The work of Beig \cite{Beig} shows that $N(r=0)\sim
e^{kt}$ thus a plot of $\ln N$ versus $t$ should be a straight line. For $t=0$
to $t=100m$ the line is very straight, while for longer times a slight bend
does occur. The lapse at the throat at $t=100m$ is approximately
$3.4\times10^{-23}$ while at $t=1000m$ is the lapse is of the order of
$10^{-145}$.

{\bf Figure \figrfr{Horizon}.}\ The size of the apparent horizon from $t=0$ to
$t=100m$ for four different resolutions of 100, 200, 400 and 800 grid points.

\vfill\eject
%
\newbox\boxa
\newbox\image
\newbox\xlabel
\newbox\ylabel
\newbox\zlabel
\newbox\caption
\newbox\imagea
\newbox\imageb
\newbox\ylabela
\newbox\ylabelb
%
\SetNoBorder
\SetBase(51,761)
\setbox\caption=\hbox to 15cm{\hsize=15cm\vtop{
\centerline{{\bf Figure \figdef{RNCThroat}.}\ The embedding diagram.}}}
\setbox\image=\hbox{\bBoxedEPSF{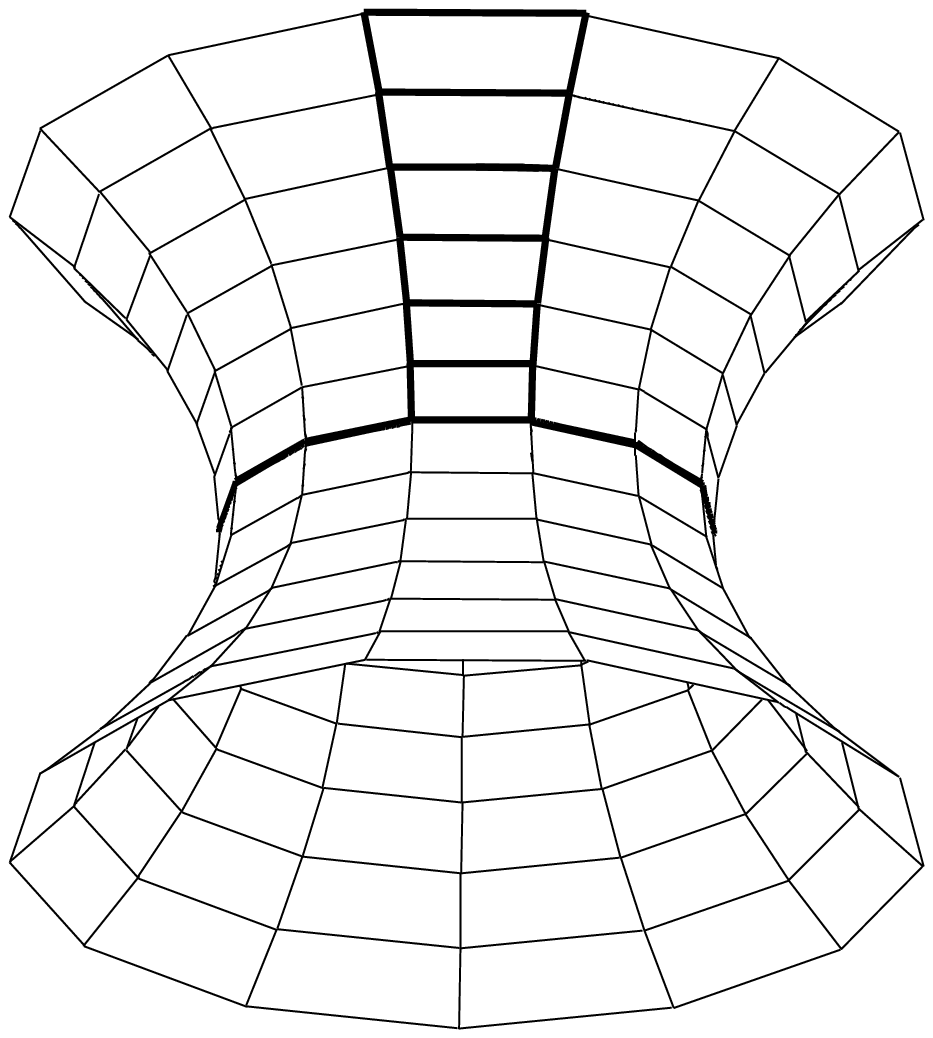 scaled 1000}}
\setbox\boxa=%
\hbox to 18.0cm{
\vtop to 18.5cm{
\Border
\atptalt( 20.0,270.0){\box\image}
\atptalt( 60.0,250.0){\box\caption}
\vfill}\hfill}
%
%
\centerline{\box\boxa}\vfill\eject
\SetNoBorder
\SetBase(51,761)
\setbox\caption=\hbox to 15cm{\hsize=15cm\vtop{
\centerline{{\bf Figure \figdef{RNCLadder}.}\ The ladder lattice.}}}
\setbox\image=\hbox{\bBoxedEPSF{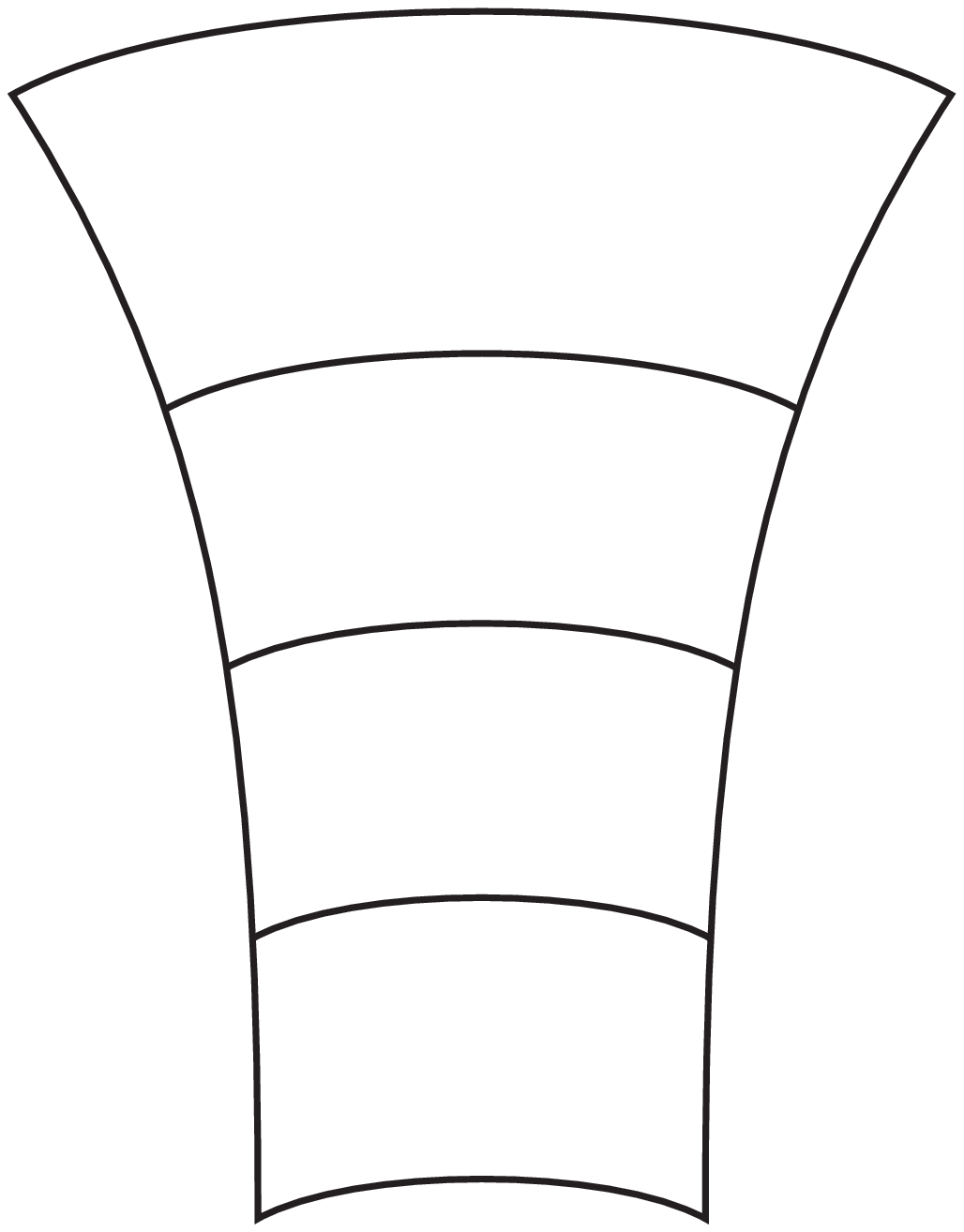 scaled 1000}}
\setbox\boxa=%
\hbox to 18.0cm{
\vtop to 18.5cm{
\Border
\atptalt(125.0,330.0){\box\image}
\atptalt( 60.0,250.0){\box\caption}
\vfill}\hfill}
%
%
\centerline{\box\boxa}\vfill\eject
\SetNoBorder
\SetBase(51,761)
\setbox\caption=\hbox to 15cm{\hsize=15cm\vtop{
\centerline{{\bf Figure \figdef{RNCCell}.}\ The typical computational cell.}}}
\setbox\image=\hbox{\bBoxedEPSF{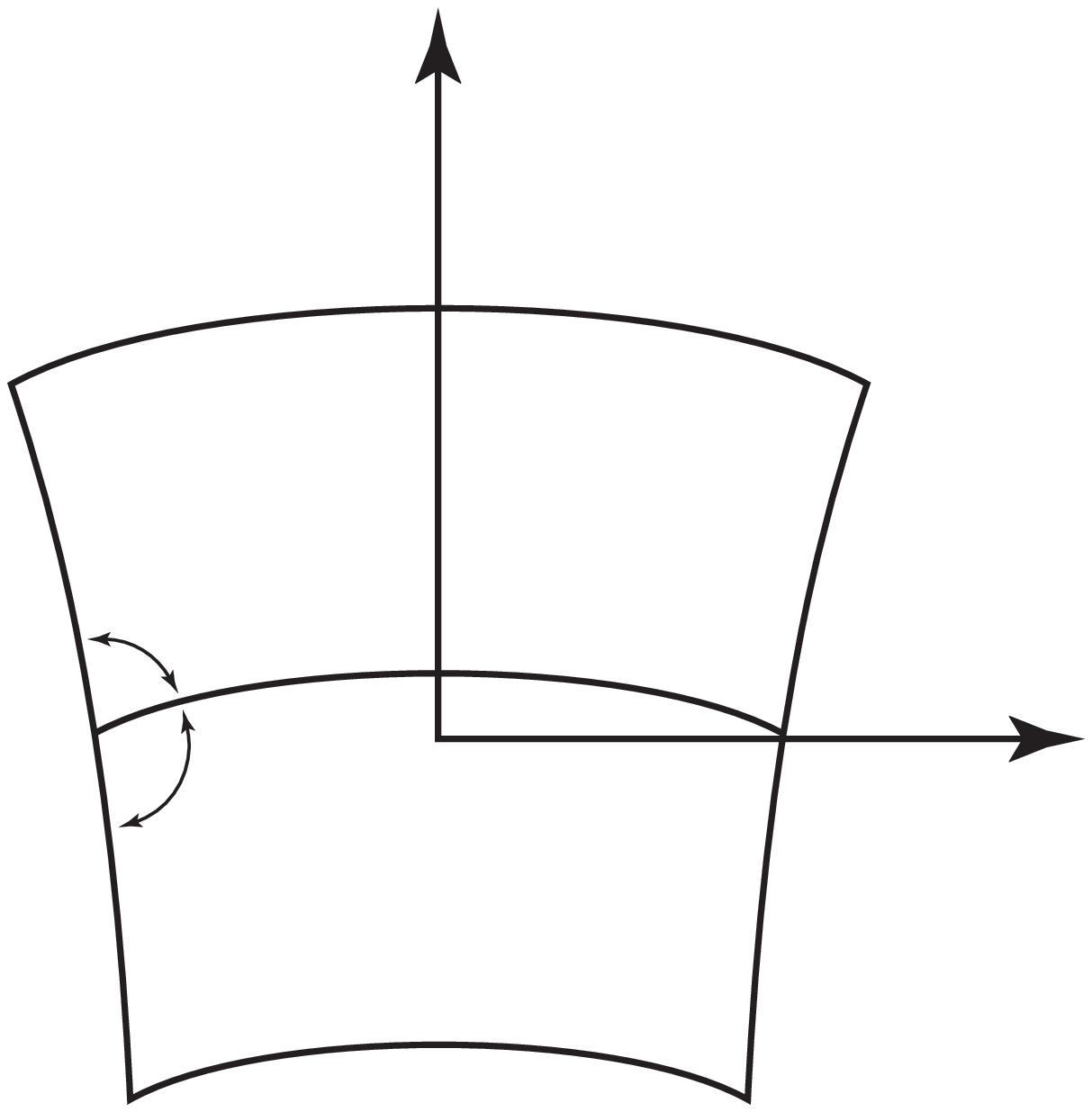 scaled 1000}}
\def\Lstrut{\vrule height 15pt depth 2pt width 0pt\relax}
\setbox\boxa=%
\hbox to 18.0cm{
\vtop to 18.5cm{
\Border\MyBig
\atptalt(130.0,320.0){\box\image}
\atptalt( 60.0,250.0){\box\caption}
\atptalt(495.0,460.0){$X$}
\atptalt(300.0,685.0){$Z$}
\atptalt(320.0,355.0){$L{\Lstrut}^{\m}_{xx}$}
\atptalt(330.0,475.0){$L{\Lstrut}^{\o}_{xx}$}
\atptalt(355.0,590.0){$L{\Lstrut}^{\p}_{xx}$}
\atptalt(131.0,372.0){$L{\Lstrut}^{\m}_{zz}$}
\atptalt(112.0,480.0){$L{\Lstrut}^{\p}_{zz}$}
\atptalt(130.0,315.0){$(2^\m)$}
\atptalt(390.0,315.0){$(1^\m)$}
\atptalt(122.0,430.0){$(2)$}
\atptalt(400.0,410.0){$(1)$}
\atptalt( 90.0,540.0){$(2^\p)$}
\atptalt(425.0,540.0){$(1^\p)$}
\atptalt(180.0,477.0){$\theta^\p$}
\atptalt(195.0,405.0){$\theta^\m$}
\vfill}\hfill}
%
%
\centerline{\box\boxa}\vfill\eject
\SetNoBorder
\SetBase(51,761)
\setbox\caption=\hbox to 15cm{\hsize=15cm\vtop{
\centerline{{\bf Figure \figdef{RNCBianchi}.}\ Cells for the Bianchi identities.}}}
\setbox\image=\hbox{\bBoxedEPSF{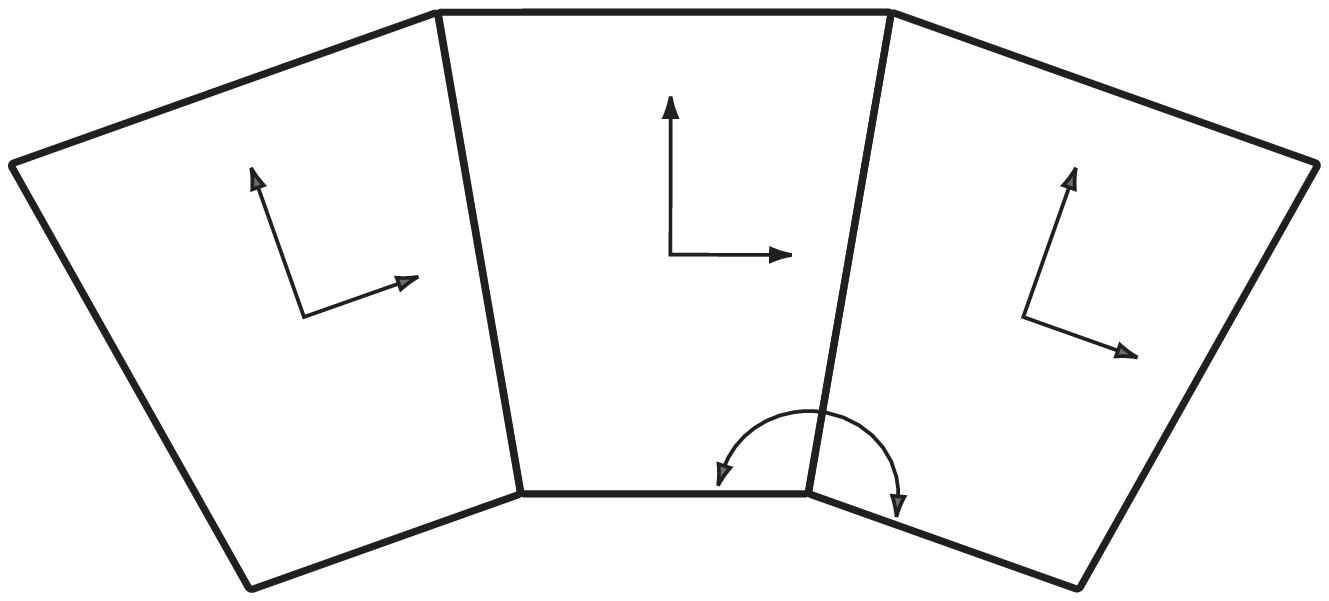 scaled 1000}}
\setbox\boxa=%
\hbox to 18.0cm{
\vtop to 18.5cm{
\Border\MyBig
\atptalt( 80.0,350.0){\box\image}
\atptalt( 60.0,250.0){\box\caption}
\atptalt(305.0,460.0){$x$}
\atptalt(285.0,500.0){$z$}
\atptalt( 95.0,530.0){$\sigma^{\l}$}
\atptalt(260.0,545.0){$\sigma^{\o}$}
\atptalt(410.0,530.0){$\sigma^{\r}$}
\atptalt(345.0,390.0){$\pi+\Delta\theta$}
\vfill}\hfill}
%
%
\centerline{\box\boxa}\vfill\eject
\SetNoBorder
\SetBase(51,761)
%
\setbox\imagea=\hbox{\bBoxedEPSF{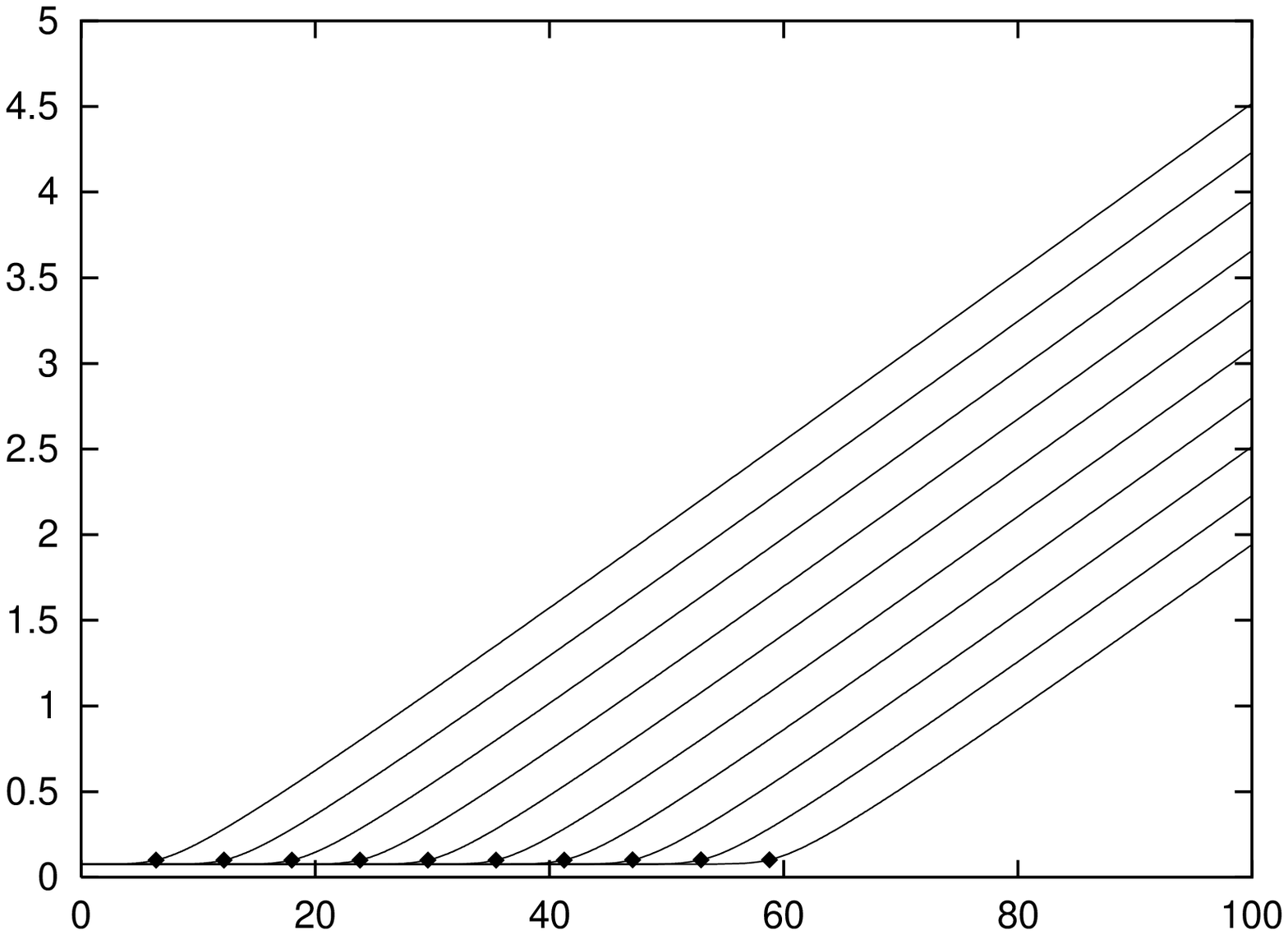 scaled 750}}
\setbox\imageb=\hbox{\bBoxedEPSF{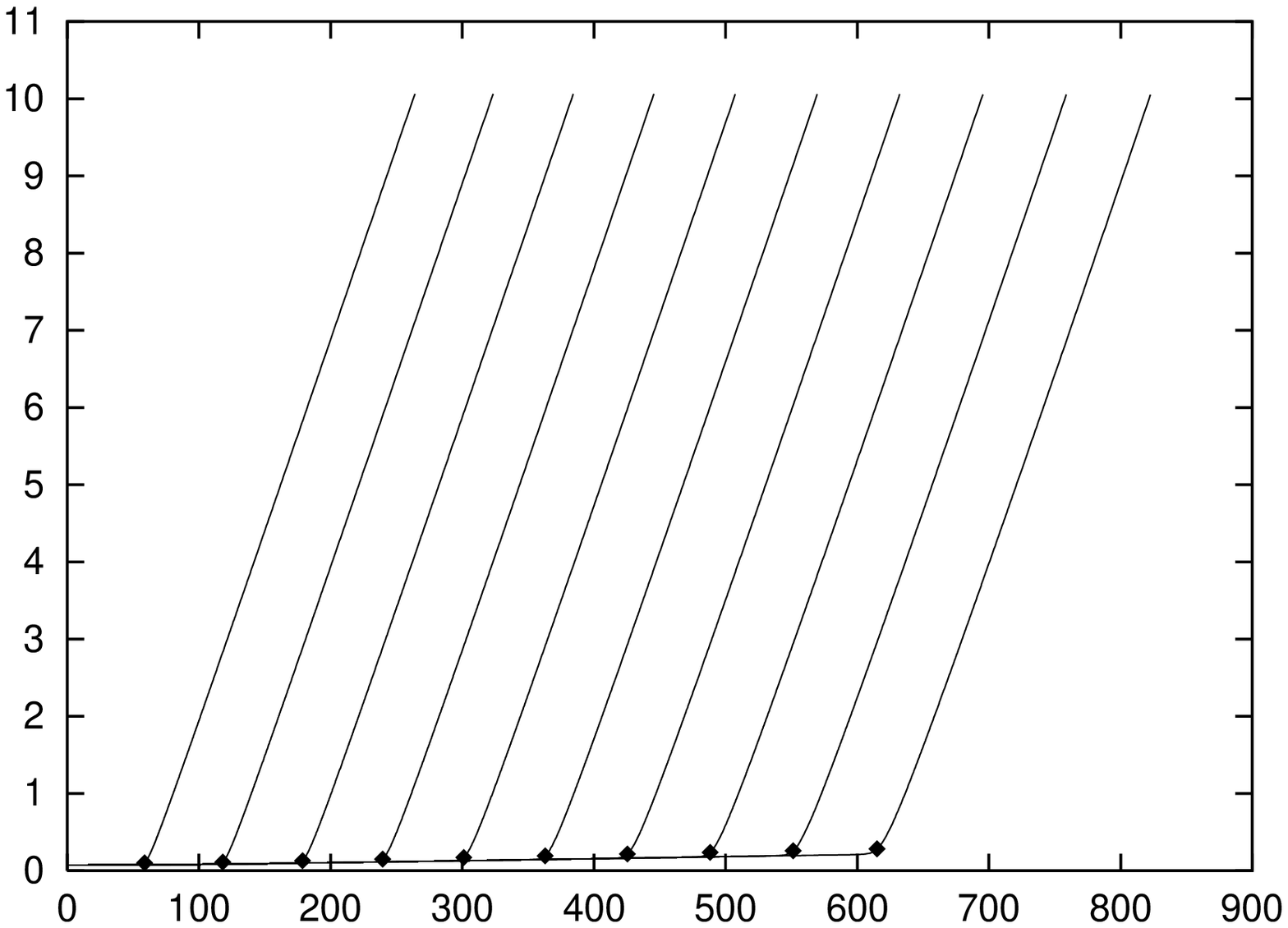 scaled 750}}
\setbox\ylabela=\hbox{\MyBig$L_{xx}$}
\setbox\ylabelb=\hbox{\MyBig$L_{xx}$}
\setbox\ylabela=\hbox{\rotl\ylabela}
\setbox\ylabelb=\hbox{\rotl\ylabelb}
\setbox\boxa=%
\hbox to 18.0cm{
\vtop to 18.5cm{
\Border
\atptalt( 90.0,490.0){\box\imagea}
\atptalt( 90.0,170.0){\box\imageb}
\atptalt(240.0,475.0){\MyBig Proper distance}
\atptalt(240.0,155.0){\MyBig Proper distance}
\atptalt( 80.0,625.0){\box\ylabela}
\atptalt( 80.0,305.0){\box\ylabelb}
\atptalt(210.0,790.0){\MyBig Figure \figdef{LxxRunCD}. Leg lengths}
\vfill}\hfill}
%
%
\centerline{\box\boxa}\vfill\eject
\SetNoBorder
\SetBase(51,761)
%
\setbox\imagea=\hbox{\bBoxedEPSF{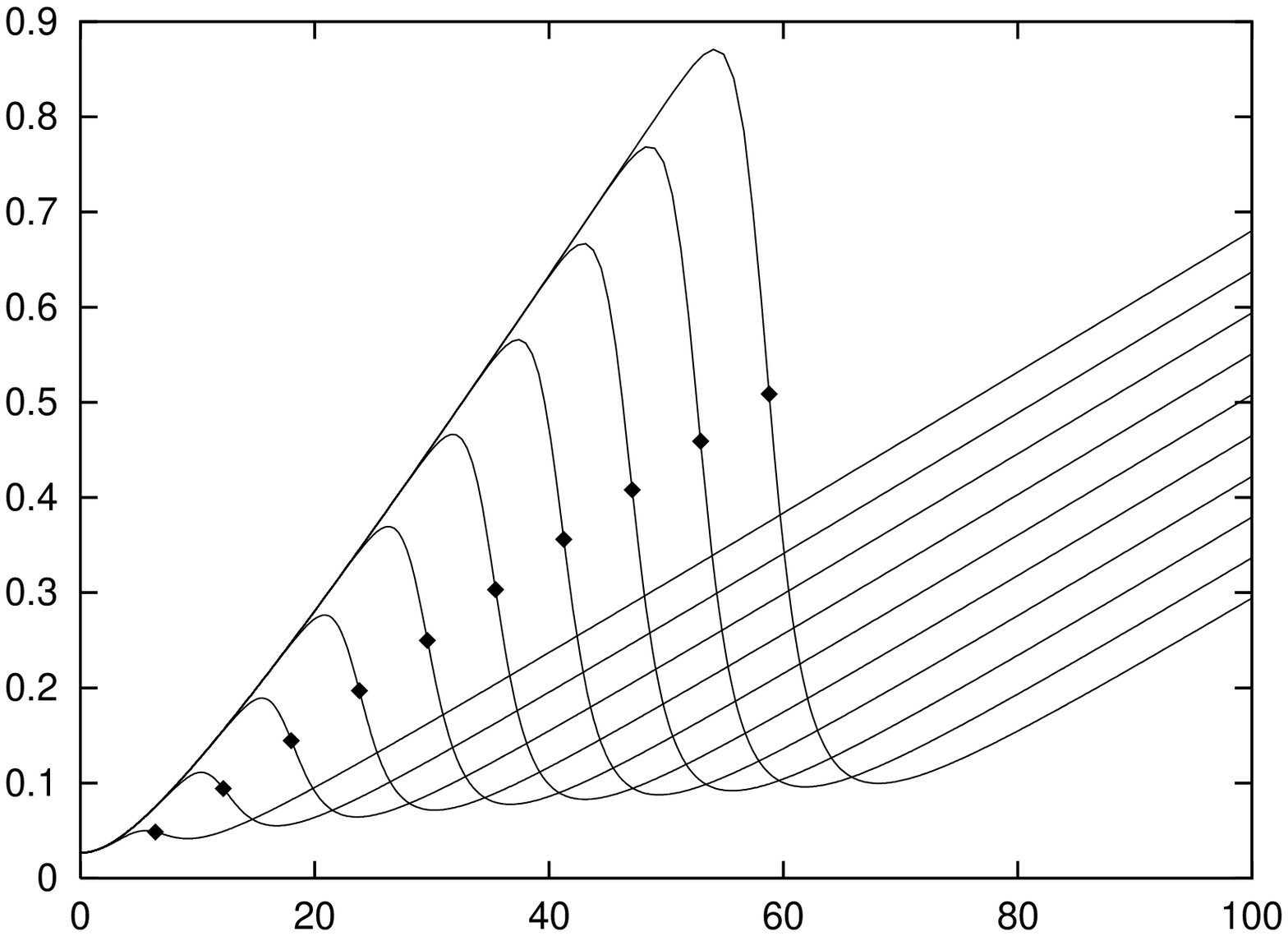 scaled 750}}
\setbox\imageb=\hbox{\bBoxedEPSF{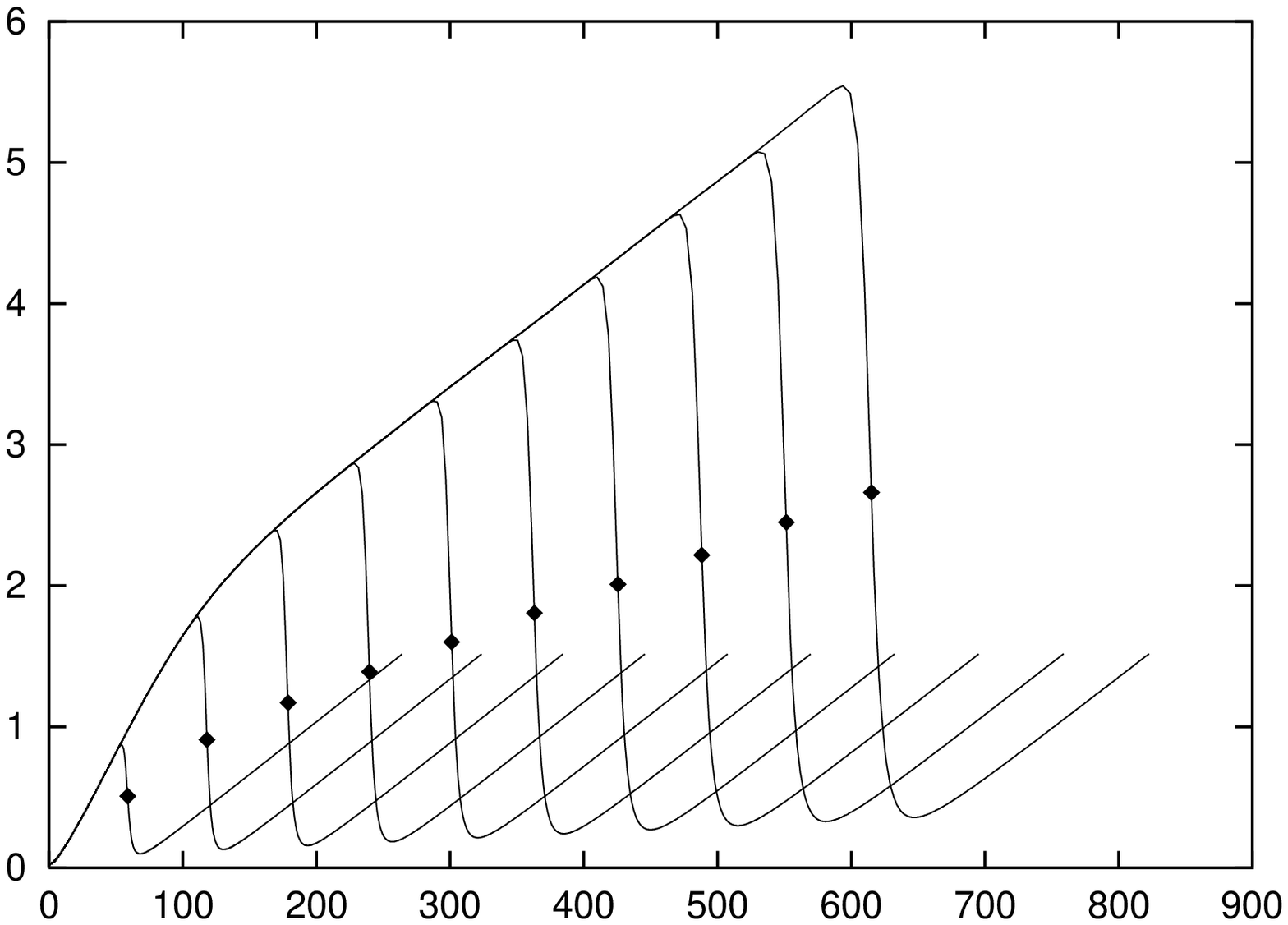 scaled 750}}
\setbox\ylabela=\hbox{\MyBig$L_{zz}$}
\setbox\ylabelb=\hbox{\MyBig$L_{zz}$}
\setbox\ylabela=\hbox{\rotl\ylabela}
\setbox\ylabelb=\hbox{\rotl\ylabelb}
\setbox\boxa=%
\hbox to 18.0cm{
\vtop to 18.5cm{
\Border
\atptalt( 90.0,490.0){\box\imagea}
\atptalt( 90.0,170.0){\box\imageb}
\atptalt(240.0,475.0){\MyBig Proper distance}
\atptalt(240.0,155.0){\MyBig Proper distance}
\atptalt( 80.0,625.0){\box\ylabela}
\atptalt( 80.0,305.0){\box\ylabelb}
\atptalt(210.0,790.0){\MyBig Figure \figdef{LzzRunCD}. Leg lengths}
\vfill}\hfill}
%
%
\centerline{\box\boxa}\vfill\eject
\SetNoBorder
\SetBase(51,761)
\setbox\imagea=\hbox{\bBoxedEPSF{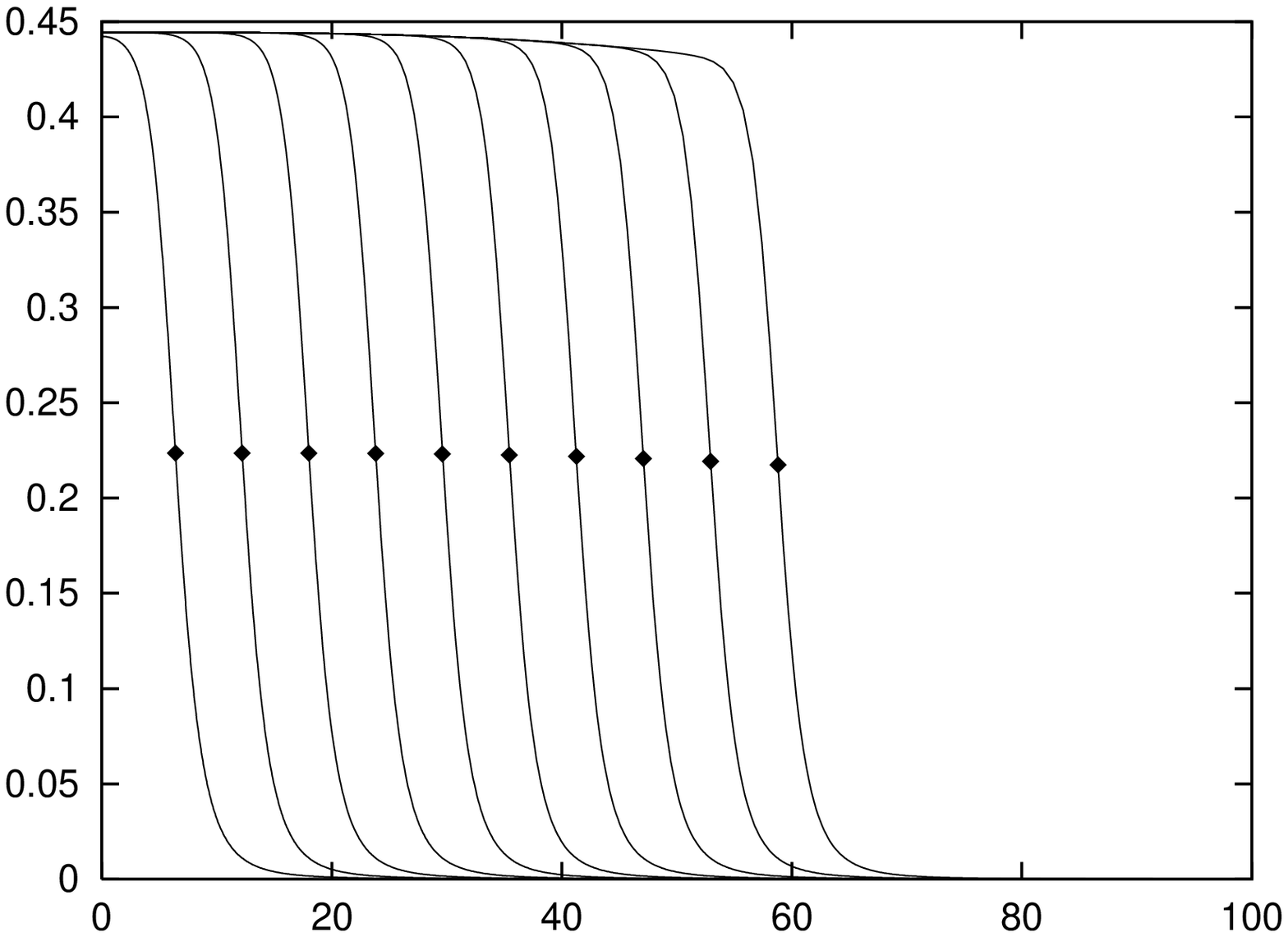 scaled 750}}
\setbox\imageb=\hbox{\bBoxedEPSF{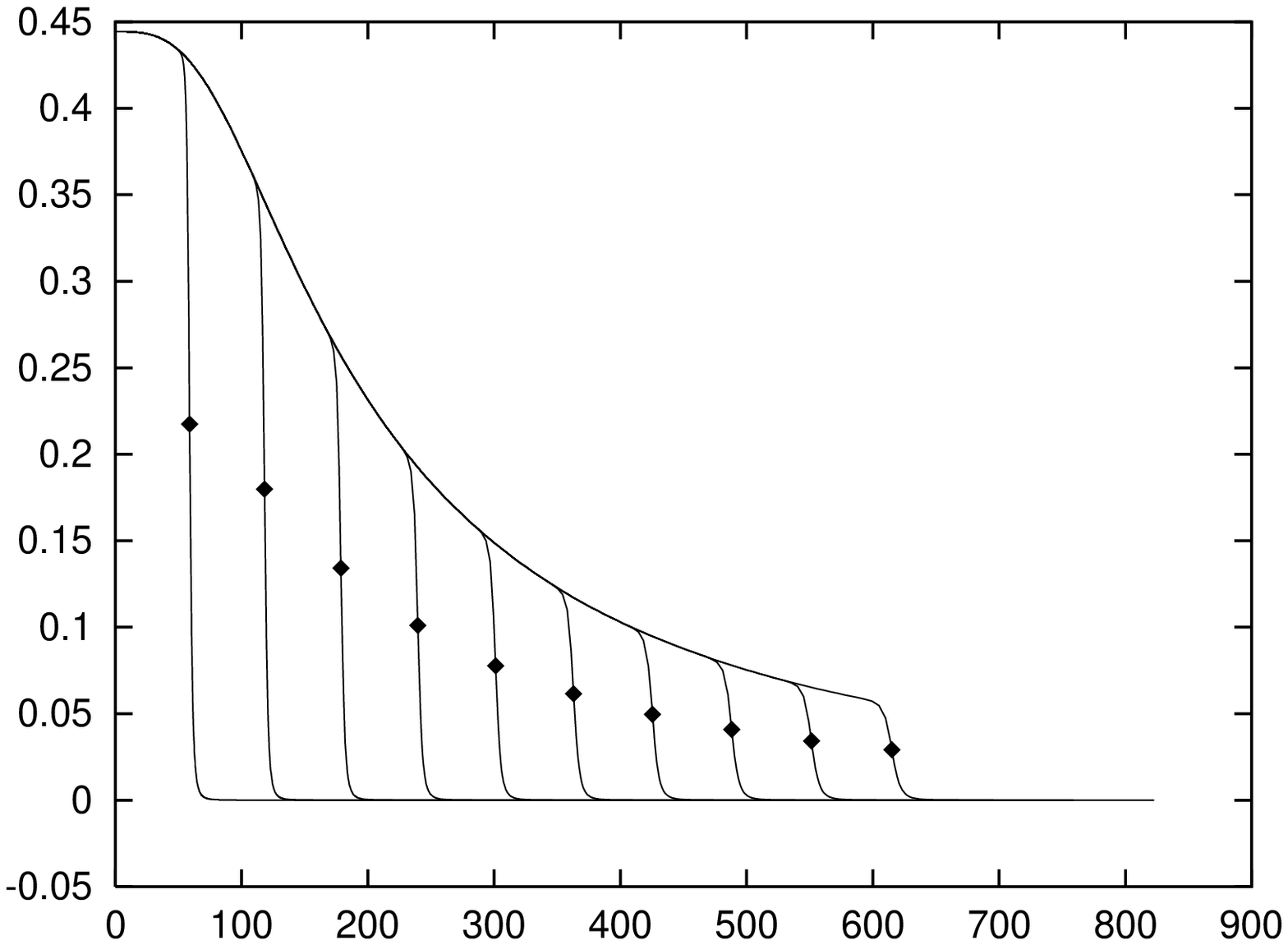 scaled 750}}
\setbox\ylabela=\hbox{\MyBig$R_{xyxy}$}
\setbox\ylabelb=\hbox{\MyBig$R_{xyxy}$}
\setbox\ylabela=\hbox{\rotl\ylabela}
\setbox\ylabelb=\hbox{\rotl\ylabelb}
\setbox\boxa=%
\hbox to 18.0cm{
\vtop to 18.5cm{
\Border
\atptalt( 90.0,490.0){\box\imagea}
\atptalt( 90.0,170.0){\box\imageb}
\atptalt(240.0,475.0){\MyBig Proper distance}
\atptalt(240.0,155.0){\MyBig Proper distance}
\atptalt( 80.0,610.0){\box\ylabela}
\atptalt( 80.0,290.0){\box\ylabelb}
\atptalt(185.0,790.0){\MyBig Figure \figdef{RxyxyRunCD}. Riemann curvatures}
\vfill}\hfill}
%
%
\centerline{\box\boxa}\vfill\eject
\SetNoBorder
\SetBase(51,761)
\setbox\imagea=\hbox{\bBoxedEPSF{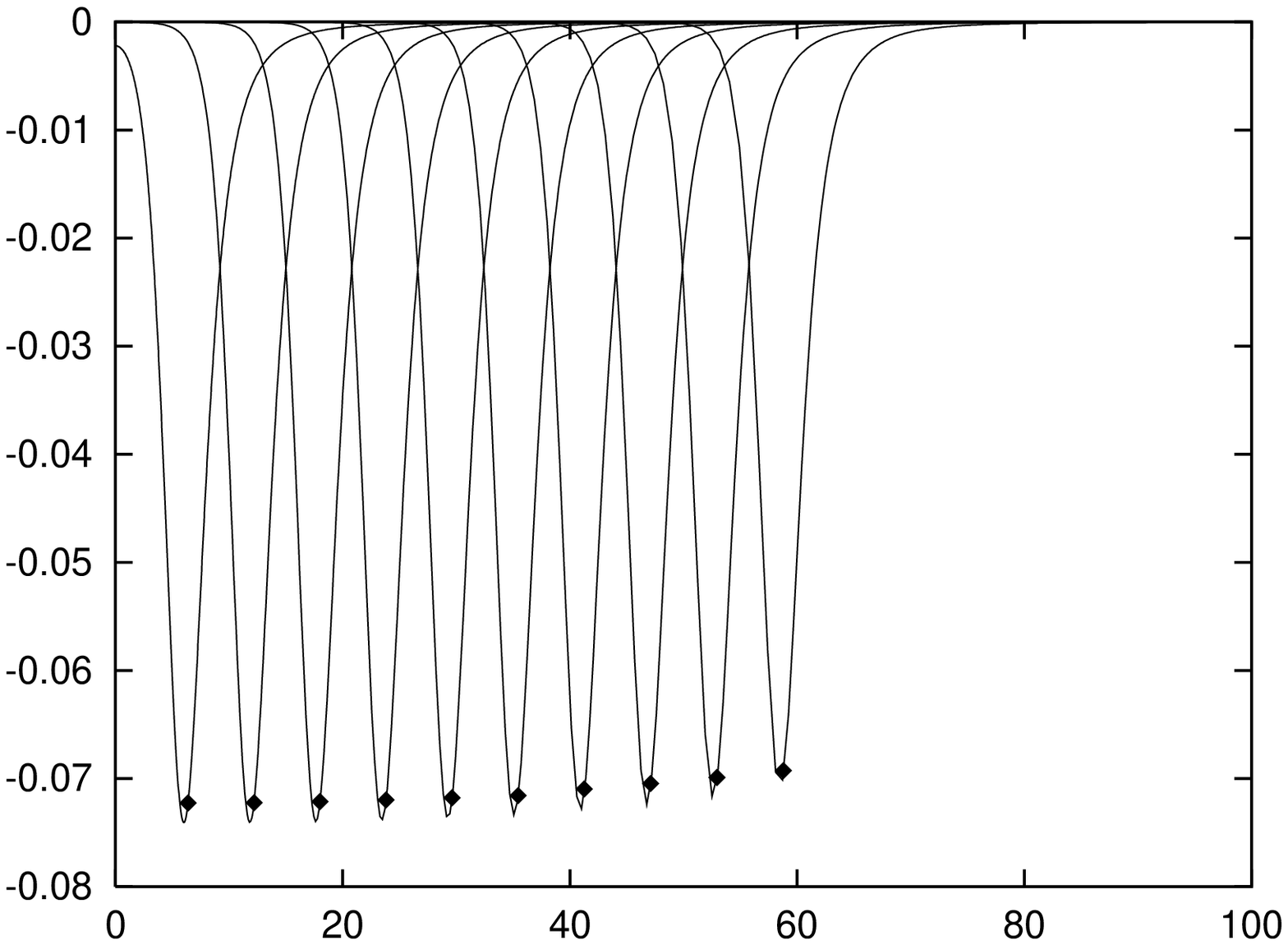 scaled 750}}
\setbox\imageb=\hbox{\bBoxedEPSF{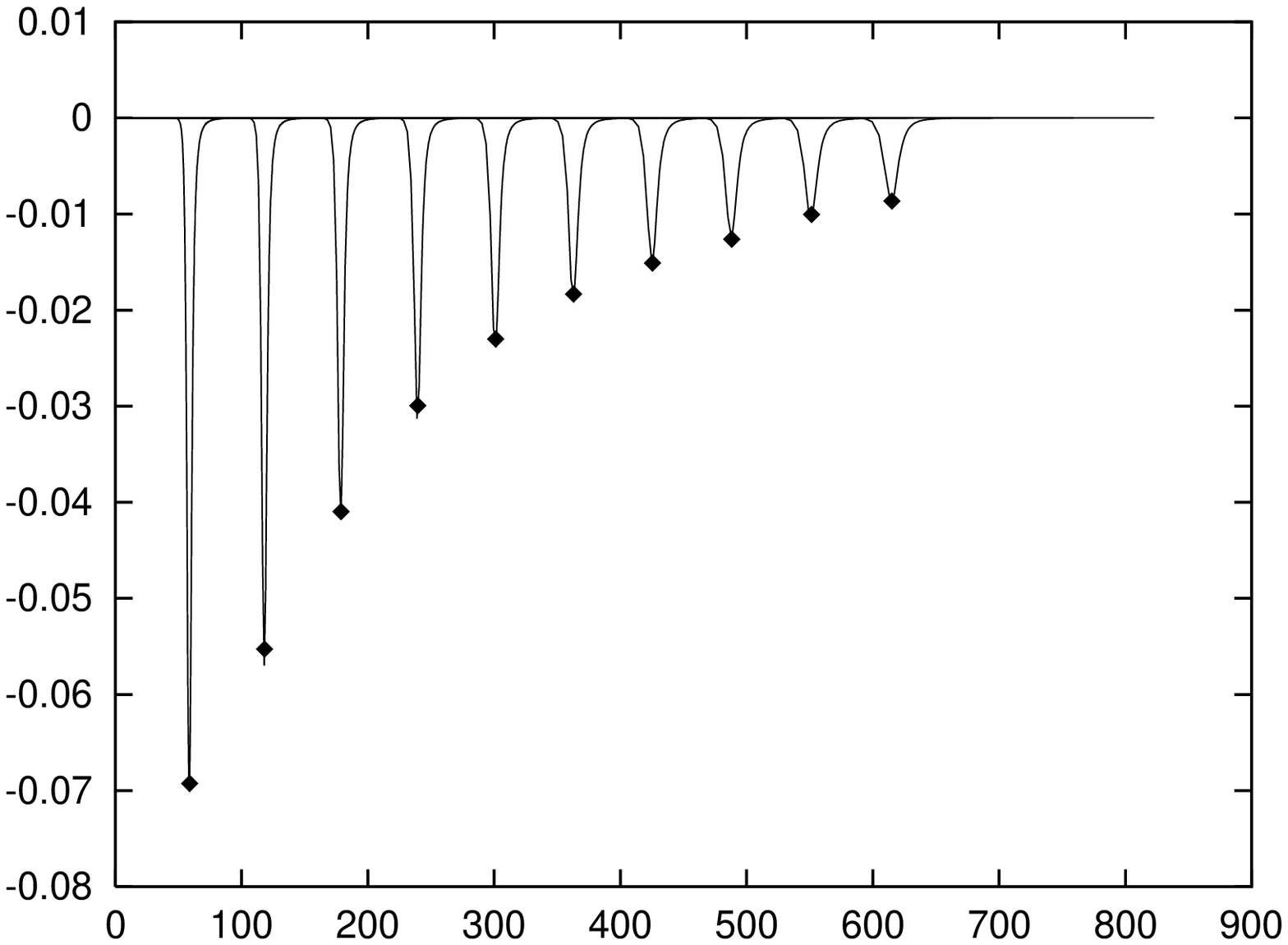 scaled 750}}
\setbox\ylabela=\hbox{\MyBig$R_{xzxz}$}
\setbox\ylabelb=\hbox{\MyBig$R_{xzxz}$}
\setbox\ylabela=\hbox{\rotl\ylabela}
\setbox\ylabelb=\hbox{\rotl\ylabelb}
\setbox\boxa=%
\hbox to 18.0cm{
\vtop to 18.5cm{
\Border
\atptalt( 90.0,490.0){\box\imagea}
\atptalt( 90.0,170.0){\box\imageb}
\atptalt(240.0,475.0){\MyBig Proper distance}
\atptalt(240.0,155.0){\MyBig Proper distance}
\atptalt( 80.0,610.0){\box\ylabela}
\atptalt( 80.0,290.0){\box\ylabelb}
\atptalt(185.0,790.0){\MyBig Figure \figdef{RxzxzRunCD}. Riemann curvatures}
\vfill}\hfill}
%
%
\centerline{\box\boxa}\vfill\eject
\SetNoBorder
\SetBase(51,761)
\setbox\imagea=\hbox{\bBoxedEPSF{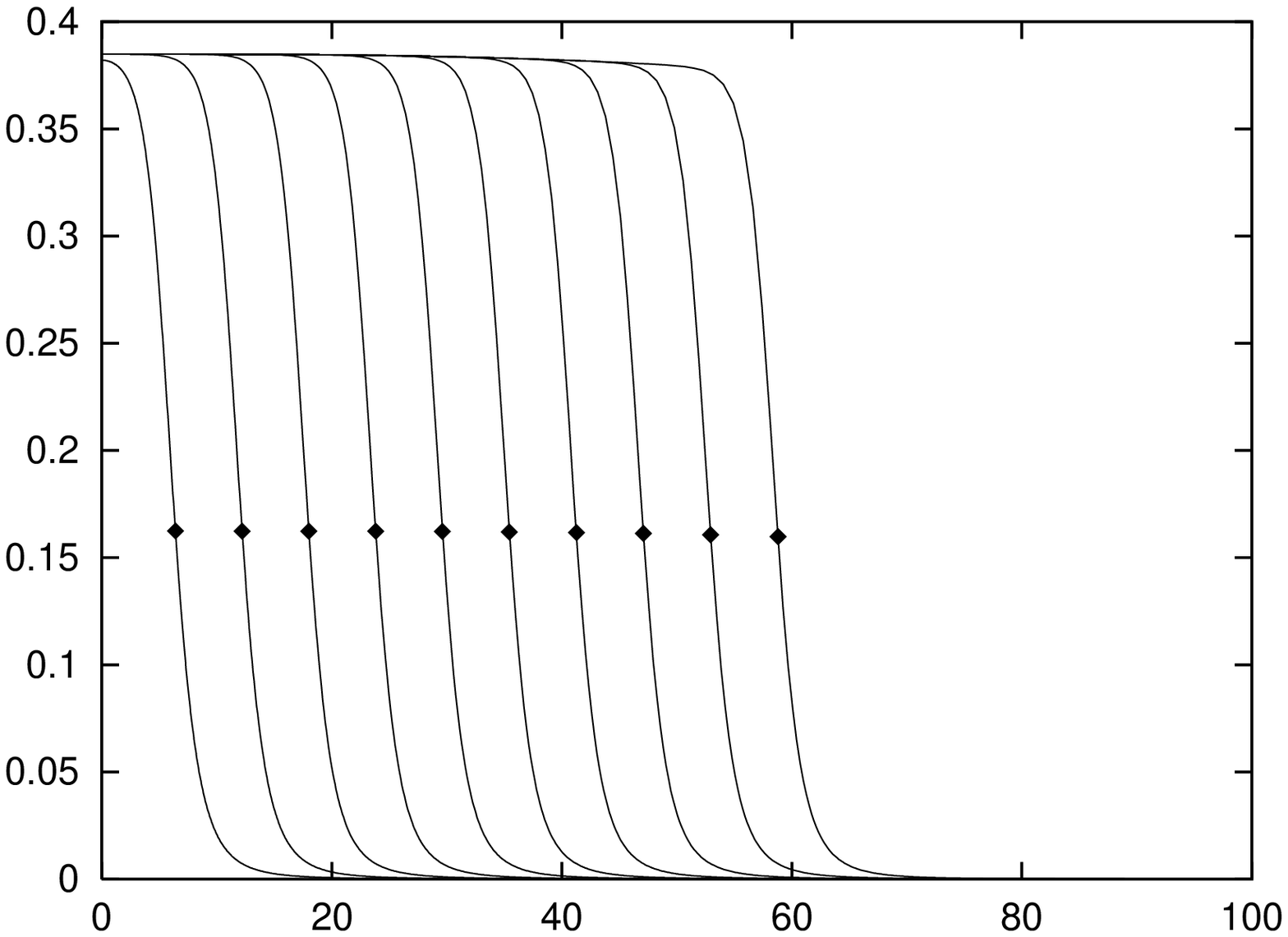 scaled 750}}
\setbox\imageb=\hbox{\bBoxedEPSF{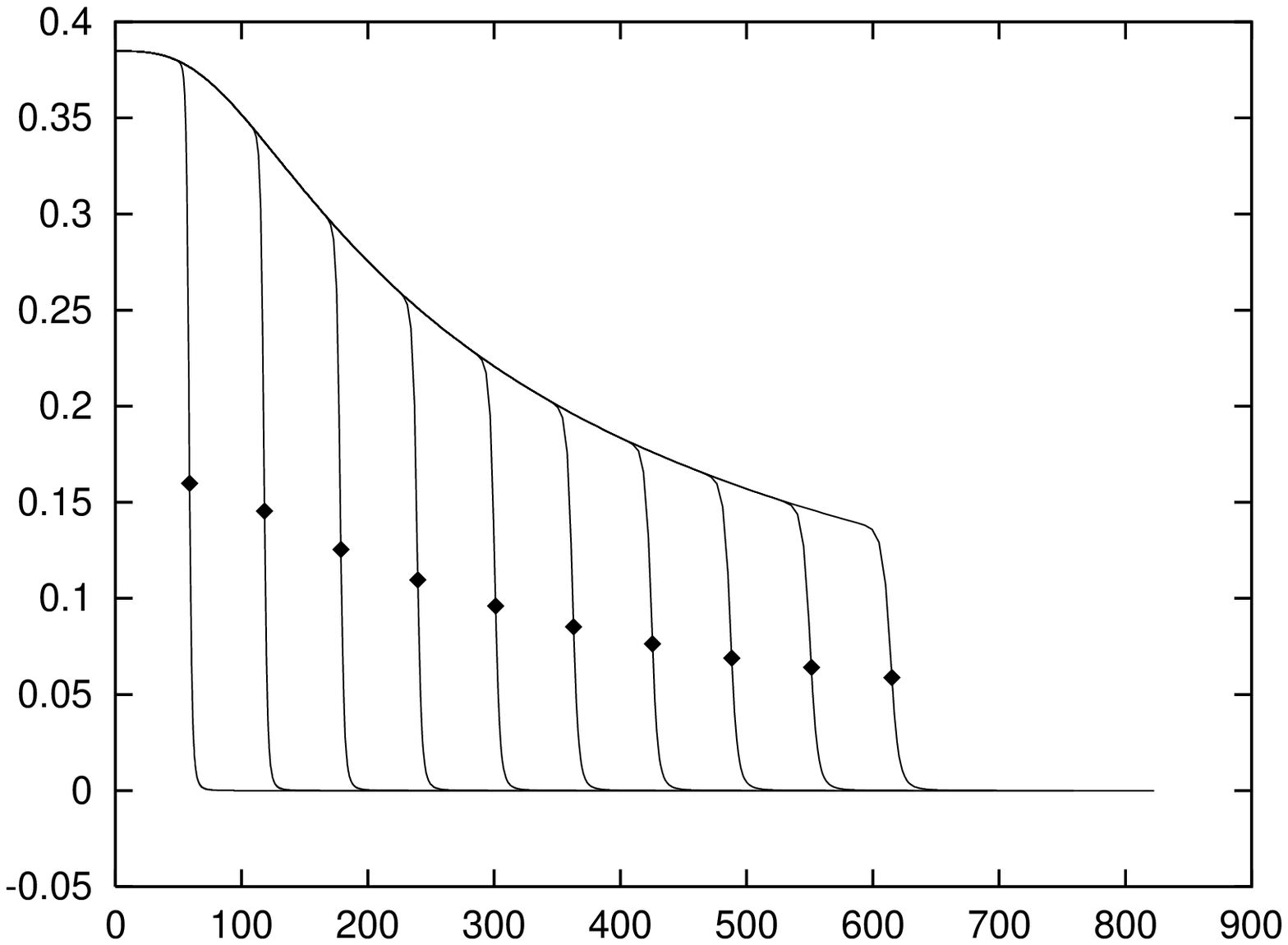 scaled 750}}
\setbox\ylabela=\hbox{\MyBig$K_{xx}$}
\setbox\ylabelb=\hbox{\MyBig$K_{xx}$}
\setbox\ylabela=\hbox{\rotl\ylabela}
\setbox\ylabelb=\hbox{\rotl\ylabelb}
\setbox\boxa=%
\hbox to 18.0cm{
\vtop to 18.5cm{
\Border
\atptalt( 90.0,490.0){\box\imagea}
\atptalt( 90.0,170.0){\box\imageb}
\atptalt(240.0,475.0){\MyBig Proper distance}
\atptalt(240.0,155.0){\MyBig Proper distance}
\atptalt( 80.0,620.0){\box\ylabela}
\atptalt( 80.0,300.0){\box\ylabelb}
\atptalt(185.0,790.0){\MyBig Figure \figdef{KxxRunCD}. Extrinsic curvatures}
\vfill}\hfill}
%
%
\centerline{\box\boxa}\vfill\eject
\SetNoBorder
\SetBase(51,761)
\setbox\imagea=\hbox{\bBoxedEPSF{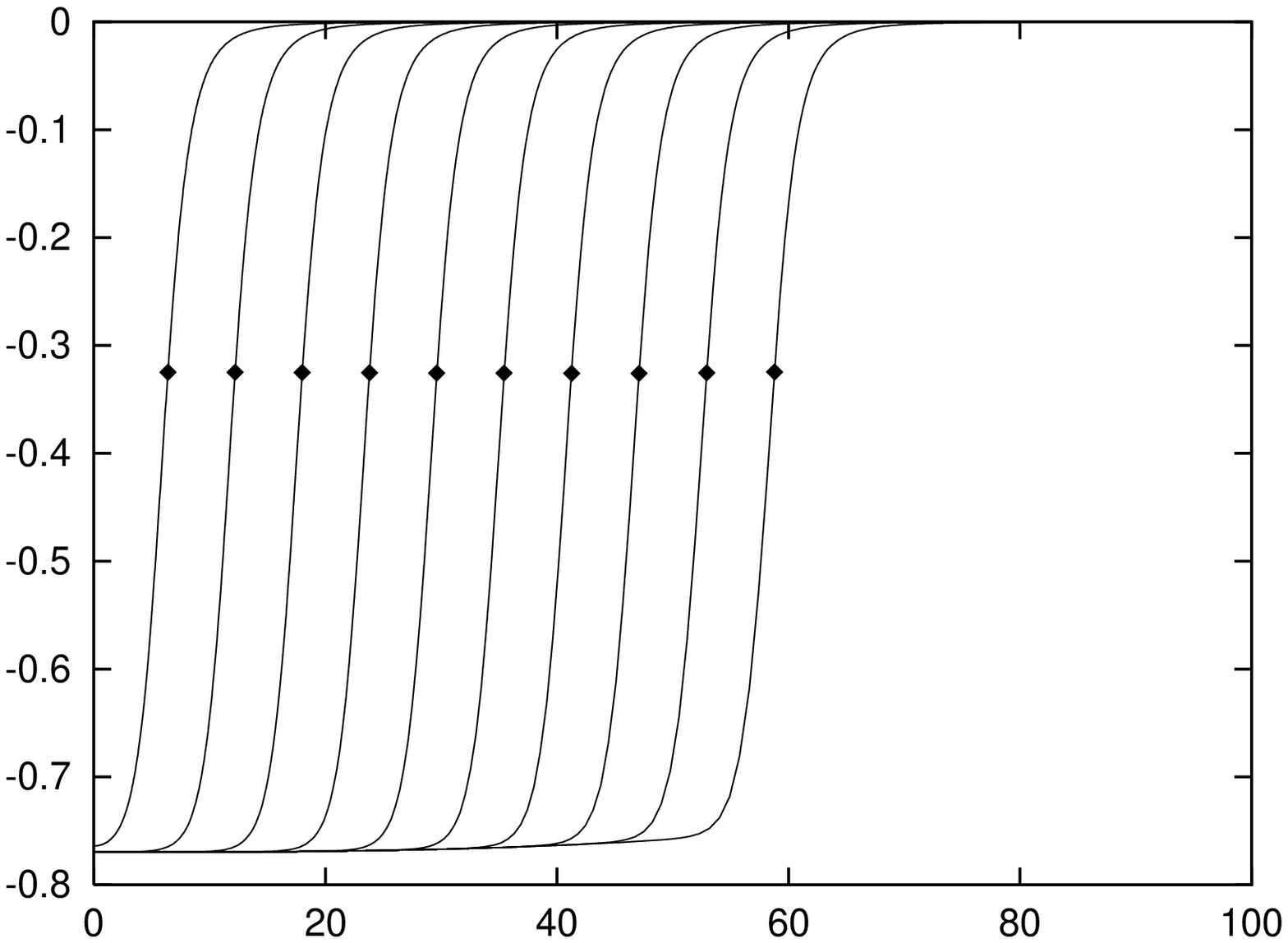 scaled 750}}
\setbox\imageb=\hbox{\bBoxedEPSF{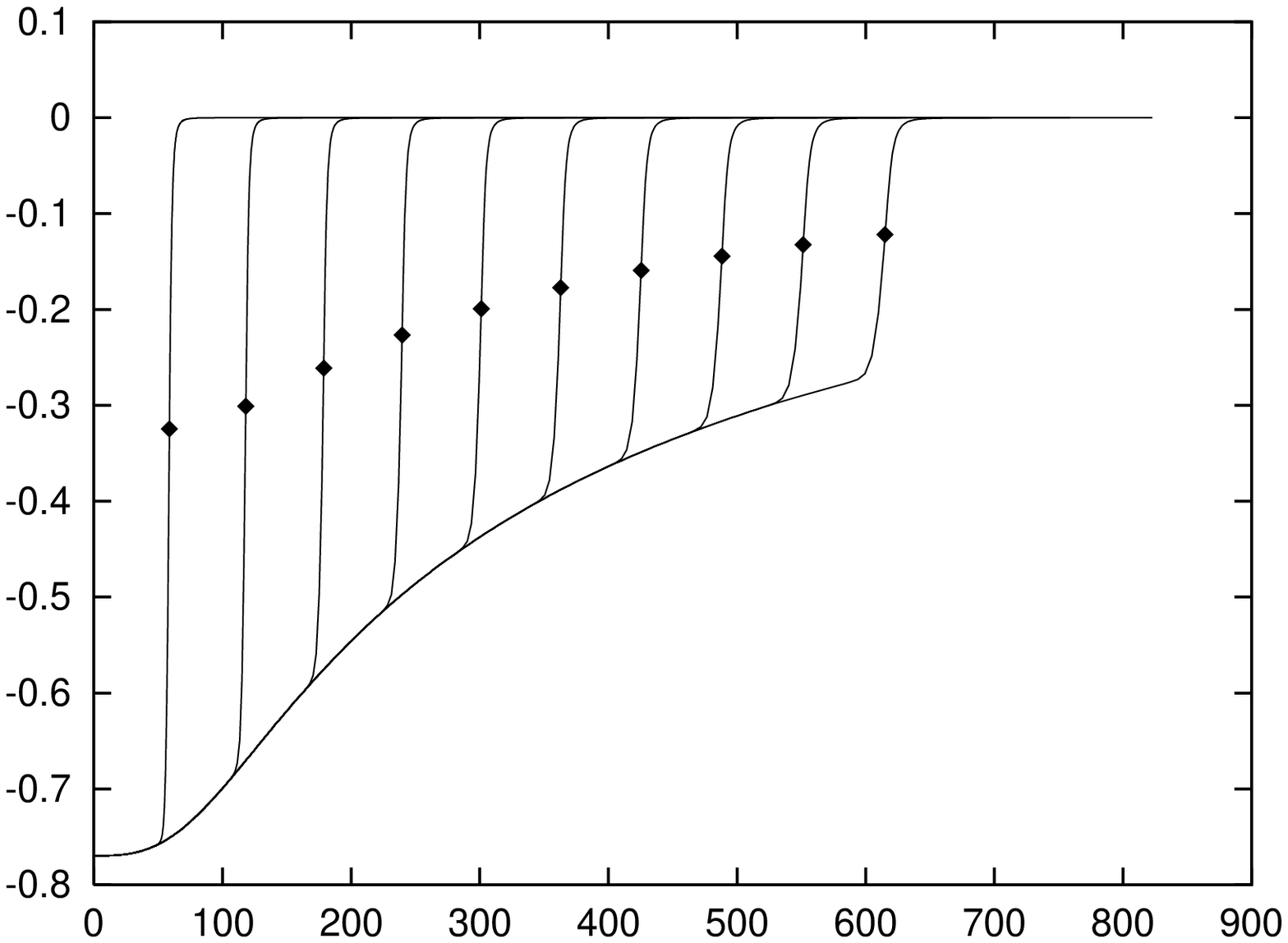 scaled 750}}
\setbox\ylabela=\hbox{\MyBig$K_{zz}$}
\setbox\ylabelb=\hbox{\MyBig$K_{zz}$}
\setbox\ylabela=\hbox{\rotl\ylabela}
\setbox\ylabelb=\hbox{\rotl\ylabelb}
\setbox\boxa=%
\hbox to 18.0cm{
\vtop to 18.5cm{
\Border
\atptalt( 90.0,490.0){\box\imagea}
\atptalt( 90.0,170.0){\box\imageb}
\atptalt(240.0,475.0){\MyBig Proper distance}
\atptalt(240.0,155.0){\MyBig Proper distance}
\atptalt( 80.0,620.0){\box\ylabela}
\atptalt( 80.0,300.0){\box\ylabelb}
\atptalt(185.0,790.0){\MyBig Figure \figdef{KzzRunCD}. Extrinsic curvatures}
\vfill}\hfill}
%
%
\centerline{\box\boxa}\vfill\eject
\SetNoBorder
\SetBase(51,761)
\setbox\imagea=\hbox{\bBoxedEPSF{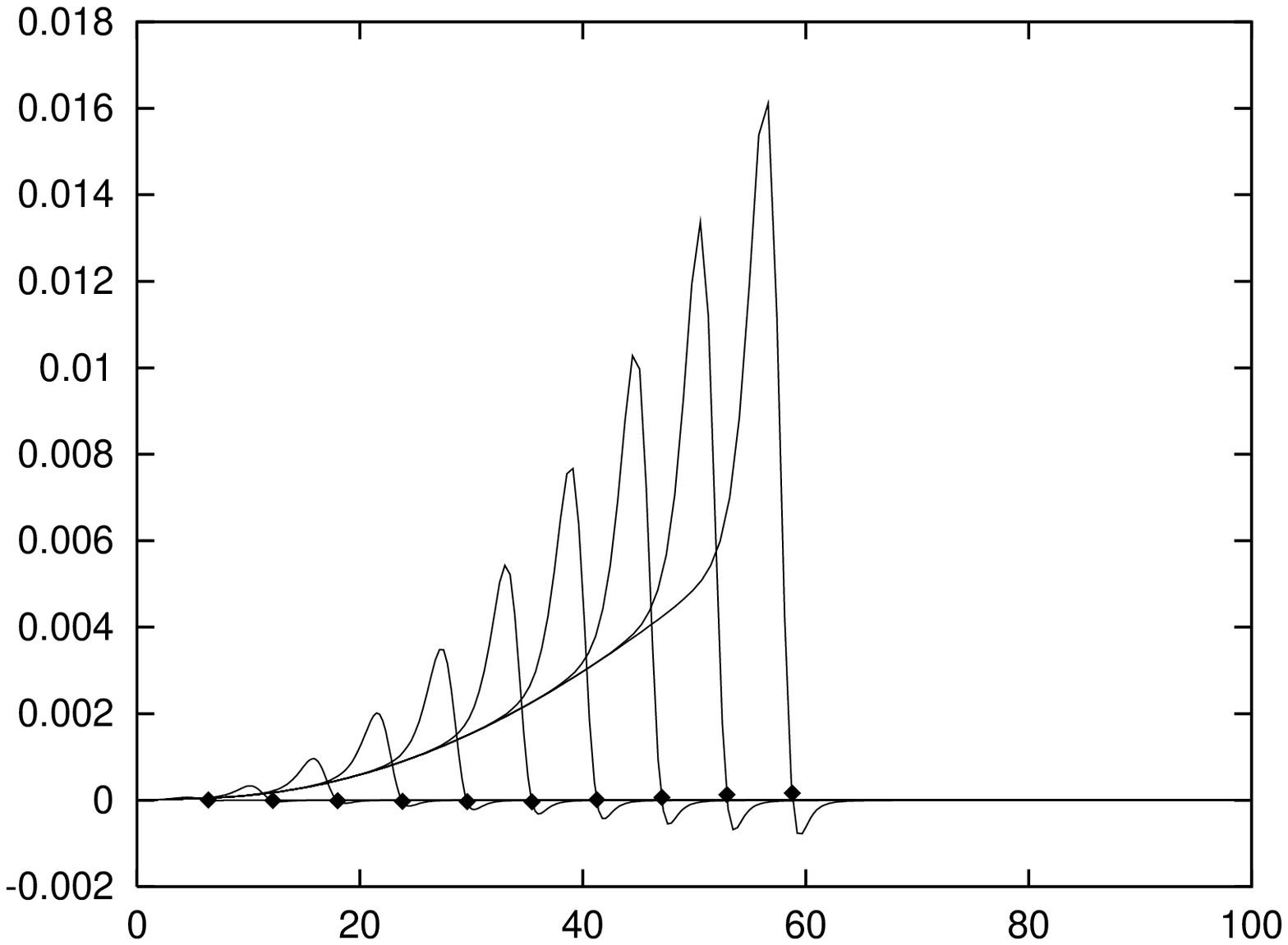 scaled 750}}
\setbox\imageb=\hbox{\bBoxedEPSF{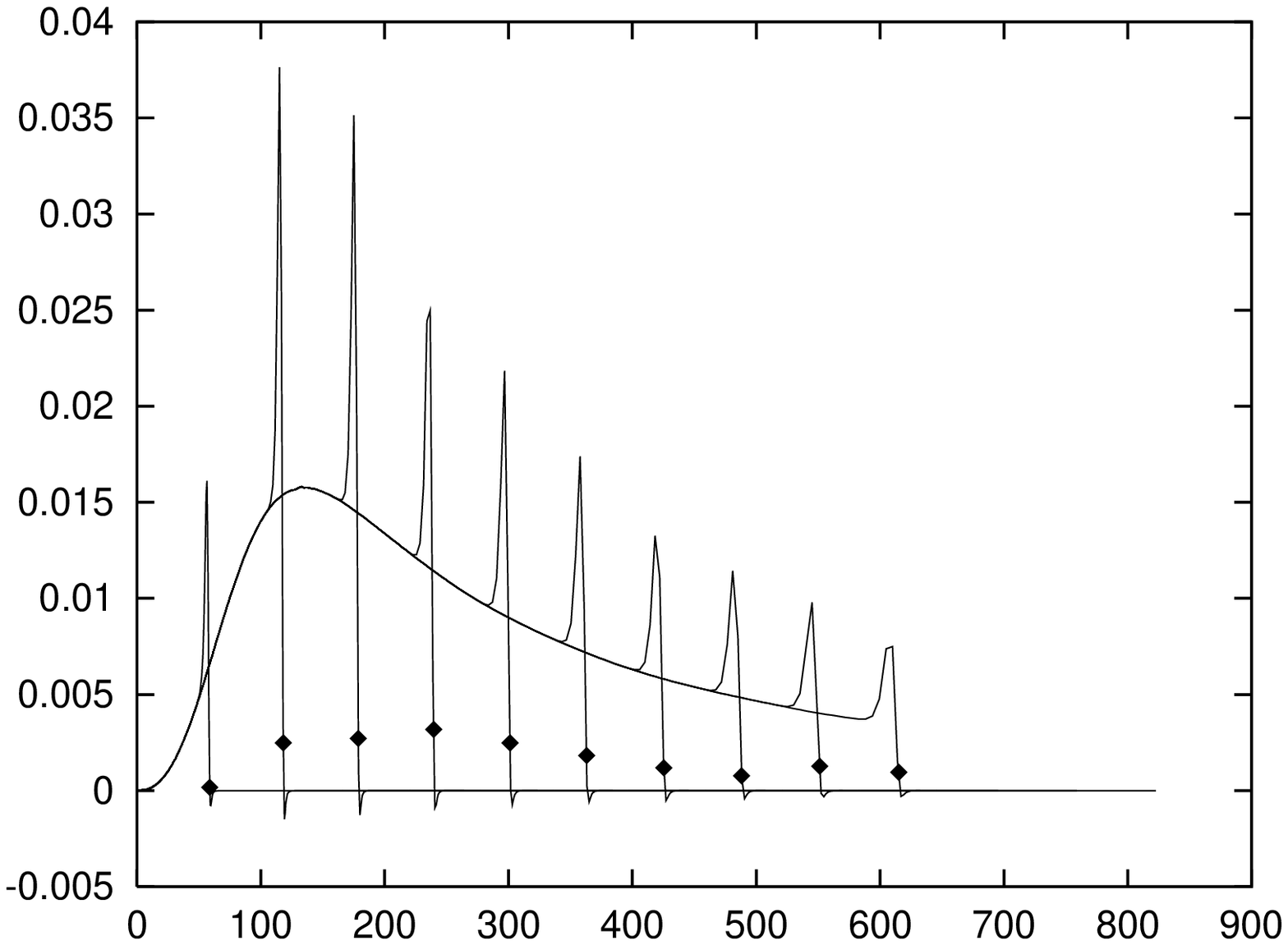 scaled 750}}
\setbox\ylabela=\hbox{\MyBig Hamiltonian}
\setbox\ylabelb=\hbox{\MyBig Hamiltonian}
\setbox\ylabela=\hbox{\rotl\ylabela}
\setbox\ylabelb=\hbox{\rotl\ylabelb}
\setbox\boxa=%
\hbox to 18.0cm{
\vtop to 18.5cm{
\Border
\atptalt( 90.0,490.0){\box\imagea}
\atptalt( 90.0,170.0){\box\imageb}
\atptalt(240.0,475.0){\MyBig Proper distance}
\atptalt(240.0,155.0){\MyBig Proper distance}
\atptalt( 80.0,590.0){\box\ylabela}
\atptalt( 80.0,270.0){\box\ylabelb}
\atptalt(180.0,790.0){\MyBig Figure \figdef{HamCnstrntRunCD}. Hamiltonian constraint}
\vfill}\hfill}
%
%
\centerline{\box\boxa}\vfill\eject
\SetNoBorder
\SetBase(51,761)
\setbox\imagea=\hbox{\bBoxedEPSF{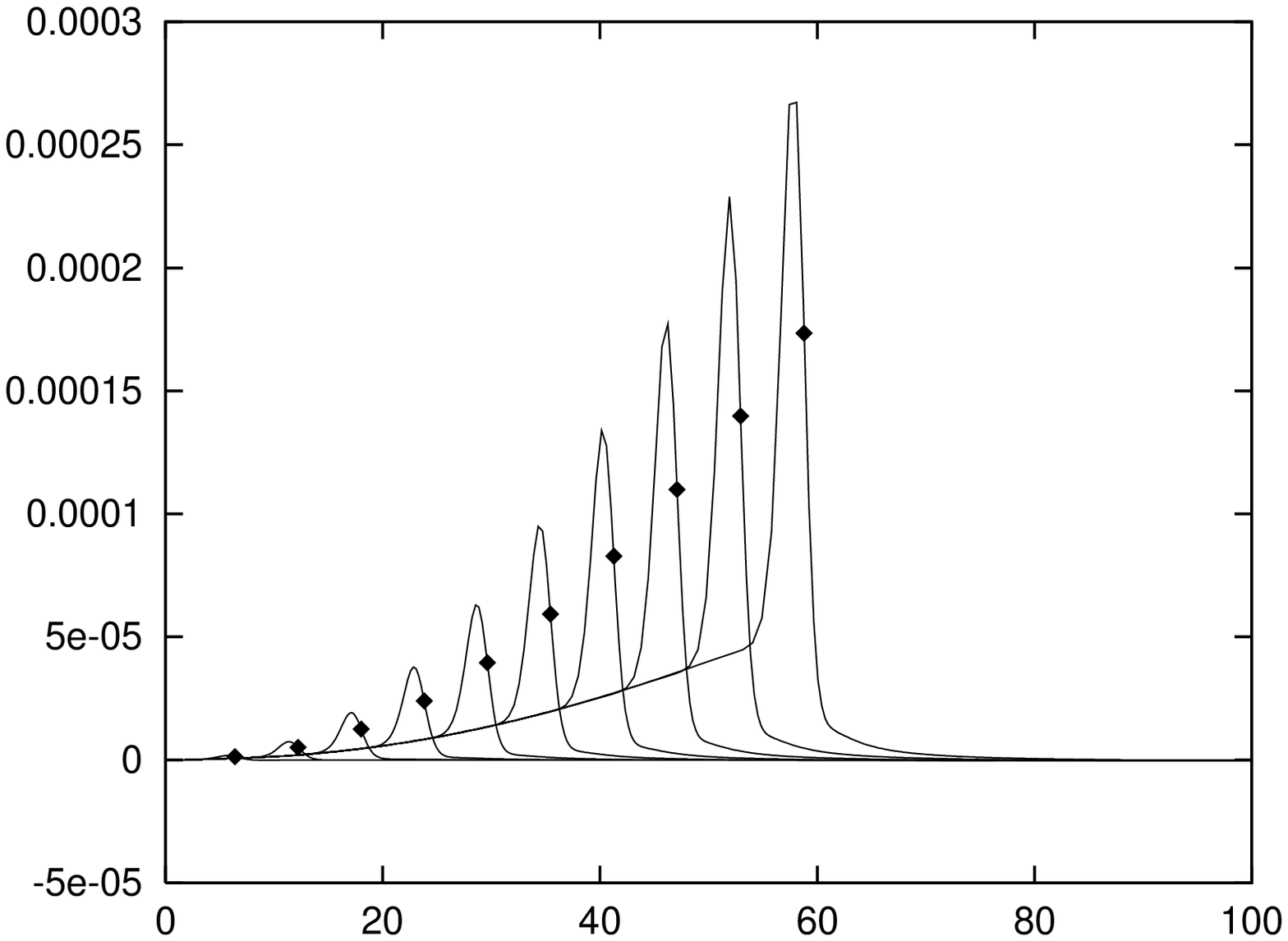 scaled 750}}
\setbox\imageb=\hbox{\bBoxedEPSF{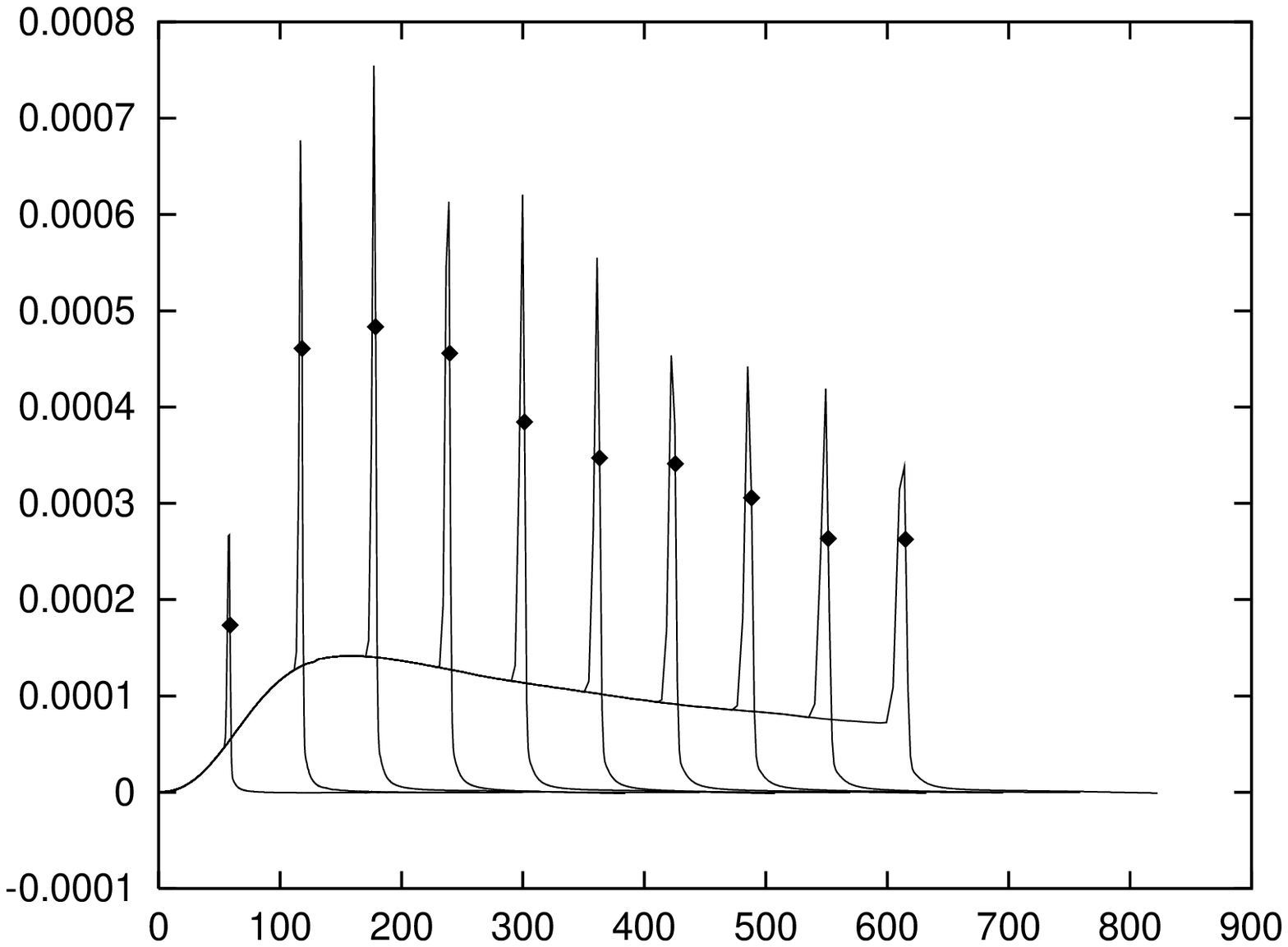 scaled 750}}
\setbox\ylabela=\hbox{\MyBig Momentum}
\setbox\ylabelb=\hbox{\MyBig Momentum}
\setbox\ylabela=\hbox{\rotl\ylabela}
\setbox\ylabelb=\hbox{\rotl\ylabelb}
\setbox\boxa=%
\hbox to 18.0cm{
\vtop to 18.5cm{
\Border
\atptalt( 90.0,490.0){\box\imagea}
\atptalt( 90.0,170.0){\box\imageb}
\atptalt(240.0,475.0){\MyBig Proper distance}
\atptalt(240.0,155.0){\MyBig Proper distance}
\atptalt( 80.0,590.0){\box\ylabela}
\atptalt( 80.0,270.0){\box\ylabelb}
\atptalt(190.0,790.0){\MyBig Figure \figdef{MomCnstrntRunCD}. Momentum constraint}
\vfill}\hfill}
%
%
\centerline{\box\boxa}\vfill\eject
\SetNoBorder
\SetBase(51,761)
\setbox\imagea=\hbox{\bBoxedEPSF{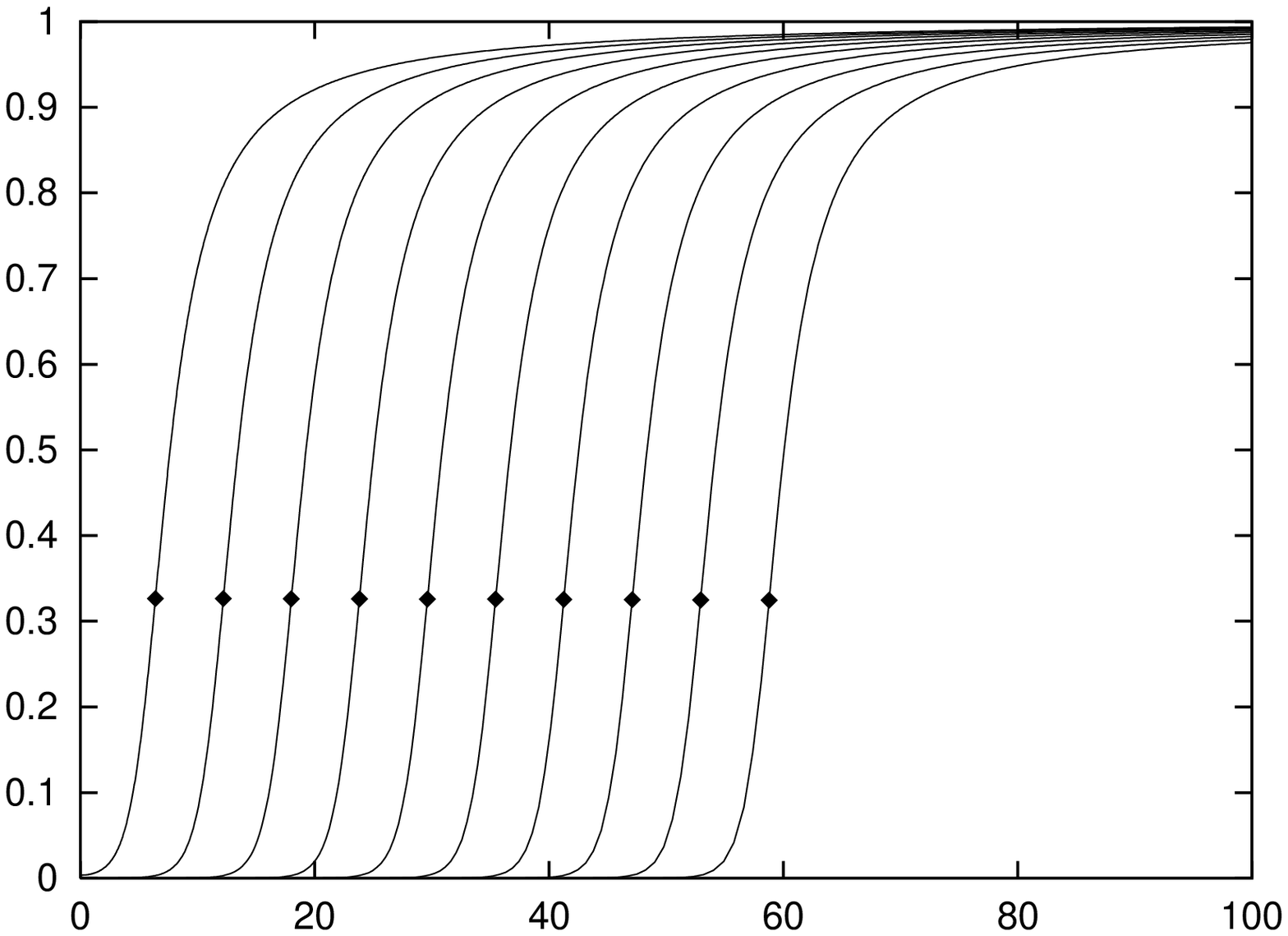 scaled 750}}
\setbox\imageb=\hbox{\bBoxedEPSF{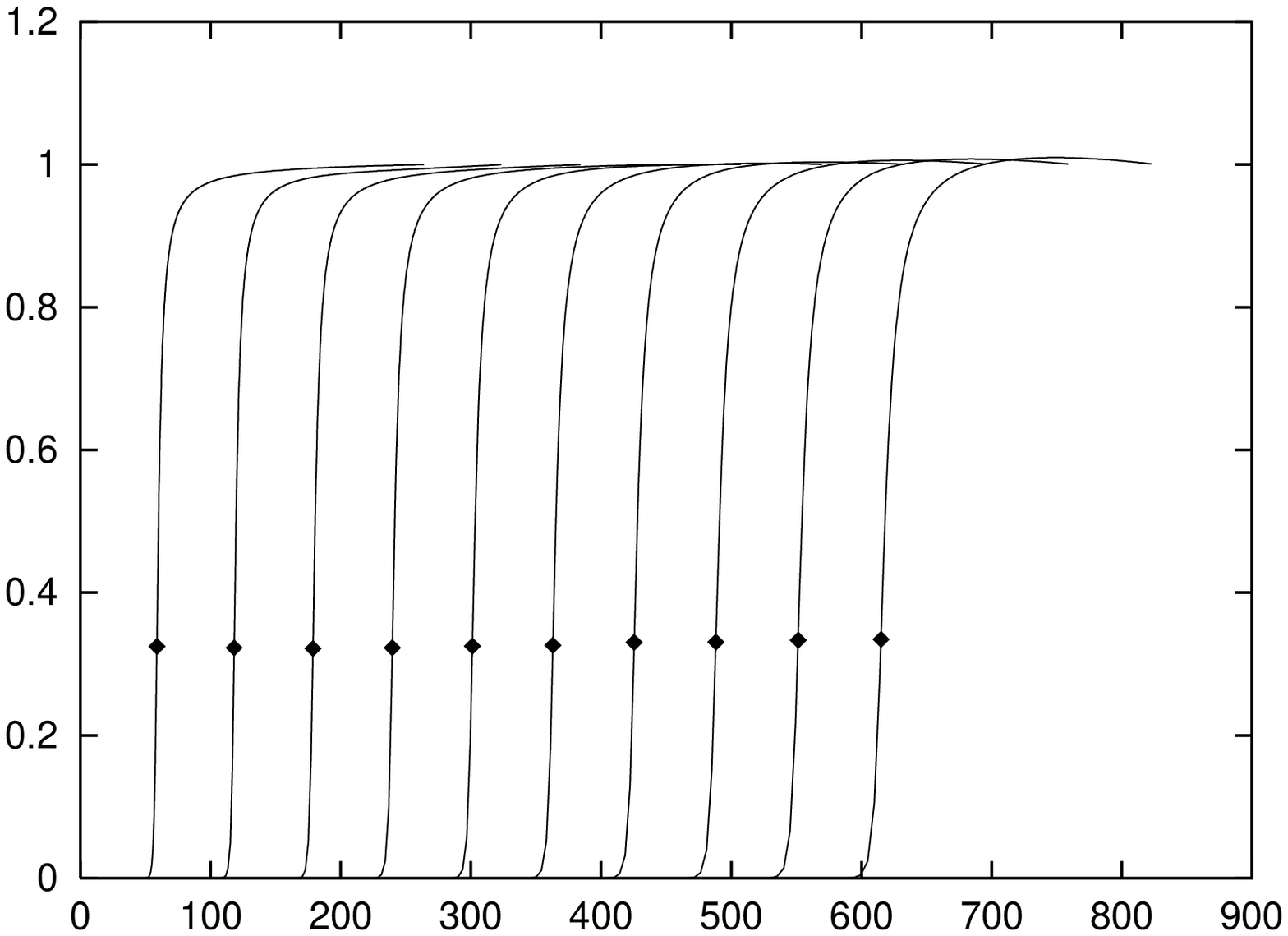 scaled 750}}
\setbox\ylabela=\hbox{\MyBig Lapse}
\setbox\ylabelb=\hbox{\MyBig Lapse}
\setbox\ylabela=\hbox{\rotl\ylabela}
\setbox\ylabelb=\hbox{\rotl\ylabelb}
\setbox\boxa=%
\hbox to 18.0cm{
\vtop to 18.5cm{
\Border
\atptalt( 90.0,490.0){\box\imagea}
\atptalt( 90.0,170.0){\box\imageb}
\atptalt(240.0,475.0){\MyBig Proper distance}
\atptalt(240.0,155.0){\MyBig Proper distance}
\atptalt( 80.0,600.0){\box\ylabela}
\atptalt( 80.0,280.0){\box\ylabelb}
\atptalt(200.0,790.0){\MyBig Figure \figdef{LapseRunCD}. Lapse function}
\vfill}\hfill}
%
%
\centerline{\box\boxa}\vfill\eject
\SetNoBorder
\SetBase(51,761)
\setbox\image=\hbox{\bBoxedEPSF{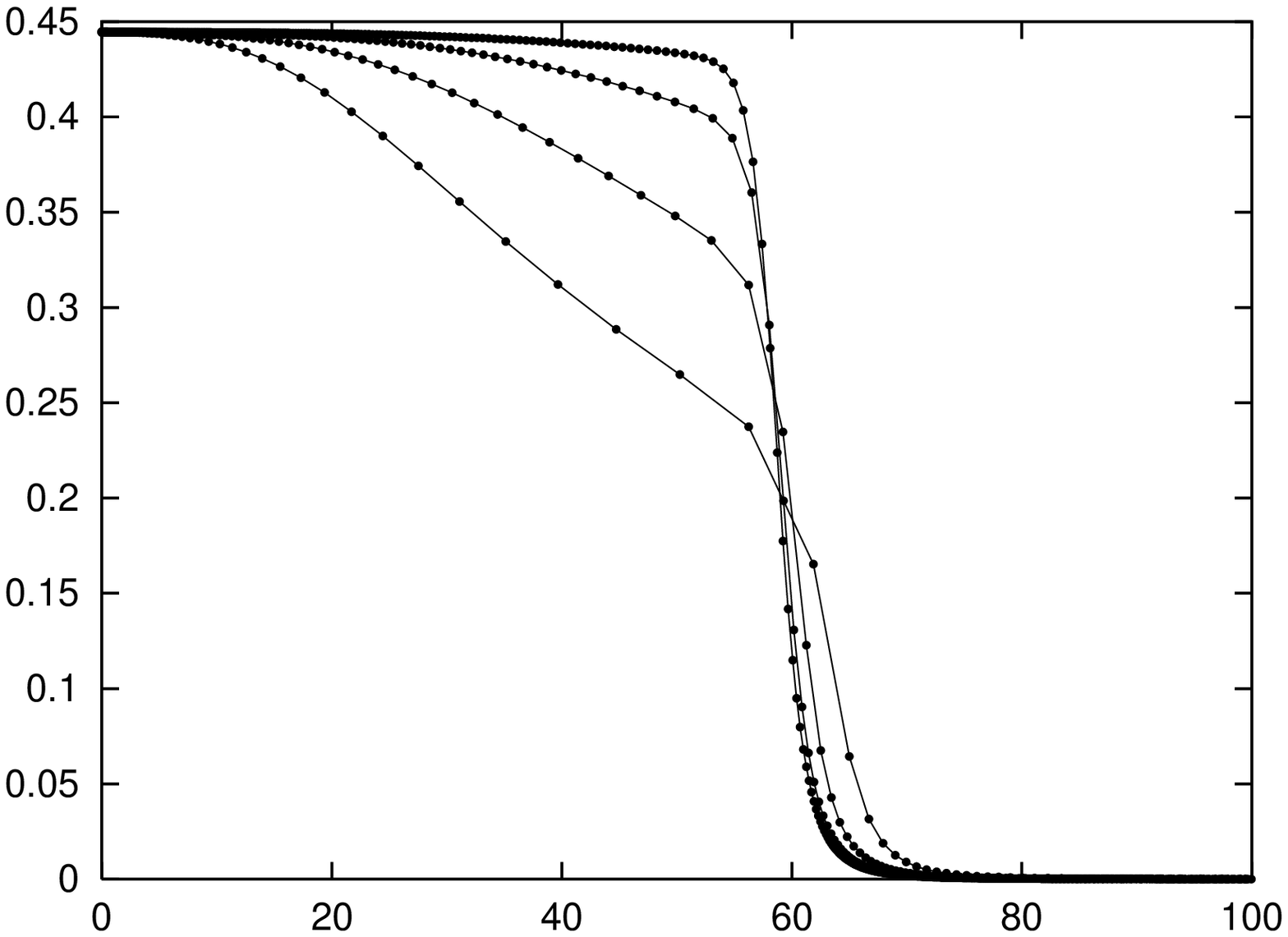 scaled 750}}
\setbox\ylabel=\hbox{\MyBig $R_{xyxy}$}
\setbox\ylabel=\hbox{\rotl\ylabel}
\setbox\boxa=%
\hbox to 18.0cm{
\vtop to 18.5cm{
\Border
\atptalt( 90.0,490.0){\box\image}
\atptalt(240.0,475.0){\MyBig Proper distance}
\atptalt( 80.0,600.0){\box\ylabel}
\atptalt(175.0,790.0){\MyBig Figure \figdef{Resolution}. Resolution}
\vfill}\hfill}
%
%
\centerline{\box\boxa}\vfill\eject
\SetNoBorder
\SetBase(51,761)
\setbox\imagea=\hbox{\bBoxedEPSF{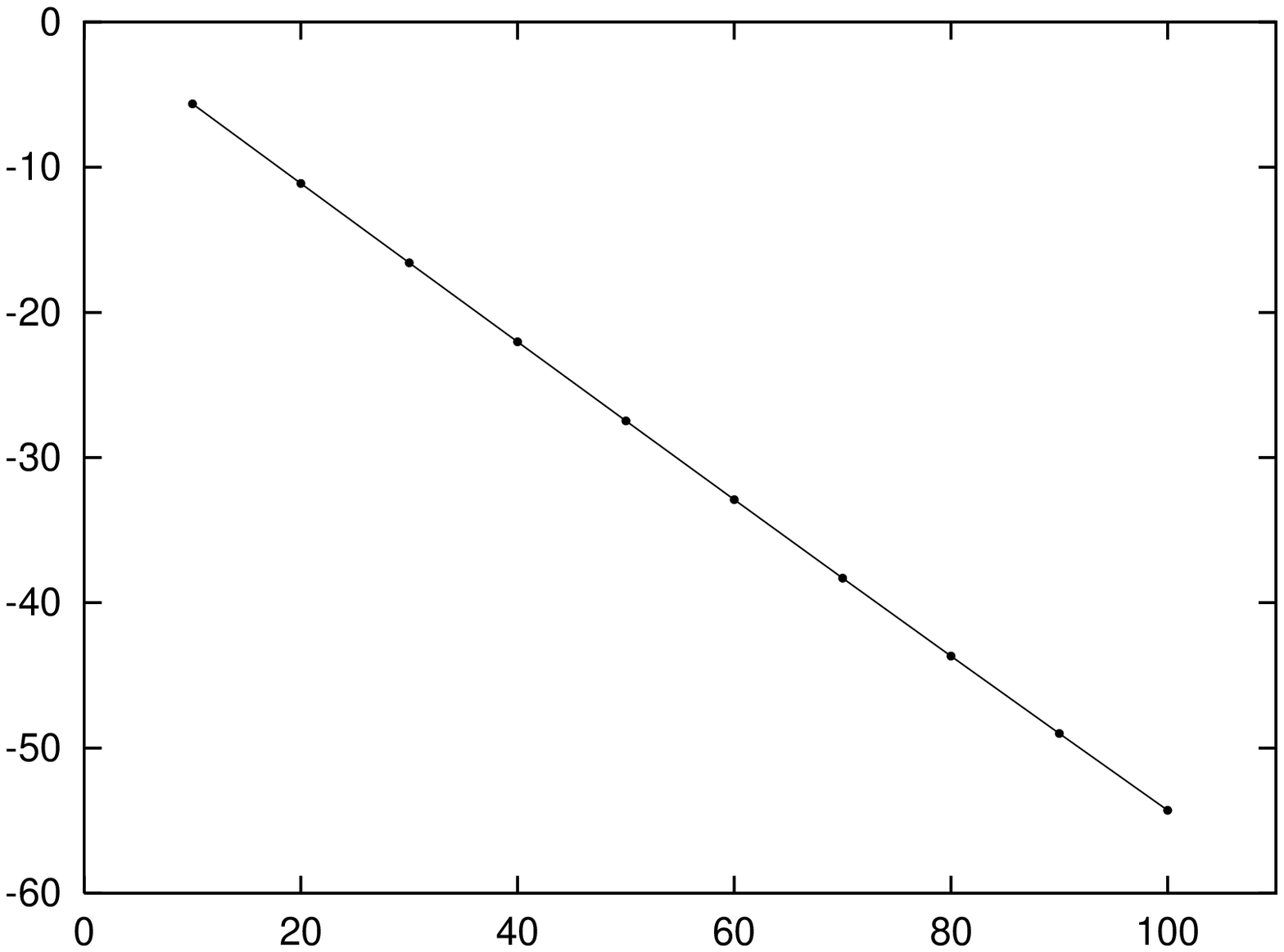 scaled 750}}
\setbox\imageb=\hbox{\bBoxedEPSF{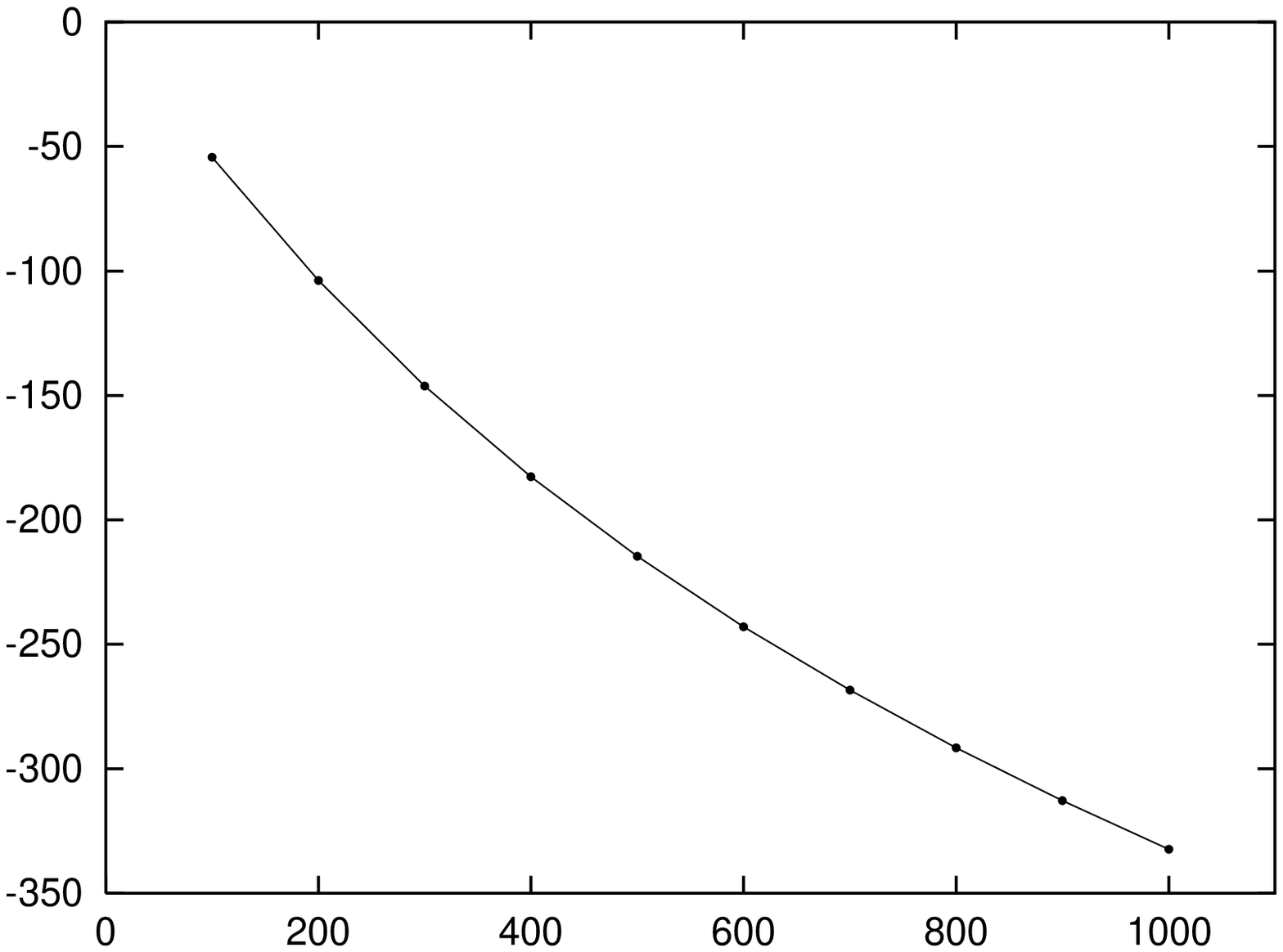 scaled 750}}
\setbox\ylabela=\hbox{\MyBig $\ln N$}
\setbox\ylabelb=\hbox{\MyBig $\ln N$}
\setbox\ylabela=\hbox{\rotl\ylabela}
\setbox\ylabelb=\hbox{\rotl\ylabelb}
\setbox\boxa=%
\hbox to 18.0cm{
\vtop to 18.5cm{
\Border
\atptalt( 90.0,490.0){\box\imagea}
\atptalt( 90.0,170.0){\box\imageb}
\atptalt(240.0,475.0){\MyBig Coordinate time}
\atptalt(240.0,155.0){\MyBig Coordinate time}
\atptalt( 80.0,600.0){\box\ylabela}
\atptalt( 80.0,280.0){\box\ylabelb}
\atptalt(175.0,790.0){\MyBig Figure \figdef{Collapse}. Collapse of the lapse}
\vfill}\hfill}
%
%
\centerline{\box\boxa}\vfill\eject
\SetNoBorder
\SetBase(51,761)
\setbox\image=\hbox{\bBoxedEPSF{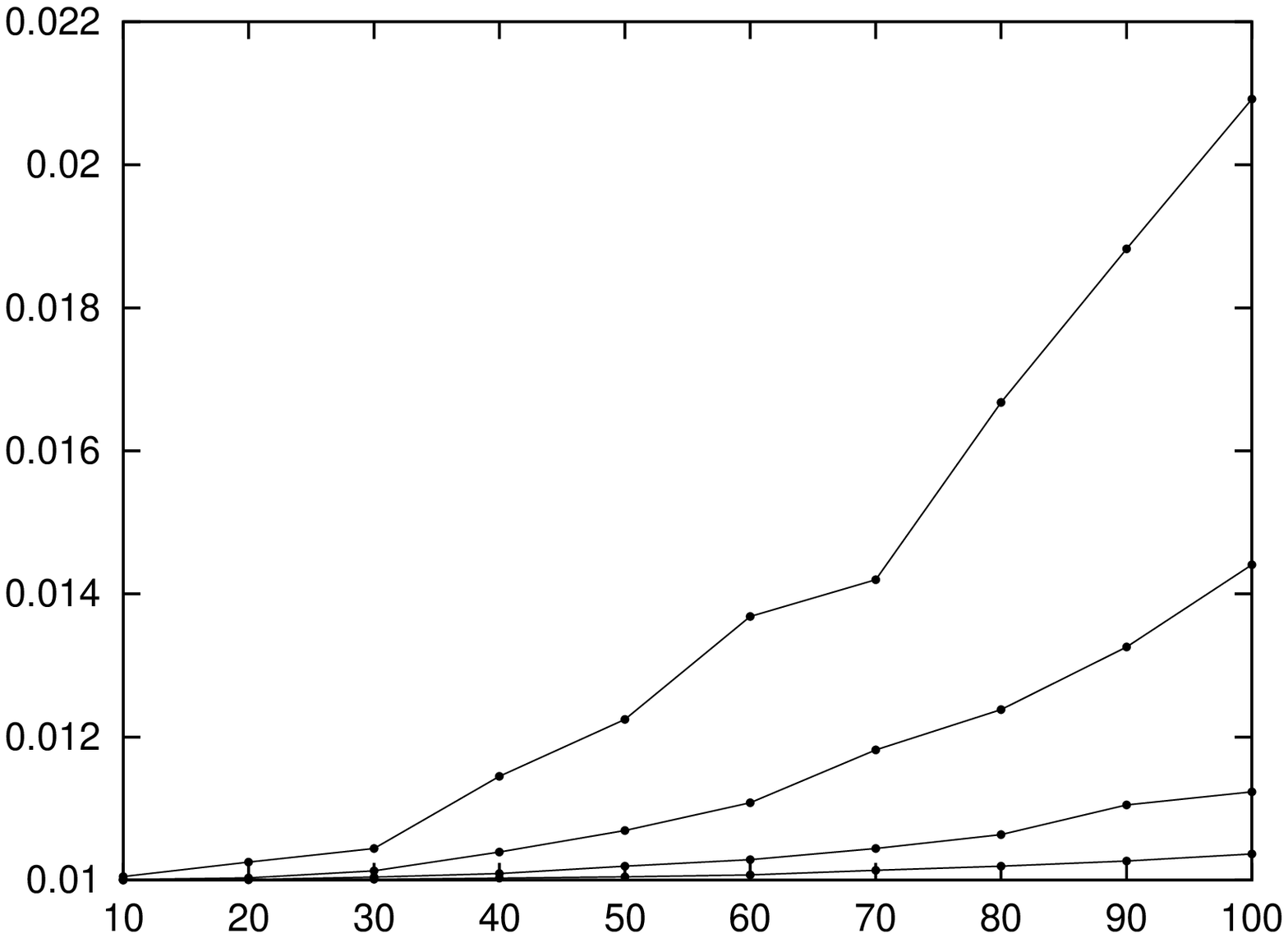 scaled 750}}
\setbox\ylabel=\hbox{\MyBig $L^2_{xx}$}
\setbox\ylabel=\hbox{\rotl\ylabel}
\setbox\boxa=%
\hbox to 18.0cm{
\vtop to 18.5cm{
\Border
\atptalt( 90.0,490.0){\box\image}
\atptalt(240.0,475.0){\MyBig Proper distance}
\atptalt( 80.0,600.0){\box\ylabel}
\atptalt(175.0,790.0){\MyBig Figure \figdef{Horizon}. The apparent horizon}
\vfill}\hfill}
%
%
\centerline{\box\boxa}\vfill\eject

\bye